\documentclass[reprint,amsmath,amssymb,aps,prb]{revtex4-2}

\usepackage{color}  									
\usepackage{graphicx}									
\makeatletter
\let\old@makecaption=\@makecaption
\usepackage{subcaption}
\let\@makecaption=\old@makecaption
\makeatother

\usepackage{dcolumn}
\usepackage{ragged2e} 									
\usepackage{float} 										
\usepackage{placeins}                                   
\restylefloat{table}

\usepackage{mathtools}			  
\usepackage{hyperref}
\hypersetup{colorlinks=true,urlcolor=blue,citecolor=blue,linkcolor=blue,breaklinks=true}
\usepackage{cleveref}
\usepackage{soul}
\usepackage{dcolumn}
\usepackage{threeparttable}
\usepackage{multirow}
\usepackage{booktabs}

\newcommand{\pw}[1]{}
\newcommand{\cw}[1]{}
\newcommand{\ak}[1]{}
\newcommand{\kh}[1]{}
\newcommand{\mpi}[1]{}
\renewcommand{\eqref}[1]{Eq.~(\ref{#1})}

\newcommand{\me}{\mathrm{e}}
\newcommand{\md}{d}
\newcommand{\mi}{i}

\usepackage{scalerel,stackengine,amsmath}

\DeclareMathAlphabet{\pazocal}{OMS}{zplm}{m}{n}         
\AtBeginDocument{ 
    \heavyrulewidth=.08em
    \lightrulewidth=.05em
    \cmidrulewidth=.03em
    \belowrulesep=.65ex
    \belowbottomsep=0pt
    \aboverulesep=.4ex
    \abovetopsep=0pt
    \cmidrulesep=\doublerulesep
    \cmidrulekern=.5em
    \defaultaddspace=.5em
}
\begin{document}
    
    
    \title{Broadening and sharpening of the Drude peak through antiferromagnetic fluctuations}
    
    
    \author{Paul Worm}
    \thanks{These authors contributed equally to this work.}
    \author{Clemens Watzenb\"ock}
    \thanks{These authors contributed equally to this work.}
    \author{Matthias Pickem}
    \author{Anna Kauch}
    \email[]{kauch@ifp.tuwien.ac.at}
    \author{Karsten Held}
    
    \affiliation{TU Wien, Institute of Solid State Physics, Vienna, Austria}
    
    
    \date{\today}
    
    \begin{abstract}
        Antiferromagnetic or charge density wave fluctuations couple with light through the recently discovered 
        $\pi$-ton contribution to the optical conductivity, and quite generically constitute the dominant vertex corrections in low-dimensional correlated electron systems. 
        Here we study the arguably simplest version of these $\pi$-tons based on the semi-analytical random phase approximation (RPA) ladder  in the transversal particle-hole channel. The vertex corrections to the optical conductivity are calculated directly for real frequencies. We validate that the RPA qualitatively reproduces the $\pi$-ton vertex corrections to the Drude peak in the Hubbard model. Depending on the temperature we find vertex corrections to broaden or sharpen the Drude peak. 
    \end{abstract}
    
    
    \maketitle
    
    
    \section{Introduction}
\begin{figure}[t]
        \centering
        \includegraphics[width=0.5\textwidth]{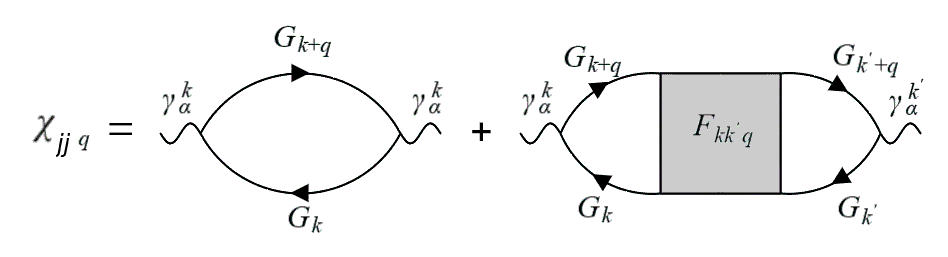}
        \caption{Diagrammatic representation of the current-current correlation function $\chi$ directly related to the optical conductivity through Eq.~(\ref{eq:opt_cond}) below, where we also introduce the indices. The diagram on the left denotes the 'bubble' contribution with one electron and one hole described by Green's functions $G$ propagating backwards and forward (in a time representation); the one on the right represents the vertex $F$ corrections which includes all Feynman diagrams connecting the upper and lower Green function line through interactions. }
        \label{fig:opt_cond}
    \end{figure}

    The optical conductivity is one of the primary information sources for the electron  dynamics in solids. However, there is no momentum resolution,
    and the interpretation of optical spectra is further bedeviled by the fact that it cannot always be interpreted in a simple one-particle picture. Exciting the solid by light leaves the number of electrons unchanged so that in a simple one-particle picture one excites a particle from an occupied to an unoccupied state, depositing the photon energy. This is the 'bubble' electron-hole contribution to the optical conductivity, see left Feynman diagram in Fig.~\ref{fig:opt_cond}. In case of a metal it leads to the Drude peak with a maximum around zero frequency ($\omega=0$) in the real part of the optical conductivity  $\sigma(\omega)$ and a broadening given by the scattering rate $\tau^{-1}$:
    \begin{equation}
    \sigma(\omega)= \Re \biggr( \frac{\sigma_0}{1- i\omega \tau}\biggr) = \frac{\sigma_0}{1+\omega^2 \tau^2}
    \label{Eq:Drude}
    \end{equation}
    Drude~\cite{Drude1900a,Drude1900b} derived Eq.~(\ref{Eq:Drude}) with  the DC conductivity given by $\sigma_0=n^2\tau/m^*$ [$n$: electron density; $m$: (effective) mass; $e$: elementary (electron) charge] in 1900 even prior to the invention of quantum mechanics  which, as shown by Sommerfeld \cite{Sommerfeld1928},
    does not alter Eq.~(\ref{Eq:Drude}) for free electrons.
    
    Besides bandstructure effects such as additional interband transitions and momentum-dependent (group) velocities (or dipole matrix elements) $\gamma$ in the solid, there are many-body corrections to the Drude-Sommerfeld model due to the interaction of the excited electron and hole with each other and the rest of the solid. In terms of Feynman diagrams these are known as vertex corrections to the bare bubble contribution of an independent (quasi-)electron and (quasi-)hole, as the vertex $F$ in the second Feynman diagram of  Fig.~\ref{fig:opt_cond} comprises all kinds of such interactions. 
    
    These many-body vertex corrections can give rise to shifts and additional peaks in the optical conductivity. Arguably  most prominent is the exciton
    \cite{Frenkel1931,Wannier37}. The importance of vertex corrections for the DC conductivity has also been recognized early on for metals with low density of electrons, where umklapp scattering is suppressed at low temperature because of a small Fermi surface~\cite{Fukuyama1997,Peierls1930,Kontani2008, Maslov2017}.
    
    An exciton can simply be understood as the binding of the electron and hole  through their attractive Coulomb interaction in a semiconductor. For strongly correlated electron systems such as transition metal oxides, it is known that the  dominant vertex contributions are often  antiferromagnetic (AFM) or charge density wave (CDW) fluctuations which have a wave vector $\mathbf{k}=(\pi,\pi,\ldots)$ \cite{Keimer2013,Aeppli2014}. Since (optical) light can only excite an electron and hole with a total momentum $\mathbf{q}=0$, such fluctuations can, in contrast to the exciton or ferromagnetic fluctuations, not directly couple to the electron-hole pair.
    
\begin{figure}[t]
        \centering
        \includegraphics[width=0.5\textwidth]{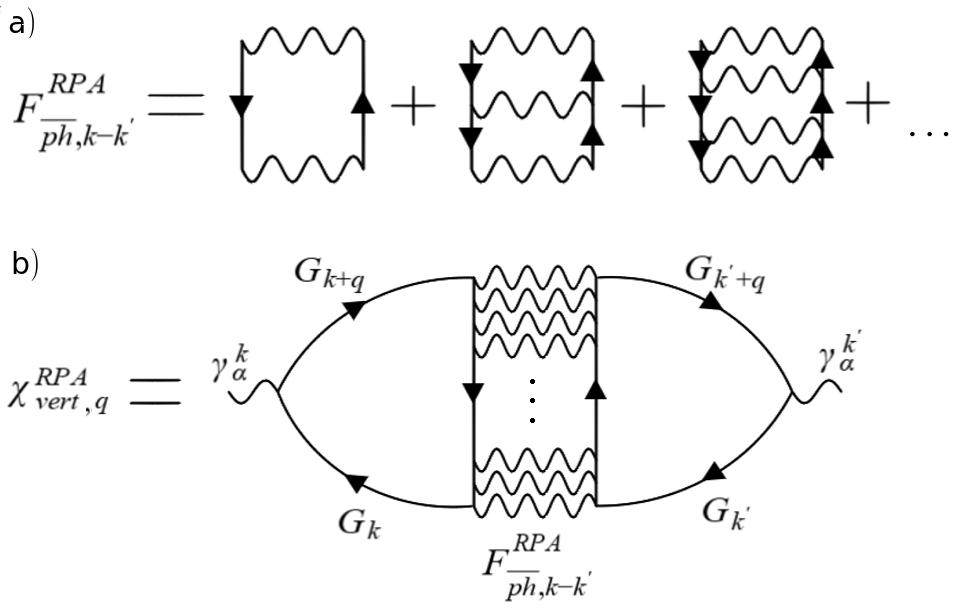}
        \caption{Diagrammatic representation of the $\pi$-ton contribution to the optical conductivity. The RPA ladder in the transversal particle-hole ($\overline{ph}$) channel shown in (a) generates AFM and CDW fluctuations and contributes to the optical conductivity or current-current correlation function shown in (b) at the momenta $\mathbf{k}-\mathbf{k'}=(\pi,\pi,\ldots)$ and $\mathbf{q}=0$.}
        \label{fig:rpa_ladder}
    \end{figure}

    As was discovered in Ref.~\onlinecite{pi_ton},  a more complicated process involving at least two (quasi-)electrons and two (quasi-)holes is necessary. The arguably easiest way to understand these vertex corrections to the optical conductivity, coined $\pi$-tons~\cite{pi_ton},  is the random phase approximation (RPA) visualized in Fig.~\ref{fig:rpa_ladder}. There are two particle-hole pairs coupled to the incoming and outgoing light---glued together by AFM and CDW fluctuations in the so-called transversal particle-hole channel.
    While earlier, pioneering  studies~\cite{Fukuyama1997, Clarke1993,Wrobel2001,Essler2001,Jeckelmann2003,Kontani2006,Gull2009a,Chubukov2014,Tremblay2011,Mravlje2018} reported vertex corrections to the optical conductivity, the  importance of $\pi$-ton contributions was only realized recently and generically occurs in various models of correlated electrons~\cite{pi_ton,Pudleiner2019,Astleithner2020}. 
    These papers used quite involved numerical techniques such as the parquet approximation~\cite{Bickers2004,Yang2009,Li2016,victory} or the parquet dynamical vertex approximation~\cite{Valli2015,Rohringer_Diagrammatic_extensions}~\footnote{The numerical bottleneck is in particle the memory consumption of a general vertex depending on three frequencies and three momenta, which might be mitigated in the future by an efficient sparse parameterization thereof~\cite{Honerkamp2018,Shinaoka2020,Li2020,Eckhardt2020,Krien2020}}, which include also many contributions beyond the RPA  in the transversal particle-hole channel and require a cumbersome and error-prone analytical continuation from Matsubara to real frequencies.
    
    In this paper, we hence use a simplified approach and study the  RPA $\pi$-ton contribution of Fig.~\ref{fig:rpa_ladder} directly for real frequencies with a simple metallic Green's function. We show that the optical conductivity with these  $\pi$-ton vertex corrections included is still described qualitatively by the Drude peak  in the metallic phase of the Hubbard model. While weak localization vertex corrections~\cite{Altshuler1985}  which originate from a particle-particle ladder for disordered systems, are known to broaden the Drude peak, we find that $\pi$-ton contributions---depending on the temperature---either broaden or sharpen the Drude peak compared to the expected width from the single-particle scattering time determined by the self-energy. 
    Similar temperature dependent broadening and sharpening is also brought by $\pi$-ton contributions in parquet dynamical vertex approximation (parquet D$\Gamma$A) which we show for comparison as well.  

  Using a simplified, almost toy-model setup for calculating the ladder vertex corrections in the $\overline{ph}$ channel has several advantages: (i) We analyze only a selected class of diagrams built with a featureless Green's function which allows us to recognize what changes these diagrams imply for the  optical conductivity, which without vertex corrections is  featureless and described by a Drude peak. (ii) The RPA-ladder vertex has a simplified structure in frequency and momenta, which makes the calculation numerically less demanding and also feasible directly on the real frequency axis with sufficient momentum resolution. (iii) The parameters of the calculation, i.e. the one-particle scattering rate, temperature, effective interaction, can be changed independently and used as knobs to turn for identifying what has the biggest effect on the vertex corrections.
   
  The paper is organized as follows: In Sec.~\ref{sec:method} we introduce the model and describe in detail the calculation scheme for the vertex corrections to the optical conductivity on the real frequency axis, which requires to properly account for altogether five branch cuts;  a detailed derivation is given in Appendix~\ref{app:derivations}. In Secs.~\ref{sec:results}A and B we present the numerical results of the optical conductivity with two types of vertex corrections: RPA and Ornstein-Zernike form. For numerical details of the computations, see Appendix~\ref{app:numerical}.  In the final part, Sec.~\ref{sec:results}C, we compare the results obtained with those of the parquet D$\Gamma$A method. We summarize the results in Sec.~\ref{sec:conclusions}.
  The main part is supplemented further by the Ornstein-Zernike analysis at high temperatures in Appendix \ref{app:OZ_high_temp}; an overview of possible RPA-ladder diagrams in Appendix~\ref{app:rpa_ladders};  the Schwinger-Dyson equation in Appendix~\ref{app:Schwinger_Dyson_real}
  for recalculating the self-energy with results thus obtained  shown in Appendix ~\ref{app:sde_selfenergy}.

    \section{Model and method}
    \label{sec:method}
    We consider the single-orbital Hubbard
    model,
    \begin{equation}
    \label{Eq:hamiltonian}
    \pazocal{H} =  \sum_{ij\sigma} t_{ij} c^\dagger_{j\sigma}c^{\phantom\dagger}_{i\sigma} + U \sum_{i } {n}_{i \uparrow}{n}_{i \downarrow},
    \end{equation}
    with $c_{i\sigma}$ ($c^{\dagger}_{i\sigma}$) denoting the fermionic  annihilation (creation) operator at site $i$ with spin $\sigma$; $n_i=c^\dagger_{i\sigma}c^{\phantom\dagger}_{i\sigma}$ is the particle number operator.
    The two terms of the Hamiltonian are  the on-site Coulomb interaction $U$ and the one-particle   hopping amplitudes $t_{ij}$ from site $i$ to site $j$. We consider a square lattice with nearest-neighbour hopping only and the dispersion relation is thus given by 
    $\epsilon_{\textbf{k}} =-2 \ t\ \big( \text{cos}(k_x) + \text{cos}(k_y) \big),$ with $-t$ being the hopping amplitude between nearest neighbors. In the following we set the $t\equiv1$ as the energy unit, $k_B\equiv 1$ for the unit of temperature and the lattice constant is $a\equiv1$. The frequency unit is set equal to the energy unit by setting $\hbar=1$.   Since we are interested in the regime with strong antiferromagnetic correlations, we consider the half-filled case with an average of $n=1$ electrons per site.

    \subsection{Optical conductivity}
    \label{sec:opt_cond}
    We are interested in the real part of the optical conductivity  
    \begin{equation}
    \label{eq:opt_cond}
    \sigma(\omega)  = \pazocal{P} \frac{\Im (\chi^{{\bf q}=0}_{jj,\omega})}{\omega}-\pi \delta(\omega) \left[ \Re (\chi^{{\bf q}=0}_{jj,\omega})-\frac{e^2 n_{q=0}}{m} \right],
    \end{equation} 
    where $\chi^{\bf q}_{jj,\omega}$ is  the current-current correlation function. The (purely static) diamagnetic part $\pi \delta(\omega) [ \Re (\chi^{{\bf q}=0}_{jj,\omega})-\frac{q^2 n_{q=0}}{m} ]$ vanishes because $U(1)$ gauge invariance implies for longitudinal electrical fields a prefect cancellation between the paramagnetic current-current correlation-function at $\omega=0$ and diamagnetic term \cite[Ch.~7.3.3]{Rammer1998QuantumTransportTheory}.
    
    The current-current correlation function can be diagrammatically represented in the following way~(for a derivation see e.g.~Ref.~\onlinecite{Tremblay2011}):
    \begin{equation}
    \label{eq:current_current_correlator_F}
    \begin{split}
    \chi^{\bf q}_{jj,\omega} = & -\frac{2}{\beta} \sum_{k} \gamma_{\alpha}^{\textbf{kq}}\gamma_{\alpha}^{\textbf{k}(-\textbf{q})}  G_{k+q} G_{k} \\ & -\frac{2}{\beta^2} \sum_{kk'} \gamma_{\alpha}^{\textbf{kq}} \gamma_{\alpha}^{\textbf{k'}(-\textbf{q})} G_{k} G_{k+q} F_{kk'q} G_{k'} G_{k'+q} \\ \equiv & \; \; \; \chi_{\rm{bub}} + \chi_{\rm vert}.\end{split}
    \end{equation}
    Here, we use the combined frequency and momentum indices $k=({\bf k},\nu)$ with fermionic Matsubara frequency $\nu$ and $q=({\bf q},\omega)$ with bosonic Matsubara frequency $\omega$;   $\beta=1/T$ is the inverse temperature and we implicitly assume a factor of $1/N$ in front of every momentum sum, $N$ being the number of lattice sites or $\textbf{k}$-points.  We also set the electron charge $e\equiv1$, thus avoiding a global prefactor $e^2/(\hbar^2 a^3)$ in our dimensionless units. The vertex of the electron-light interaction, $\gamma_{\alpha}^{\textbf{kq}}$, is for  ${\bf q}=0$ given in the Peierls approximation~\cite{Peierls1933} by $\gamma_{\alpha}^{\textbf{kq}=0}=\frac{\partial \epsilon_{\textbf{k}}}{\partial \textbf{k}_{\alpha}}\equiv\gamma_{\alpha}^{\textbf{k}}$, with $\alpha=x,y$ denoting the direction in space. In general $\chi^{\bf q}_{jj,\omega}$ depends also on $\alpha$, but since we are interested in the square lattice, the results do not depend on whether we choose $\alpha=x$ or $\alpha=y$ and drop the index $\alpha$.

    The first term in \eqref{eq:current_current_correlator_F} is often denoted as 'bubble' diagram $\chi_{\rm bub}$ (see Fig.~\ref{fig:opt_cond}) and contains only the Green's functions $G_k$. The second term are the vertex corrections $\chi_{\rm vert}$ with $F_{kk'q}$ being the full two-particle vertex in the density channel~\cite{Rohringer_Diagrammatic_extensions}.

    \subsection{Calculation of the optical conductivity with vertex corrections on the real frequency axis}
    \label{sec:methods_real_freq_vertex}
    In the following we will derive analytical expressions for the vertex corrections to the optical conductivity directly for real frequencies, assuming a simplified form of the vertex. To this aim we will transform the Matsubara frequency sums in \eqref{eq:current_current_correlator_F} into real frequency integrals by contour integration. For the bubble-diagram this has been done before (see e.g.~Ref.~\onlinecite{altland_simons_2010} and the calculation in Appendix~\ref{app:derivations}) and yields
    \begin{equation}
    \begin{split}
    \chi_{\text{bub}}^{R} & (\omega,\textbf{q}=0) = \\ &-2 \sum_{\textbf{k}} \big[\gamma_{\alpha}^{\textbf{k}} \big]^2 \int_{-\infty}^{+\infty} d\nu \ \eta_{F}(\nu) A_{\nu}^{\textbf{k}} \big[ G_{\nu + \omega}^{R \: \textbf{k}}  + G_{\nu-\omega }^{A \: \textbf{k}} \big],
    \end{split}
    \label{eq:bc_ana_cont_2}
    \end{equation}
    where $\eta_{F}(\nu)$ is the Fermi distribution function, \mbox{$ G_{\nu }^{\text{R} \: \textbf{k}}$  } (\mbox{$ G_{\nu }^{\text{A} \: \textbf{k}}$  }) is the retarded (advanced) Green's function, and \mbox{$A_{\nu}^{\textbf{k}}=-\frac{1}{\pi} \Im G_{\nu}^{R \:\textbf{k}}$} is the spectral function. 
    
    In order to analogously calculate the vertex corrections on the real axis one needs to know the vertex itself for real frequencies. This is generally not the case, but since we want to specifically address the $\pi$-ton contributions, we assume a simplified form of the full vertex
    \begin{equation}
    \label{eq:simplified_vertex}
    F_{kk'q}\equiv F_{k-k'},  
    \end{equation}
    which only depends on the total frequency-momentum $k-k'$ in the $\overline{ph}$-channel (this is precisely the form of the RPA ladder vertex in Fig.~\ref{fig:rpa_ladder}).
    In this way the complex structure of $F$ is the same as that of a physical susceptibility (transformed into the $\overline{ph}$-channel). This implies that $F^\textbf{k}_{z}$ is analytic in the whole complex plane except for $\Im z=0$ where it has a branch cut. The analytical continuation from the upper (lower) complex half-plane gives then a retarded (advanced) vertex $F^R$ $(F^A)$ on the real axis. The RPA-vertex (depicted in Fig.~\ref{fig:rpa_ladder}) as well as the Ornstein-Zernike form we will use later fulfill this assumption. For any vertex that fulfills \eqref{eq:simplified_vertex} the vertex contributions to the current-current correlation function in \eqref{eq:current_current_correlator_F}  can be expressed as  
    \begin{equation}
    \begin{split}
    \chi^{R}_{\text{vert}}&(\omega,\textbf{q} = 0) =\\&{2} \sum_{\textbf{k}\textbf{k}'}  \gamma_{\alpha}^\textbf{k} \gamma_{\alpha}^{\textbf{k}'} \left(
    \zeta_1^{\textbf{k}\textbf{k}'}(\omega) + 
    \zeta_{2p3}^{\textbf{k}\textbf{k}'}(\omega) 
    \right),
    \end{split}
    \label{eq:opt_cond_vert_final}
    \end{equation}
    
    where the two terms arising due to five branch cuts in the complex plane (see Appendix~\ref{app:derivations} for a figure and full derivation) are given by 
    %
    %
    %
    \begin{widetext}
        \begin{subequations}
            \label{eq:zeta_final}
            \begin{flalign}
            \zeta_1(\omega)^{R \: \textbf{k}\textbf{k}'} & = 
            \begin{aligned}[t]
            \frac{-1}{4 \pi^2} \iint\limits_{\mathbb{R}^2}  \md\nu \, \md \nu'   \: \eta_F(\nu) \, \eta_B(\nu')\big[ F_{-\nu'}^{R \: \textbf{k}-\textbf{k}'}- F_{-\nu'}^{A \: \textbf{k}-\textbf{k}'}  \big]
            & \mathrlap{\big[G_{\nu + \omega}^{R \: \textbf{k}} G_{\nu+\nu'+\omega}^{R \: \textbf{k}'} + G_{\nu-\omega}^{A \: \textbf{k}} G_{\nu+\nu'-\omega}^{A \: \textbf{k}'} \big]} \\ 
            & \mathrlap{\big[G_{\nu}^{R \: \textbf{k}} G_{\nu'+\nu}^{R \: \textbf{k}'}-G_{\nu}^{A \: \textbf{k}} G_{\nu'+\nu}^{A \: \textbf{k}'} \big],}
            \label{eq:zeta_1}
            \end{aligned} 
            \\
            \zeta_{2p3}(\omega)^{R \: \textbf{k}\textbf{k}'} & =  
            \begin{aligned}[t]
            \frac{i}{2 \pi} \iint\limits_{\mathbb{R}^2}  \md \nu \, \md \nu'  \: \eta_F(\nu) \, \eta_F(\nu') A_{\nu'}^{\textbf{k}'} \bigg[ & \big[ G_{\nu + \omega}^{R \: \textbf{k}} G_{\nu'+\omega}^{R \: \textbf{k}'} + G_{\nu-\omega}^{A \: \textbf{k}} G_{\nu'-\omega}^{A \: \textbf{k}'} \big] \big[G_{\nu}^{R \: \textbf{k}} F_{\nu-\nu'}^{R \: \textbf{k}-\textbf{k}'} - G_{\nu}^{A \: \textbf{k}} F_{\nu-\nu'}^{A \: \textbf{k}-\textbf{k}'} \big] + \\ 
            & (-2\pi i) A_{\nu}^{\textbf{k}} \big[ G_{\nu'+\omega}^{R \: \textbf{k}'} G_{\nu-\omega}^{A \: \textbf{k}} F_{\nu-\nu'-\omega}^{A \: \textbf{k}-\textbf{k}'} + G_{\nu'-\omega}^{A \: \textbf{k}'} G_{\nu+\omega}^{R \: \textbf{k}} F_{\nu-\nu'+\omega}^{R \: \textbf{k}-\textbf{k}'} 
            \big]
            \bigg].
            \end{aligned}
            \end{flalign}
        \end{subequations}
    \end{widetext}
    It is advantageous to perform the real frequency integrals in  Eqs.~(\ref{eq:opt_cond_vert_final}) and (\ref{eq:zeta_final})  instead of the Matsubara sums in \eqref{eq:current_current_correlator_F}. One would otherwise have to transform  the optical conductivity into real frequencies by means of a (numerical) analytic continuation. While there are several methods for this purpose \cite{Max_Ent_Jarrell, SOM1, SpM, SpM_YOSHIMI2019, Tikhonov_Regularization}, very fine features in the resulting real frequency spectrum tend to be blurred by all of them, especially, in the presence of noise/box-effects. In the worst case (e.g. when applied to data with correlated noise without proper treatment of the co-variance matrix\cite{Max_Ent_Jarrell, Kappl_2020}) it can lead to unphysical artifacts in the real frequency spectrum. Our approach avoids these problems altogether.

    \subsection{RPA ladder in the $\overline{ph}$-channel}
    \label{sec:rpa_vertex}
    
    In the parquet D$\Gamma$A calculations of Ref.~\onlinecite{pi_ton} the dominant contributions to the vertex corrections for the square lattice Hubbard model were identified as coming from the transversal particle-hole channel, usually denoted as $\overline{ph}$. A general diagram in this category can be very complicated, but it can be represented as a ladder in terms of the irreducible vertex in this channel, $\Gamma_{\overline{ph}}$, which includes  all kind of diagrams (fully irreducible and insertions from the other channels). These diagrams are the same diagrams that make the magnetic susceptibility diverge in RPA and  single-shot ladder D$\Gamma$A~\cite{ladder_DGA}, only the building block $\Gamma_{\overline{ph}}$  is different; also the DMFT susceptibility is calculated from such a ladder (in the particle-hole channel)~\cite{kotliar_dmft,Kunes2011}.  To qualitatively understand their effect on the optical conductivity, we consider here only the simplest form of the ladder diagram, namely the RPA series with $\Gamma_{\overline{ph}}=-U$ in the dominating magnetic channel. The contribution of these diagrams to the vertex $F$ in the $\overline{ph}$-channel (illustrated in Fig.~\ref{fig:rpa_ladder}a) is given by
    \begin{equation}
    F_{\overline{ph},k-k'}^{\text{RPA}} 
    = \frac{{U}^2 \chi_{k-k'}^0}{1-{U} \ \chi_{k-k'}^0} = U^2 \chi_{q=k-k'}^{\text{RPA}},
    \label{eq:RPA_effective_vertex}
    \end{equation} 
    with $\chi_{q}^0 = -\frac{1}{\beta}\sum_{k} G_k G_{k+q}$ (for the discussion of all RPA-ladder contributions to $F$ in the $ph$ and $\overline{ph}$ channels see Appendix~\ref{app:rpa_ladders}). The same RPA-series, only without the ${U}^2$ prefactor, constitutes the RPA susceptibility  $\chi_{q}^{\text{RPA}}$.  Since this susceptibility diverges for ${\bf q}=(\pi,\pi)$ at a finite temperature in the square lattice Hubbard model, the results we obtain cannot be treated quantitatively or even compared to parquet D$\Gamma$A calculations for the same value of $U$. The full frequency and momentum dependent irreducible vertex $\Gamma_{\overline{ph}}$ is replaced here with a constant value of $-U$. This makes $U$ an effective parameter that we choose to be lower than the "bare" interaction (which we denote with $U_0$ in the following) due to screening effects, originating from other scattering channels, that are not included in the RPA approach. In the following we will choose the (temperature independent) value of $U$ such, that in the temperature range considered we will be close to the AFM phase transition.
    
    In Fig.~\ref{fig:rpa_ladder}b we illustrate how the $\overline{ph}$ RPA-ladder diagrams contribute to the  optical conductivity. It is important to note, that if one considers the ladder in the longitudinal particle-hole channel, only  $F_{q}^{\text{RPA}}$ could contribute. This contribution is first of all restricted to ${\bf q}=0$, because of the restriction on the incoming light momentum and secondly it will vanish because if the vertex is not dependent on $\mathbf{k}$ and $\mathbf{k}'$: the $\gamma_{\alpha}^{\mathbf{k}}$ vertices are antisymmetric in $\mathbf{k}$ and $\mathbf{k}'$ and will thus cause the contribution to vanish. This is also the reason why the first order in $U$ vanishes and is disregarded in \cref{eq:RPA_effective_vertex}. Building the ladder in the $\overline{ph}$-channel allows for contributions from all momenta $\mathbf{k}-\mathbf{k}'$, since $k$ and $k'$ are summed over. In particular, it allows for contributions from $\mathbf{k}-\mathbf{k}'=(\pi,\pi)$, where the ladder becomes very large. With the  $\mathbf{k}-\mathbf{k}'$ dependence of the vertex, the contribution is also  not nullified by the antisymmetry  of $\gamma_{\alpha}^{\textbf{k}}$.

    
    \subsection{Effective vertex in the vicinity of the antiferromagnetic phase}
    \label{sec:OZ_vertex}
    Close to the AFM phase transition, when fluctuations become critical, the susceptibility can be approximated by the universal Ornstein-Zernike (or Hertz-Millis-Moriya) correlation function~\cite{RevModPhys.79.1015,Hertz,Millis,Moriya} of the following form
    \begin{equation}
    \label{eq:OZ}
    \chi_{\textbf{q}, \omega}^{\rm{OZ}} = \frac{\widetilde{A}}{\xi^{-2} + (\textbf{q}-\textbf{Q})^2 +  \lambda|\omega|}, 
    \end{equation}
    where in case of AFM fluctuations ${\bf Q}=(\pi,\pi)$ and $\xi$ is the magnetic correlation length, $\widetilde{A}$ and $\lambda$ are further parameters, and $\omega$ is the bosonic Matsubara frequency. Upon approaching the phase transition one can extract the correlation length $\xi$ (as well as the two other parameters) by fitting the expression~(\ref{eq:OZ}) to the susceptibility. The fit parameters will in general be functions of both temperature $T$ and interaction $U$. 
    
    In analogy to the RPA approximation of the vertex, where we assumed $F_{\overline{ph},k-k'}^{\text{RPA}}=U^2 \chi_{k-k'}^{\rm{RPA}}$, we can make a step further and assume an Ornstein-Zernike form of the vertex:
    \begin{equation}
    \label{eq:OZ_vertex}
    F_{\textbf{k}-\textbf{k}', \nu-\nu'}^{\rm{OZ}} = \frac{A}{\xi^{-2} + (\textbf{k}-\textbf{k}'-\textbf{Q})^2 +  \lambda|\nu-\nu'|}, 
    \end{equation}
    with $A=U^2\widetilde{A}$. 
    This expression also allows for direct analytic continuation, since it is dependent only on one Matsubara frequency (difference) $\nu-\nu'$. The analytically continued expression is as following 
    \begin{equation}
    F^{R/A}_{\textbf{k}-\textbf{k}', \nu-\nu'} = \frac{A}{ \xi^{-2}+(\textbf{k}-\textbf{k}' - \textbf{Q})^2 
        \mp  i \lambda(\nu -\nu')},
    \end{equation}
    where $\nu$, $\nu'$ are now real frequencies.

\subsection{Relation to vertex corrections coming from electron-phonon scattering}
\begin{figure}
        \centering
        \includegraphics[width=1\linewidth]{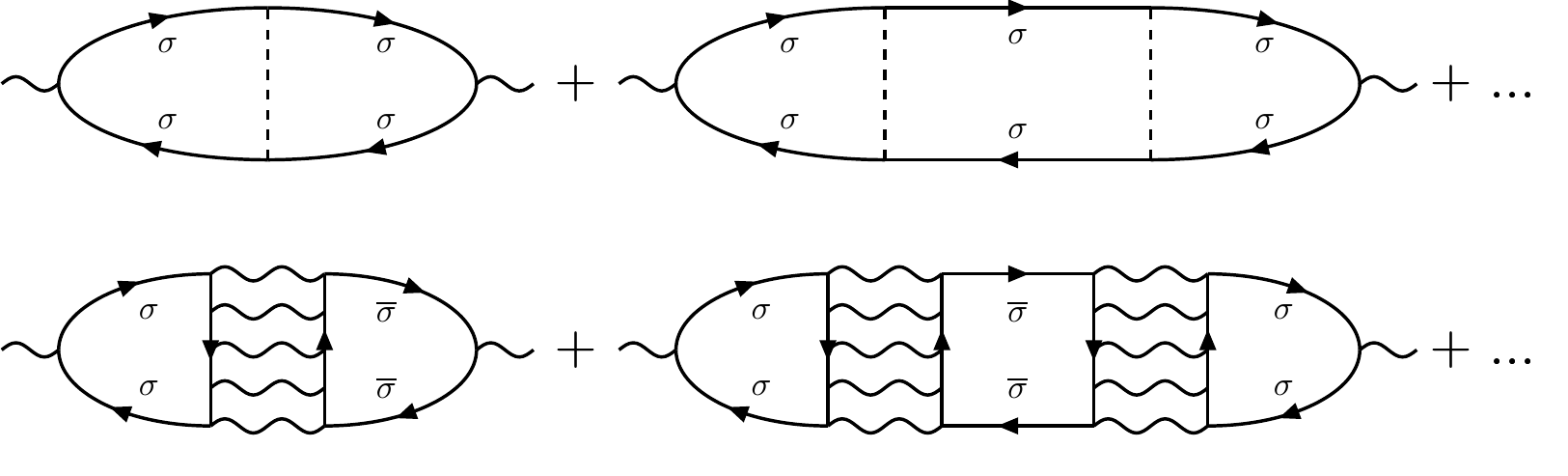}
        \caption{Upper panel: diagrams representing the ladder of electron-phonon interactions. Lower panel: diagrams representing $ph$-ladders of multiple $\pi$-ton contributions. Please note, that for electron-electron interaction diagrams that $\pi$-ton diagrams are composed of, each closed loop corresponds to an additional minus sign -- contrary to diagrams representing scattering on phonons, which all come with the same sign. }
        \label{fig:el_ph}
    \end{figure}

The diagrams in $\pi$-ton contributions to optical conductivity, as illustrated in Fig.~\ref{fig:rpa_ladder}, can also be interpreted in terms of the exchange of an effective boson. In this sense they are similar to electron-phonon scattering~\cite{Mahan2000} illustrated in the upper panel of Fig.~\ref{fig:el_ph}. In the case of scattering on (acoustic) phonons it is necessary to take into account the entire ladder of electron-phonon scatterings to obtain the dominant vertex correction~\cite{Mahan1966}. In the case of $\pi$-tons we can also construct such ladder of multiple boson exchanges (as illustrated in the lower panel of Fig.~\ref{fig:el_ph}). However, as we have learnt from Ref.~\onlinecite{pi_ton}, it is the first term of this ladder that gives the dominant contribution, and in this work we take only this term in our calculations. The second term in the lower panel of Fig.~\ref{fig:el_ph}, as well as all further terms, belong to the $ph$-reducible channel, which was shown to give almost no contribution in the parquet approach of Ref.~\onlinecite{pi_ton}. This is confirmed by the authors of Ref.~\cite{Simard2020}, who also considered a 'double $\pi$-ton', which is exactly the second term of Fig.~\ref{fig:el_ph} (lower panel). The reason likely lies in the differences between the properties of electron-phonon vs. electron-electron scattering: (i) there is a sign change connected with the number of fermionic loops -- electron-phonon ladders have no additional closed loops, contrary to purely electronic diagrams, where we add a closed loop (and therefore an additional minus sign) with every new rung of the $ph$-ladder; (ii) the ladder of phonons connects only electrons with the same spin, whereas for $\pi$-tons the spins alternate; (iii) last but not least, multiple $\pi$-tons are connected by two Green's function lines that are offset by a momentum ${\bf q}=0$ and not by the strong nesting vector ${\bf q}=(\pi,\pi)$ in the $\overline{ph}$ ladders.      
 
    \subsection{Approximate form of the self-energy}

    \label{sec:self_energy}
\begin{figure}
        \centering
        \includegraphics[width=1\linewidth]{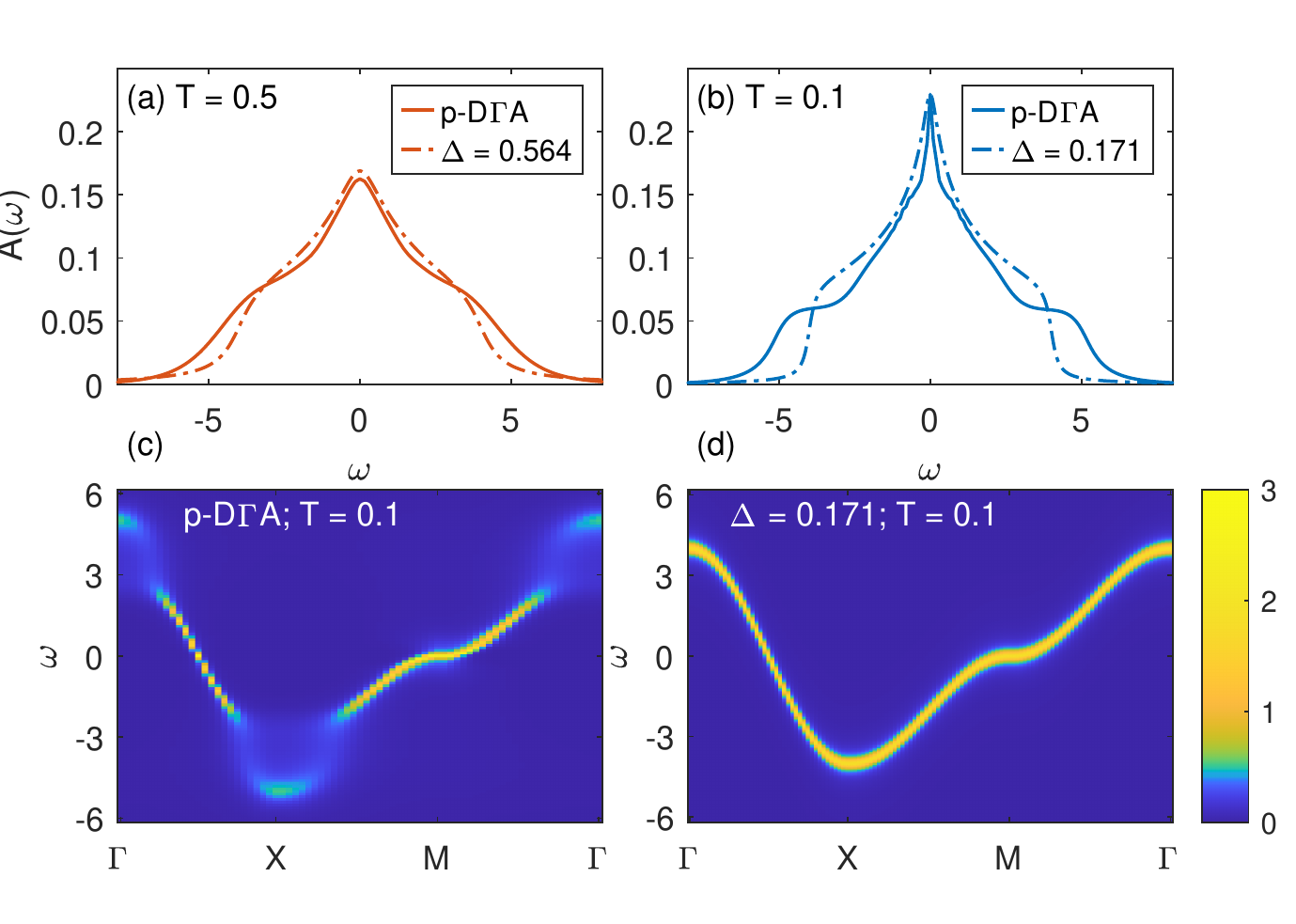}
        
        \caption{Upper panel: spectral function for (a) $T = 0.5$ and (b) $T=0.1$ from p-D$\Gamma$A with $U_0=4$ (solid line) and a model with constant self-energy $-i\Delta_{0}$ (dashed line). Lower panel: intensity plot of the momentum resolved spectral function along a path in the Brillouin zone obtained with p-D$\Gamma$A self-energy (left) and constant self-energy $-i\Delta_{0}$ (right) for $T = 0.1$.}
        
        \label{fig:spectral_functions}
    \end{figure}

    The derived formalism for calculating optical conductivity on the real axis is valid for any Green's function that is known on the real frequency axis. Expressed through the self-energy $\Sigma_{\nu}^\textbf{k}$, the  retarded/advanced (R/A) Green's functions are given by:
    \begin{equation}
    \label{eq:gf}    
    G^{R/A \: \textbf{k}}_{\nu}=\frac{1}{\nu -\epsilon_{\bf k}-\Sigma^{R/A\: \textbf{k}}_{\nu}  \pm i 0^+ }.
    \end{equation}
    The self-energy must then be known on the real frequency axis as well. Since in the current work we are specifically interested in the effect of vertex corrections on the Drude peak we take a simple, frequency and momentum independent form:
    \begin{equation}
    \label{eq:sigma}
    \Sigma^{{R/A}\: \textbf{k}}_{\nu} = \mp i\Delta_0.
    \end{equation}
    In Fig.~\ref{fig:spectral_functions} we show the local and non-local spectral function $A^{\bf{k}}_{\nu}=-\frac{1}{\pi}\Im G^{R \: \textbf{k}}_{\nu}$ from a parquet D$\Gamma$A (p-D$\Gamma$A) calculation  of Ref.~\onlinecite{pi_ton} (at the interaction value of $U_0=4$) analytically continued to real frequencies with the maximum entropy method. In the two top plots we show the local spectral function $A_{\nu}$  from p-D$\Gamma$A for two different temperatures, together with  $A_{\nu}$ obtained from the model with the constant self-energy \eqref{eq:sigma}, which gives  a Lorentzian of width $\Delta$ for $A_{\nu}$. The difference  is mainly visible in: (i) The absence of largely broadened waterfall-like  structures at $|\omega|\gtrsim 2.5$ in Fig.~\ref{fig:spectral_functions} in case for the constant self-energy. These are precursors of the Hubbard bands and of little relevance for the low frequency Drude peak which we focus on. (ii) A temperature dependence of the quasi-particle peak. In order to account for this temperature dependence, we use in the following a temperature dependent $\Delta(T)$ which we obtain from a $T^2$ fit to the p-D$\Gamma$A data (for details of this fit see Appendix~\ref{app:dga_fit}). This is motivated by the known $T^2$ dependence of the Fermi-liquid scattering rate. In our fit the  $T^2$-dependence is accompanied by an additional temperature independent offset. 
    
    \section{Results}
    \label{sec:results}

   \begin{figure}
        \centering
        \includegraphics[width=0.5\textwidth]{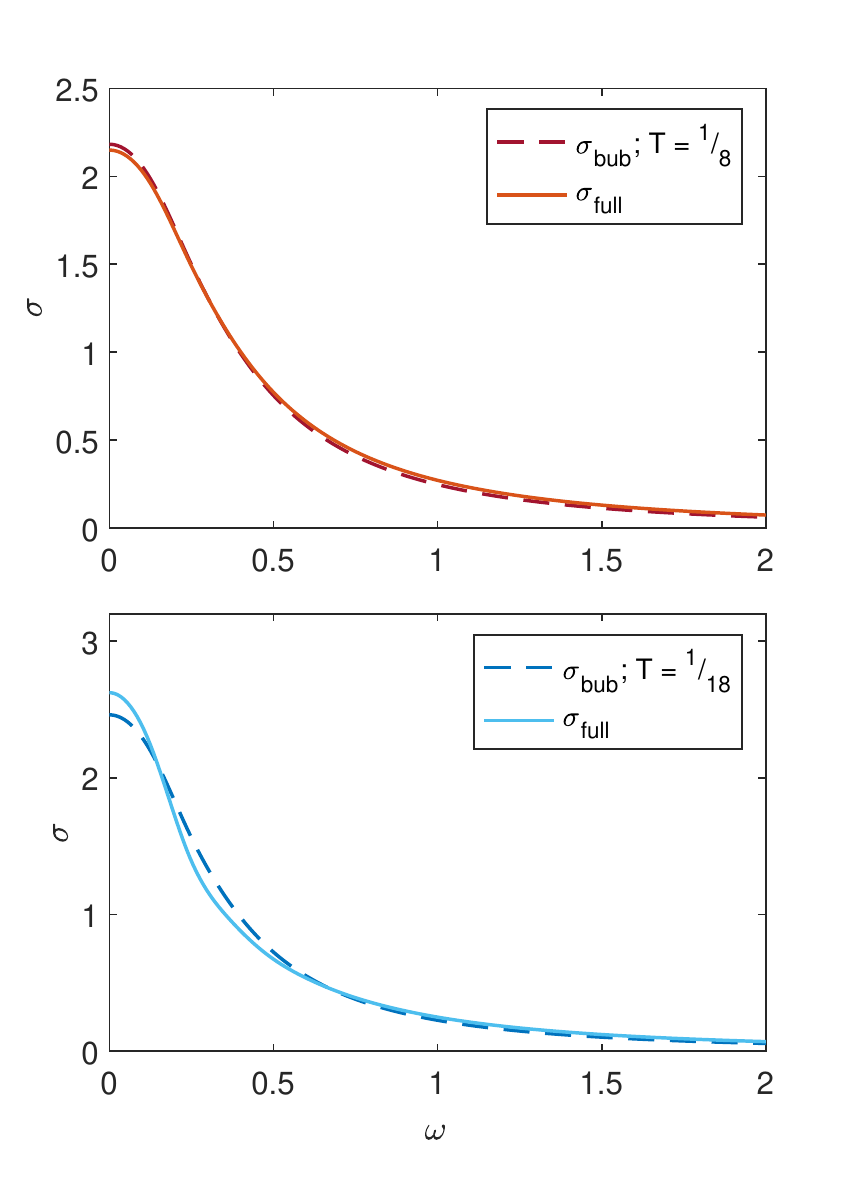}
        \caption{Optical conductivity $\sigma(\omega)$  at two different temperatures with ($\sigma_{\rm full}$) and without ($\sigma_{\rm bub}$) vertex corrections from RPA the ladder in the $\overline{ph}$ channel. The effective interaction in the RPA-ladder is $U=1.9$.}
        \label{fig:Drude_broadening}
    \end{figure}
    
       \begin{figure*}
        \centering
        \includegraphics[width=\textwidth]{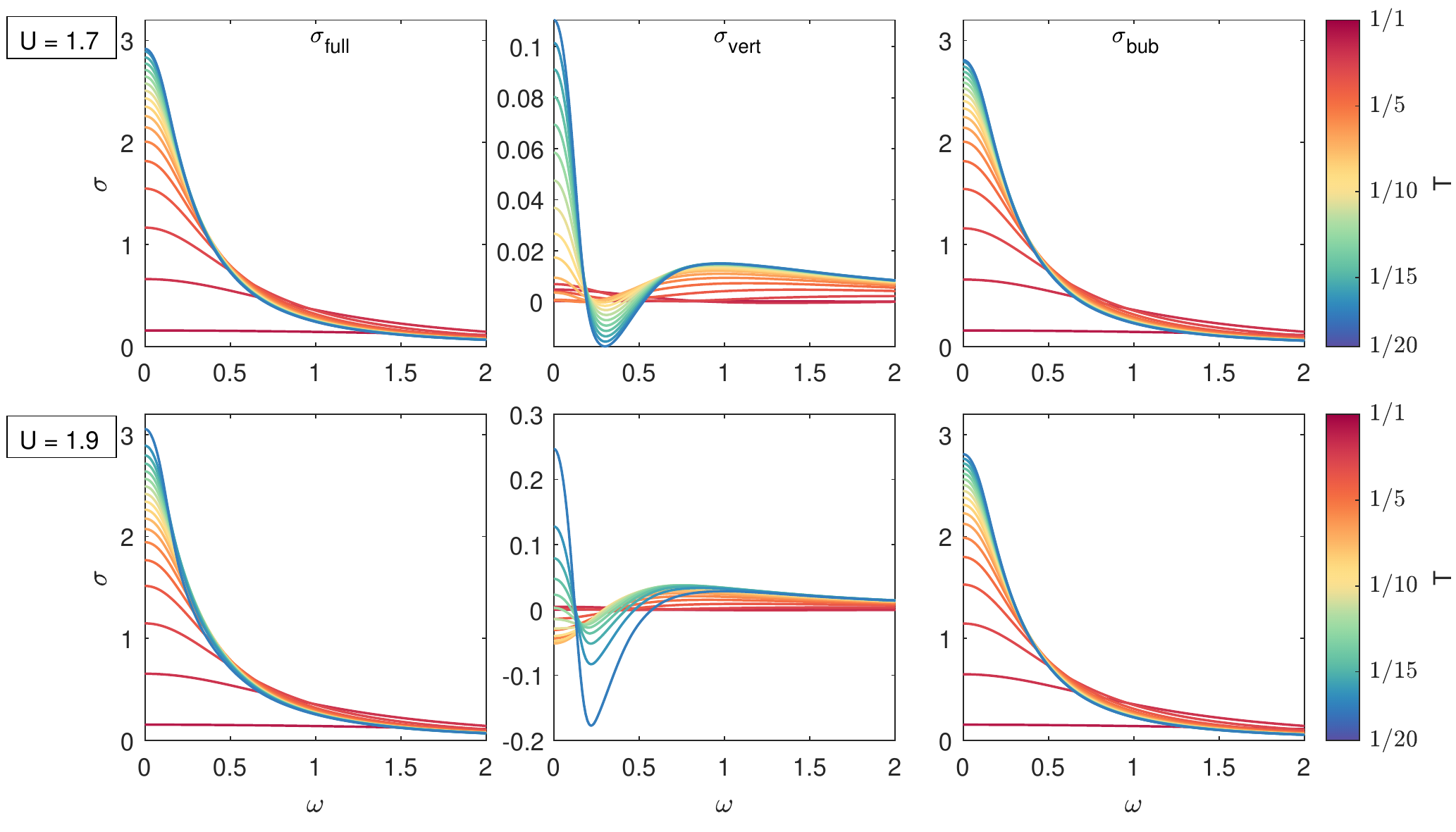}
        \caption{Full optical conductivity $\sigma_{\rm full}$ (left) including the vertex corrections in the RPA ladder $\sigma_{\rm vert}$  (middle) and the bare particle-hole bubble $\sigma_{\rm bub}$ (right) at two different effective interactions $U=1.7$ (top) and $U=1.9$ (bottom) and various temperatures $T$ (color-coded according to the bar on the right with a step-size between the different temperatures $\sim 1/T = \beta$).}
        \label{fig:RPA_temp}
    \end{figure*}
    
    The focus of our calculations is on the vertex corrections to the Drude peak that are generated by $\pi$-ton contributions. As already discussed, the simplest such contribution  is the RPA ladder in the $\overline{ph}$-channel shown in Fig.~\ref{fig:rpa_ladder}. 
    In the following we will show results for the real part of the optical conductivity $\sigma(\omega)=\Im \chi_{jj}(\omega)/\omega$ calculated directly on the real axis using \eqref{eq:opt_cond_vert_final} with a simplified two-particle vertex: either the RPA ladder of \eqref{eq:RPA_effective_vertex} or the Ornstein-Zernike form (\ref{eq:OZ_vertex}). The Green's function is of the form \eqref{eq:gf} with a Fermi-liquid-like  temperature dependence of the scattering rate $\Delta_0(T)\sim T^2$, however with a constant offset. Specifically we use $\Delta_0(T)= 0.1547 + 1.637\,T^2$, which we obtained from a $T^2$ fit to p-D$\Gamma$A see Appendix \ref{app:dga_fit} \footnote{The value of the interaction in p-D$\Gamma$A is $U_0=4$, whereas here we use different values of $U=1.7-1.9$. The reason is that the RPA series is a good approximation only for much smaller interaction values, and it grossly overestimates the AFM fluctuations. To represent in our approximate approach the intermediate coupling regime of $U_0=4$, where $\pi$-tons were observed, we hence need to use an effective value of $U<4$ in the RPA ladder, as it is done e.g. in the TPSC approach~\cite{Vilk1994}.}. This featureless  form of the Green's function corresponds to the metallic phase of the Hubbard model~\footnote{The fit we use will no longer be valid for the low temperature regime, $T\lesssim 0.05$, where we expect the pseudogap behavior in the Hubbard model~\cite{Schaefer2015-2,schaefer2020tracking} .} and gives per construction a Drude peak in the bubble part of the optical conductivity. Additional features in the bubble part, coming from frequency dependence of the self-energy (see e.g. Ref.~\cite{Berthod2013}), are neglected since we focus on the effect of vertex corrections on the two-particle level exclusively. In Appendix~\ref{app:sde_selfenergy} we show that our conclusions are not changed if we assume a frequency dependent self-energy.
    
    In the final subsection we also qualitatively compare the  vertex corrections originating from the RPA ladder with constant self-energy with those from the $\overline{ph}$-channel of the fully-fledged p-D$\Gamma$A calculation.

    \subsection{Vertex corrections originating from  RPA ladders in the transversal particle-hole  channel} 
    \label{ch:char_RPA_ladder}

    \begin{figure}
        \centering
        \includegraphics[width=1.0\linewidth]{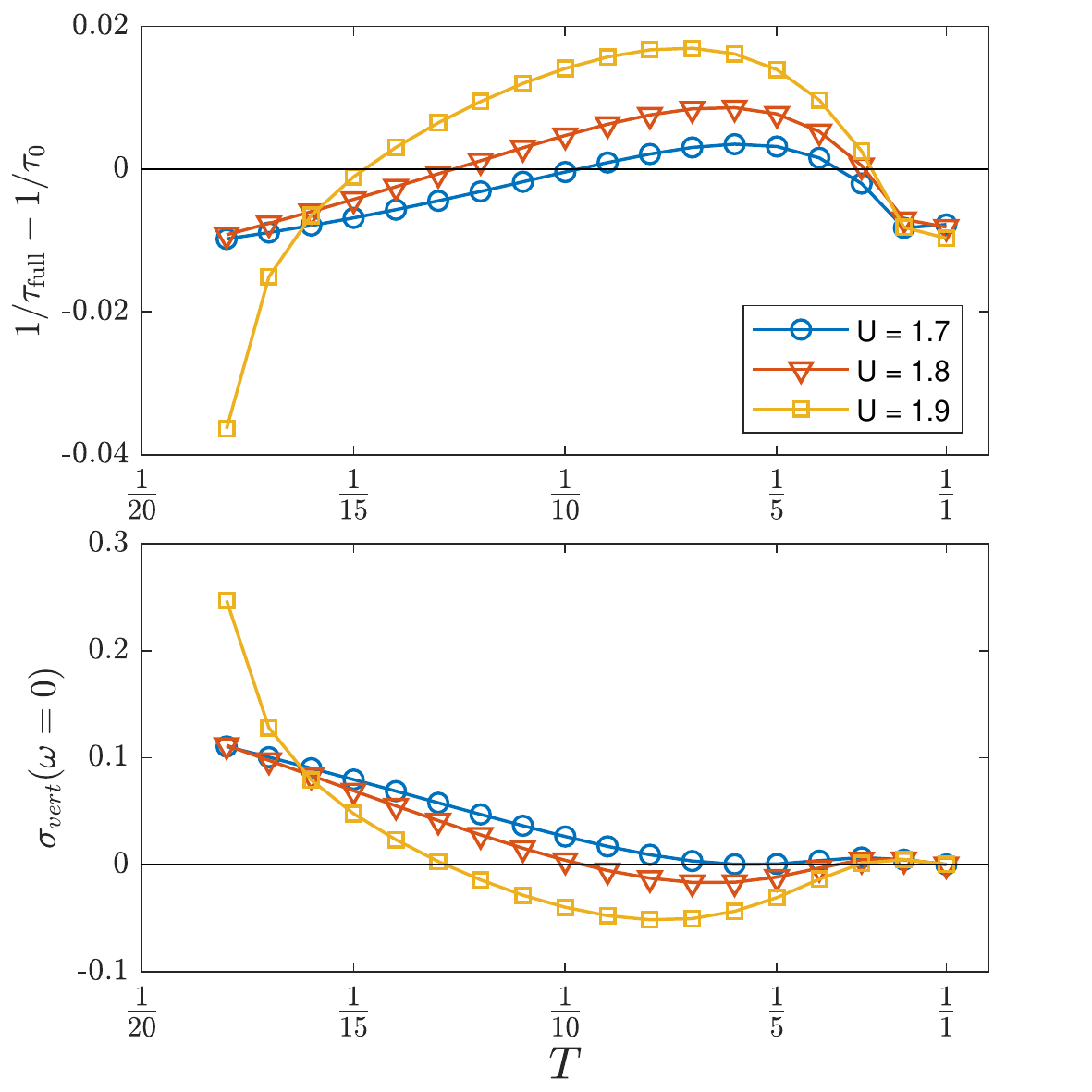}
        \caption{Top: Difference between the width of the Drude peak including vertex corrections ($1/\tau_{\rm{full}}$) and   the width  of the Drude peak using the bubble contribution only ($1/\tau_{\rm{0}}=2\Delta_0$)  as a function of temperature (on a scale linear in $1/T=\beta$) for three different values of $U$. Bottom: vertex contribution $\sigma_{\rm vert}(\omega=0)$ to the DC conductivity as a function of temperature for the same $U$ values.}
        \label{fig:brodening_vs_T}
    \end{figure}

    \begin{figure*}
        \centering
        \includegraphics[width=1\textwidth]{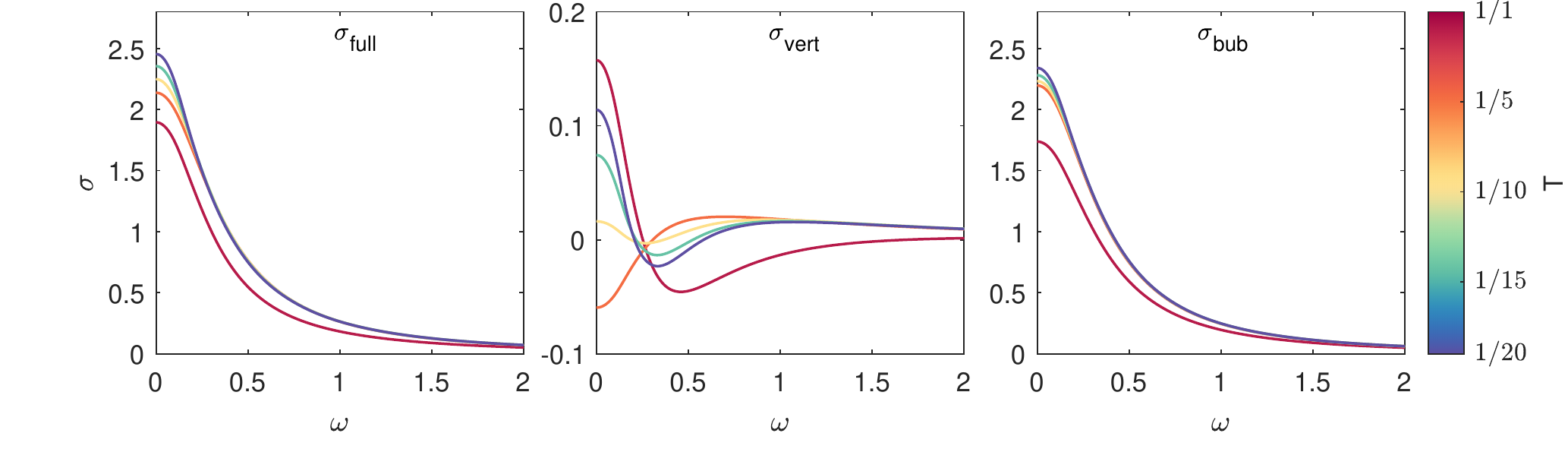}
        \caption{Optical conductivity $\sigma_{\rm full}$ (left) including RPA vertex corrections in the transversal particle-hole channel, shown for different temperatures while keeping the self-energy $\Sigma^R = -i \Delta_0= - 0.18\mi$ constant; $U=1.8$. In the middle and right column the respective summands, $\sigma_{\rm vert}$ and $\sigma_{\rm bub}$, are shown.}
        \label{fig:T_scan}
    \end{figure*}

    \begin{figure}
        \centering
        \includegraphics[width=1.0\linewidth]{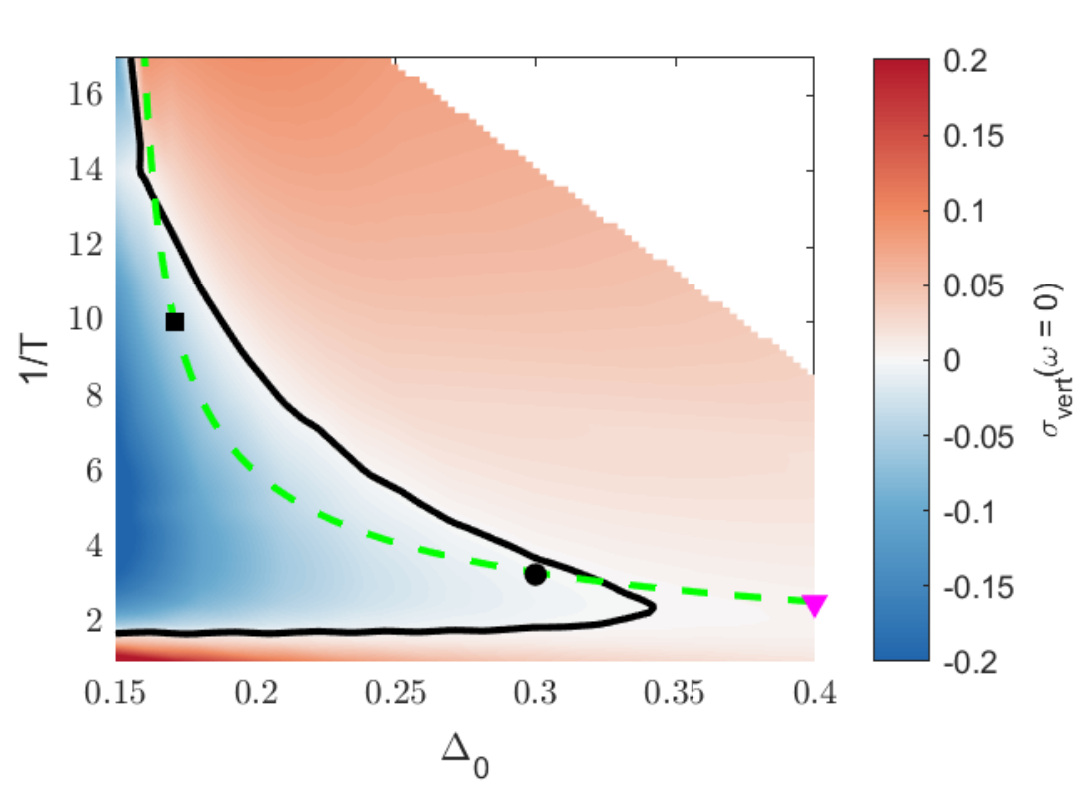}
        \caption{RPA vertex contribution to the DC conductivity $\sigma_{\rm{vert}}(\omega = 0)$ -- color coded -- as a function of inverse temperature $1/T$ and the one-particle scattering rate $\Delta_0$ treated as independent parameters, for $U = 1.9$. Dashed green line: $\Delta_0(T)= 0.1547 + 1.637\,T^2$ as used in Figs.~\ref{fig:Drude_broadening}-\ref{fig:brodening_vs_T}. Thick black line: boundary between positive and negative values of $\sigma_{\rm{vert}}(\omega = 0)$. Black box: parameter point of Fig.~\ref{fig:OZ_opt_cond_xi_scan}. Black circle and magenta triangle: parameter points discussed further in Appendix~\ref{app:OZ_high_temp}. }
        \label{fig:contour}
    \end{figure}
    
    In Fig.~\ref{fig:Drude_broadening} we show the effect of the RPA-ladder vertex corrections on the Drude peak for two different temperatures. The bubble contribution to the optical conductivity  $\sigma_{\rm bub}$ is a Drude peak per construction, since we used a simplified Green's function with a constant imaginary part of the self-energy, depending only on temperature: $\Sigma_\nu^{R\;\mathbf{k}}(T) = - i\Delta_0(T)$. Our first observation is that including the vertex corrections does not change the overall shape of the peak and the full optical conductivity  $\sigma_{\rm full}=\sigma_{\rm bub} + \sigma_{\rm vert}$ can still be fitted by a (Lorentzian) Drude peak. Both for the bubble as well as for the full vertex-corrected conductivity the Drude peak becomes sharper with decreasing temperature, as to be expected for a metal.  Surprisingly however,  it depends on the temperature, whether the vertex corrections broaden the Drude peak ($T=1/8$, upper panel of Fig.~\ref{fig:Drude_broadening}) or sharpen it ($T=1/18$, lower panel of Fig.~\ref{fig:Drude_broadening}). The vertex corrections are relatively small and their magnitude depends on the value of the effective interaction $U$ used in the RPA ladder (for the results presented in Fig.~\ref{fig:Drude_broadening}, $U=1.9$). We observe however similar behavior for other values of $U$.

    In  Fig.~\ref{fig:RPA_temp} we analyze  the temperature dependence in more detail. In the left column  we show the full optical conductivity $\sigma_{\rm full}=\sigma_{\rm bub}+\sigma_{\rm vert}$ with RPA ladder vertex corrections for different temperatures and two values of the effective interaction $U=1.7$ and $U=1.9$. These values of the effective interaction $U$ were chosen such that we are close to the divergence of the RPA-ladder, which indicates the AFM phase transition. This happens with our choice of $\Delta_0$ for $U=1.9$ at $T_c=1/19$ and for $U=1.7$ at $T_c<1/60$. The optical conductivity has the form of a Drude peak for all temperatures that we studied and the peak is broadened with increasing temperature. The latter is mainly due to the monotonic temperature dependence of the bubble contribution $\sigma_{\rm bub}$, shown in the right column of Fig.~\ref{fig:RPA_temp}. Indeed that is exactly as to be expected for a Fermi-liquid-like metal. The RPA ladder vertex corrections  $\sigma_{\rm vert}$ (plotted in the middle column of Fig.~\ref{fig:RPA_temp}) show however a strongly non-monotonic behavior. They broaden the Drude peak in the intermediate temperature regime (e.g., yellowish colors in Fig.~\ref{fig:RPA_temp} bottom), but lead to its sharpening for high temperatures (reddish colors in Fig.~\ref{fig:RPA_temp}) and in particular for low temperatures (bluish colors). The latter is achieved by a vertex correction $\sigma_{\rm vert}(\omega)$ that is positive at the lowest frequencies and negative around $\omega\sim 0.3$ (Fig.~\ref{fig:RPA_temp} top) respectively $\omega\sim 0.2$  (Fig.~\ref{fig:RPA_temp} bottom). 
    
    For larger frequencies around $\omega\sim 1$ the vertex corrections are always positive, independent of temperature. Together with the broadening and sharpening, this implies that in this region there are corrections to the Lorentzian-shape of the Drude peak.   However, in this frequency region  there will also be further contributions to the optical conductivity such as interband transitions (in multi-band systems), transitions to Hubbard bands, and our assumption of a constant self-energy is no longer justified since the Fermi liquid also implies a term $\Im \Sigma_\nu^{R\;\mathbf{k}} \sim - \nu^2$, i.e., an increase of the one-particle scattering rate with frequency \cite{Berthod2013}.  Anyhow, except for this feature  which is difficult to entangle from other effects and to observe in experiment, the RPA-ladder vertex corrections do not alter the Lorentzian form of the  Drude peak, which is an important result by itself, besides of the non-monotonous temperature dependence of the broadening or sharpening.

    \begin{figure*}
        \centering
        \includegraphics[width=1\textwidth]{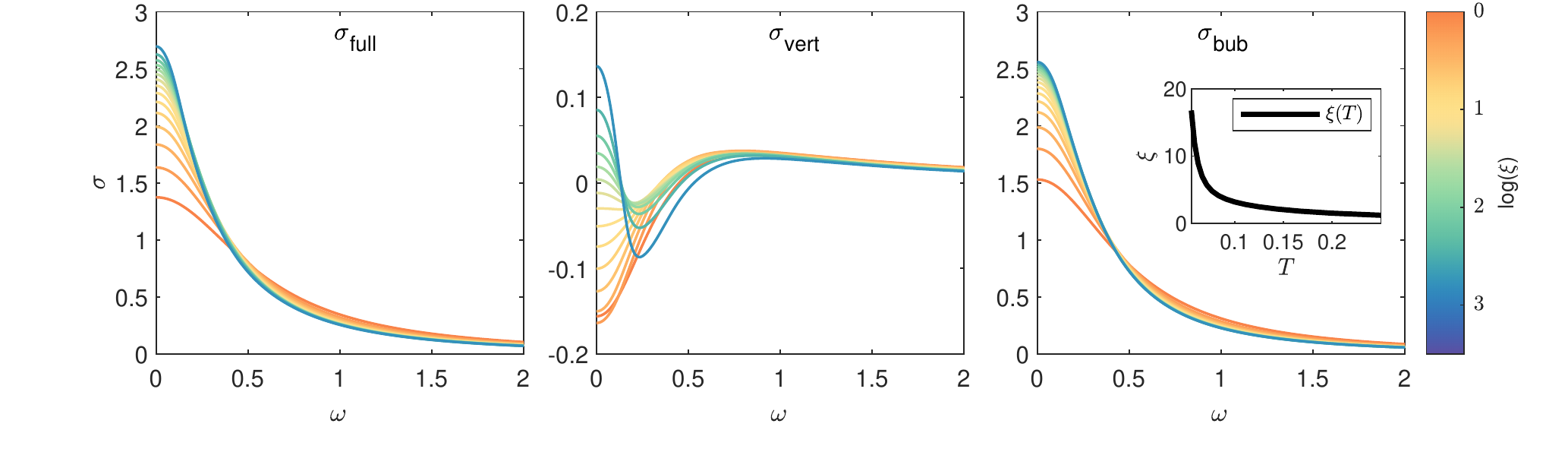}
        \caption{Optical conductivity $\sigma_{\rm full}$ (left) now with the  Ornstein-Zernike vertex corrections in the transversal particle-hole channel, shown for different temperature-dependent correlation lengths $\xi(T)$ (color coded on a logarithmic scale as in the bar on the right hand side).  In the middle and right column the respective summands, $\sigma_{\rm vert}$ and $\sigma_{\rm bub}$, are shown. Inset: $\xi(T)$ as fitted to the RPA-vertex at $U=1.9$. }
        \label{fig:OZ_opt_cond}
    \end{figure*}

    \begin{figure*}
        \centering
        \includegraphics[width=1\textwidth]{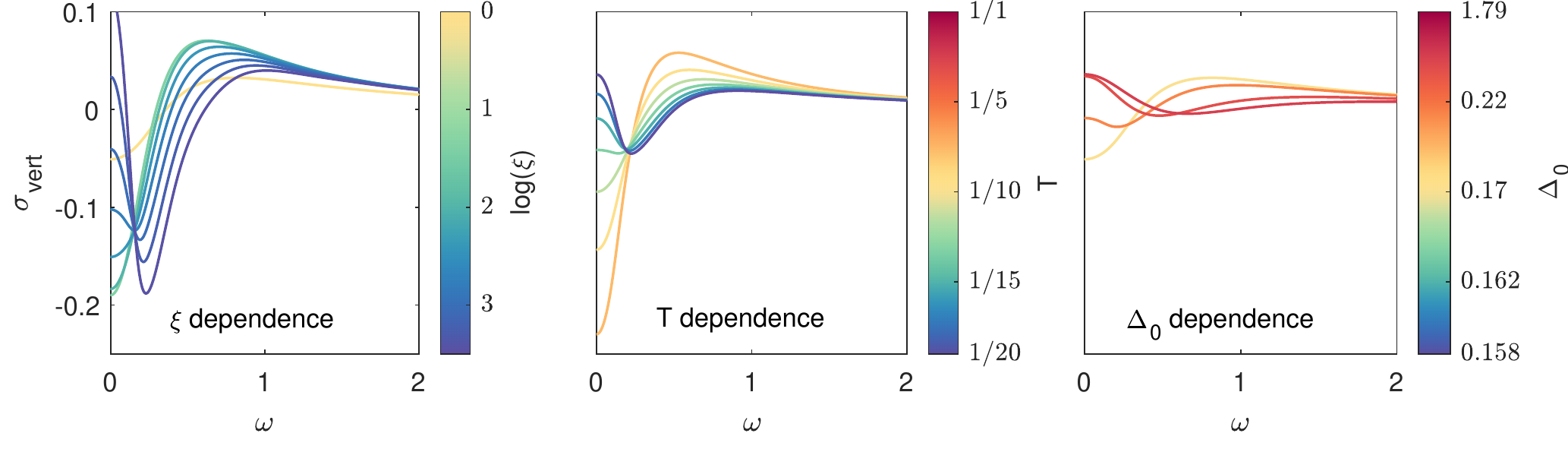}
        \caption{Vertex corrections to the optical conductivity $\sigma_{\rm vert}$ changing the correlation lengths $\xi$ (left), the temperature $T$ (middle), and the single-particle broadening $\Delta_0$ (right) in the OZ calculation, while keeping the other parameters fixed at $T = 1/10$, $\Delta_0 = 0.17$ (these parameters correspond to the black square in Fig.~\ref{fig:contour} ). For the middle plot $\xi = 5.0$, $\lambda = 1.3$ and $A = 1.6$ are fixed; for the other two $\xi = 3.18$, $\lambda = 0.93$ and $A = 1.6$.}
        \label{fig:OZ_opt_cond_xi_scan}
    \end{figure*}
    
    This broadening and sharpening of the Drude peak is better illustrated in Fig.~\ref{fig:brodening_vs_T} (top panel) where we plot the difference between the full width of the peak $1/\tau_{\rm full}$ and the width of the Drude peak from the bubble contribution $1/\tau_0$ as a function of temperature. Note that the Drude width $1/\tau_0$ of the numerically calculated bubble contribution agrees -- as it should for a Fermi-liquid-like metal --  with the one-particle scattering rate $\Delta_0(T)$ taken as input for the self-energy, i.e., $1/\tau_0=2\Delta_0$~\footnote{This relation holds for higher temperatures only approximately.} . 
    We clearly see three temperature regimes: (i) For $T\gtrsim1/2$ the vertex corrections slightly sharpen the Drude peak (decrease its width, \mbox{$\frac{1}{\tau_{\mathrm{full}}} - \frac{1}{\tau_{0}} <0$}) and the optical conductivity at $\omega=0$ is slightly increased (lower panel of Fig.~\ref{fig:brodening_vs_T}) by the  vertex corrections. This is also in agreement with an exact diagonalization study of vertex corrections beyond DMFT in~Ref.~\onlinecite{Mravlje2018, Vranic2020} which was restricted to the high temperature regime. (ii) In an intermediate temperature regime, whose limits depend on the interaction value $U$, the Drude peak is broadened by vertex corrections and the $\omega=0$ value (DC conductivity) is decreased.  (iii) For the lowest temperatures we could reach, the vertex corrections sharpen the Drude peak and DC conductivity is increased. All effects become stronger and the intermediate temperature regime (ii) is enhanced when increasing the effective interaction $U$, which enhances the AFM fluctuations.     

    In order to assess how important the temperature dependence of the scattering rate $\Delta_0$ is for the temperature dependence of the vertex corrections, we perform a temperature scan with a fixed $\Delta_0=0.18$ for all Green's function lines [including the ones in the RPA ladder in \eqref{eq:RPA_effective_vertex}]. That is the temperature dependence enters solely via the Fermi and Bose distribution functions both in the bubble \eqref{eq:bc_ana_cont_2} and in the vertex part \eqref{eq:opt_cond_vert_final}~\footnote{For clarity: the temperature enters in the RPA-ladder expression for the vertex,~\eqref{eq:RPA_effective_vertex},  through the Fermi distribution function present in $\chi_0$.}. In Fig.~\ref{fig:T_scan} we show the resulting RPA-ladder optical conductivity and its bubble and vertex parts for different temperatures at fixed  $\Delta_0$. The non-monotonic behavior of the vertex corrections,, including all three temperature regimes (i)-(iii), is still present as a function of temperature. We thus see that while the temperature dependence of $\Delta_0$ is quantitatively important, it is not needed for the correct qualitative behavior. 
    
    
In Fig.~\ref{fig:contour} we present the value of the $\pi$-ton RPA vertex contribution to the DC conductivity $\sigma_{\textrm{vert}}(\omega=0)$ for different (inverse) temperatures and scattering rates $\Delta_0$ treated as independent parameters. The green dashed line corresponds to the fit $\Delta_0(T)= 0.1547 + 1.637\,T^2$ obtained from p-D$\Gamma$A data as used in Figs.~\ref{fig:Drude_broadening}-\ref{fig:brodening_vs_T}. The black line denotes the sign change of $\sigma_{\textrm{vert}}(\omega=0)$, which corresponds to the boundary between broadening (negative values, blue) and sharpening (positive values, red) of the Drude peak. Note that, as in Fig.~\ref{fig:brodening_vs_T}, negative $\sigma_{\textrm{vert}}(\omega=0)$ corresponds to an increase of the two-particle scattering rate  $1/\tau_{\textrm{full}}$, whereas positive  $\sigma_{\textrm{vert}}(\omega=0)$, to a decrease of $1/\tau_{\textrm {full}}$ due to vertex corrections.

As is already clear from Fig.~\ref{fig:T_scan}, keeping a constant scattering rate $\Delta_0$ and increasing temperature, we go from sharpening to broadening and again to sharpening at high temperature, crossing the black line twice. In Fig.~\ref{fig:contour} this corresponds to moving along the green dashed line where we also cross the black line twice, which is due to the specific (quadratic with an offset) temperature dependence of $\Delta_0$, obtained from the p-D$\Gamma$A fit. For large values of $\Delta_0$ we do not see any sign changes in $\sigma_{\textrm{vert}}(\omega=0)$ any more. We also cannot achieve the double sign change by changing $\Delta_0$ at constant temperature. For the RPA vertex, the temperature dependence of the vertex corrections to the two-particle scattering rate $1/\tau_{\textrm full}$ (broadening and sharpening) is thus closely related to the temperature dependence of the one-particle scattering rate $\Delta_0$.

The conclusion from Fig.~\ref{fig:contour} is that low temperatures and a large one-particle scattering $\Delta_0$ lead to a narrowing of the Drude peak [positive (red) $\sigma_{\textrm{vert}}(\omega=0)$]. That we have two crossovers from narrowing to broadening and back again originates from the specific temperature-dependence of $\Delta_0$.

In this work we focus on two-particle vertex corrections to a featureless bubble contribution $\sigma_{\textrm{bub}}$. Including consistently vertex corrections on both one- and two-particle level is a formidable task and was already done within the numerically cumbersome parquet approach in~Ref.~\onlinecite{pi_ton}. However, in order to at least briefly assess how our results would change if we also introduced  the same vertex corrections in the calculation of one-particle scattering rate, we computed the optical conductivity including these vertex corrections also in the one-particle Green's functions in~\eqref{eq:opt_cond}. Note that self-energy (scattering) rate and vertex corrections are related by the Schwinger-Dyson equation, see Appendix~\ref{app:Schwinger_Dyson_real}.
  The results are presented in Appendix~\ref{app:sde_selfenergy} and they show that the overall behavior of the vertex corrections is not changed -- we also see temperature dependent broadening and sharpening of a slightly modified Drude peak (see Figs.~\ref{fig:opt_cond_vs_T_SDE}-      \ref{fig:brodening_vs_T_SDE}). 
    
    \begin{figure*}
        \centering
        \includegraphics[width=1\textwidth]{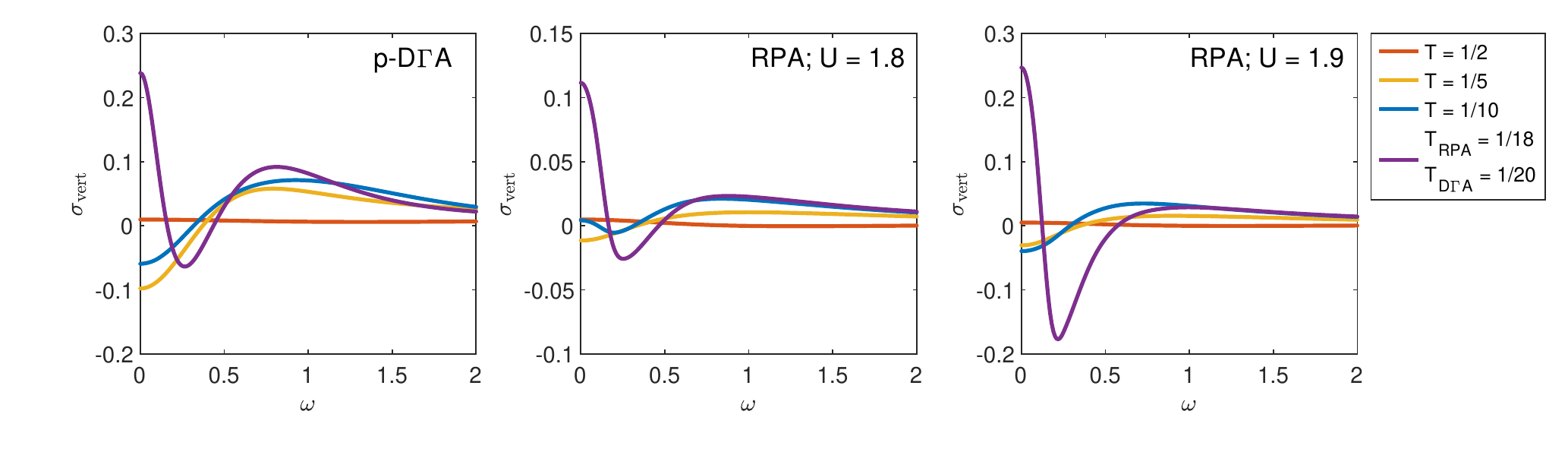}
        \caption{Comparison of the  vertex corrections in the $\overline{ ph}$-channel. Left:  analytic continuation of the p-D$\Gamma$A results of Ref.~\onlinecite{pi_ton} for the (bare) interaction $U_0=4$.  Middle and  right: RPA-ladder results for two different values of effective $U$. Different colors denote again different temperatures, see legend box. The lowest temperature shown is different for the RPA-ladder vertex, since it diverges at $T=1/20$.}
        \label{fig:pdga_vs_RPA}
    \end{figure*}
    
    \subsection{Effective vertex given by the Ornstein-Zernike form}
    \label{ch:char_OZ}
    
    In order to assess if the RPA-ladder effects are still present if we use a different form of the vertex that resembles the magnetic susceptibility, we have calculated vertex corrections with an Ornstein-Zernike (OZ) vertex of \eqref{eq:OZ_vertex}. The temperature dependence of the free parameters, $A$, $\lambda$ and the correlation length $\xi$ was obtained by fitting the RPA vertex with the OZ form~\eqref{eq:OZ_vertex}. The details of the fits including the temperature dependencies of OZ parameters are presented in Appendix~\ref{app:OZ_fits}.

    In Fig.~\ref{fig:OZ_opt_cond} we show the optical conductivity obtained from the OZ vertex~\eqref{eq:OZ_vertex}. Here, we use for the  two particle and two hole Green's functions connecting the OZ vertex in the $\overline{ph}$ channel to the incoming and outgoing light exactly the same simple frequency-independent scattering rate (self-energy) $\Delta_0(T)$ as in the RPA (where this Green's function was additionally employed for calculating the vertex itself). That is, when changing temperature, we change the Green's function lines in the bare bubble and the four  Green's function outer lines in the vertex diagram (cf. Fig.~\ref{fig:opt_cond}), but the vertex itself is not diagrammatically built from Green's function lines. Its temperature dependence enters only via the OZ parameters. The most important parameter is the AFM correlation length $\xi$ which is strongly temperature dependent as the inset of Fig.~\ref{fig:OZ_opt_cond} shows; the two other parameters can be found in Appendix~\ref{app:OZ_fits}. Knowing this dependence (here from fitting to the RPA vertex), we can translate temperature to AFM correlation length. This way we can  look at the sign and shape of vertex corrections to the optical conductivity as a function of $\xi$ in Fig.~\ref{fig:OZ_opt_cond}.
    
    Even with a simplified OZ form of the vertex, we see the same effect (now driven by changing $\xi$). The Drude peak is sharpened for large correlation lengths (corresponding to low temperatures) and broadened for smaller $\xi$'s.  We did not reach the high temperature regime since we restricted the plot to $\xi\ge1$ (i.e. $T\leq0.25$), so the other sharpening regime at high $T$ is not present in the plot.
    
    The form of the effective OZ vertex allows us to use the OZ parameters as additional knobs that we can turn independently of $U$ or $T$, for helping us understand the origin of the non-monotonic temperature-dependence of the vertex corrections. In Fig.~\ref{fig:OZ_opt_cond_xi_scan} we do exactly this: in each panel we change  one of the three relevant parameters (correlation length $\xi$, temperature $T$, and one-particle broadening $\Delta_0$) and keep the others as well as $A$ and $\lambda$ of the OZ vertex fixed. The parameters were chosen so that we move away from the point with $T=0.1$ and $\Delta_0=0.17$, denoted with a black square in Fig.~\ref{fig:contour}. Note that, since we  use a fixed OZ-vertex here, changing $\Delta_0$ or $T$ does not correspond to moving along lines in Fig.~\ref{fig:contour}, where the vertex was from RPA.  
    
    In Fig.~\ref{fig:OZ_opt_cond_xi_scan} (left panel) we see that changing $\xi$, which is the key parameter for the temperature-dependent enhancement of the antiferromagnetic correlations at $(\pi,\pi)$, changes the vertex corrections    from a dampening in the intermediate temperature range (ii) to the sharpening in the low temperature range (iii). A comparison with Fig.~\ref{fig:OZ_opt_cond} (middle panel) however also shows already that other factors must be at work as well.  
    
    The two other parameters, i.e., $T$  in Fig.~\ref{fig:OZ_opt_cond_xi_scan} (middle panel) and $\Delta_0$ (right panel), only enter in the four Green's function lines connecting the (now fixed) OZ vertex with the incoming and outgoing light. These parameters change the frequency range ($\nu-\nu'$) over which vertex contributions are included. The temperature $T$  yields also an important contribution  to the broadening in the  intermediate temperature regime (ii) and a sharpening at low temperatures (iii); whereas $\Delta_0$ is less relevant at low temperature where it hardly changes, but important for the sharpening in the high temperature regime (i). Please also note the isosbestic point~\cite{Greger2013} around $\omega\sim 0.1$. In terms of the frequency integrals \eqref{eq:zeta_final} around the branch cuts, $\zeta_{2p3}$ always yields a positive vertex correction and $\zeta_1$ a negative one (except for very high temperature, when it can become positive), but depending on $\xi$, $T$, and $\Delta_0$, the total balance is changing sign.
    
   An analogous parameter scan as done in Fig.~\ref{fig:OZ_opt_cond_xi_scan} for parameters close the low temperature regime in Fig.~\ref{fig:contour} (black square), was done for a high temperature in the vicinity of two values of $\Delta_0$ (points denoted by black circle and magenta triangle in Fig.~\ref{fig:contour}). A detailed analysis is presented in Appendix~\ref{app:OZ_high_temp}. The overall conclusion is that in case of the OZ vertex used in the high temperature regime, the vertex corrections lead to sharpening (increased value of  $\sigma_{\textrm{vert}}(\omega=0)$) only at large values of $\Delta_0$ or when setting the correlation length very small (see Fig.~\ref{fig:OZ_parameter_scan_appendix}). This was not the case for the RPA vertex. We note, that for high temperatures the correlation length description of the system provided by the OZ vertex is no longer valid and the RPA and OZ vertices differ more.

    \subsection{Relation to the parquet D$\Gamma$A results}
    \label{ch:RPA_DGA_comp}
    
    The $\pi$-ton contributions to optical conductivity, as described in Ref.~\onlinecite{pi_ton}, are not directly identifiable with ladder diagrams only. In fact, without insertions from other channels, the $\overline{ph}$-ladder diagrams would eventually diverge at $(\pi,\pi)$ when lowering the temperature in the systems with strong AFM/CDW fluctuations. These diagrams though, even if dampened by other contributions, provide the dominant part of the $\overline{ph}$-channel. It is however not obvious if the simplest approach, where the ladder is built only with a constant effective $U$ as the interaction vertex, can describe the same physics as the full p-D$\Gamma$A calculation. 
    
    In Fig.~\ref{fig:pdga_vs_RPA} we show a qualitative comparison between the  p-D$\Gamma$A vertex corrections (at $U_0=4$) and the RPA-ladder ones (at two values of the effective interaction $U=1.8$ and $U=1.9$) in a range of temperatures. Even though the Green's functions were simplified in the RPA ladders, the qualitative behavior with temperature in both approaches is similar: a sharpening at the lowest temperature and broadening in the higher temperature regime. We show here only the low frequency range, since for higher frequencies there are effects coming from the (precursors of the) Hubbard bands that are absent in the calculation with the constant self-energy. Another important difference between the  p-D$\Gamma$A and RPA results is the analytic continuation: it had to be done numerically in  p-D$\Gamma$A, whereas in RPA we derived and used expressions in real frequencies.

    \section{Conclusion}
    \label{sec:conclusions}

    Motivated by the work on $\pi$-ton contributions to the optical conductivity~\cite{pi_ton} and the importance of the AFM and CDW fluctuations for vertex corrections in strongly correlated electron systems, we have studied in this paper their influence on the Drude peak in a simplified semi-analytical calculation. To this end, we have employed the RPA ladder for studying one vertex correction that is especially effective and indeed the diagrammatic archetype for strong  AFM or CDW fluctuations.  In order to couple this ladder which becomes large for a wave vector $(\pi, \pi)$ to light with $\mathbf{q}=0$, we need to built the ladder in the transversal particle-hole channel. Here, two particle-hole pairs are created by the incoming and outgoing light, respectively. For one pair, particle and hole carry the momentum $\mathbf{k}$ and for the other $\mathbf{k}'$. This allows the coupling to AFM and CDW fluctuations at  $\mathbf{k}'-\mathbf{k}=(\pi, \pi)$. While antiferromagnetic fluctuations are quintessential for the Hubbard model, in other parameter regimes and for other lattices  
 e.g.~superconducting or ferrmognetic fluctuations may dominate and with this other channels may become important.

    In the context of high-temperature superconductivity a classification of vertex corrections into Maki~\cite{Maki1968}-Thompson~\cite{Thompson1970}  and Aslamazov-Larkin (AL)~\cite{Aslamazov1968} type is sometimes employed, also in the context of the optical, Hall and Raman conductivity \cite{Kontani1999,Kontani2008,Kampf1992,Tremblay2011,Chubukov2014}. The $\pi$-ton  or transversal particle-hole channel is a different classification: diagrams of both, the MT and AL class~\footnote{The second order in $U$ diagram of the RPA ladder is of AL, the others are of MT type; there is no first order diagram.},  contribute to the $\pi$-tons in the   $\overline{ph}$ ladder.
    
    By using a simplified vertex we have made it possible to calculate the optical conductivity with vertex corrections on the real frequency axis \footnote{For the functional renormalization group \cite{Metzner2012}, which is related to the parquet through its multiloop extensions~\cite{Kugler2017,Hille2020} a simplified quasiparticle description directly on the real axis has been recently proposed~\cite{Rohe2020}. For a p-D$\Gamma$A additionally a local vertex for real frequencies is needed, which one might obtain, e.g., through matrix product states~\cite{Evertz2017} } and we derived the necessary analytical expressions for this purpose. These expressions are general and can be used for arbitrary Green's function known on the real axis and a vertex that has the same frequency structure in the complex plane as a physical susceptibility. In this way we avoid the uncertainty that is sometimes present in the analytic continuation of numerical data obtained on the Matsubara axis      and have sufficient momentum resolution. This allows us to study changes reliably also for small vertex corrections and to understand whether the simplest RPA-ladder diagrams in the $\overline{ph}$ channel are sufficient to explain the numerical results of
       Ref.~\onlinecite{pi_ton}.

 As for the validity of our calculations, we can only expect an RPA-based approach to be at least qualitatively reliable
      at weak coupling in the metallic phase.
      By construction,  our approach is further  limited to the parameter regimes, where the $\overline{ph}$ contributions are indeed dominant, i.e. when AFM or CDW fluctuations are prevalent \footnote{
Please note that within the RPA-approach using an  effective interaction is necessary, and hence we cannot reliably treat different channels on an equal footing.}.
      As was shown in      Ref.~\onlinecite{pi_ton},  for the considered parameter range  the  $\overline{ph}$ channel is indeed the dominant vertex corrections to the optical conductivity; the second largest is from the $pp$ channel;  the $ph$ channel  is large at ${\mathbf q}=(\pi,\pi)$ but its overall contribiution to the optical conductivity with  ${\mathbf q}=0$ is weak.

    We first of all find that the RPA-ladder $\pi$-ton contributions do not alter the overall Lorentzian shape of the Drude peak in the low frequency regime--at least in the weak coupling domain of RPA. The width of the Drude peak for a given temperature is however modified by the ladder $\pi$-tons: the peak is sharpened at low and high temperatures, but broadened for intermediate temperatures.
That there are two such crossovers eventually originates from the temperature-dependence of the single-particle scattering rate $\Delta_0$ which has a larger curvature than the crossover from broadening at high $T$ and low $\Delta_0$ to sharpening at low $T$ and high $\Delta_0$ in Fig.~\ref{fig:contour}.
    The sharpening at high temperatures through vertex corrections has been observed also  in Ref.~\onlinecite{Mravlje2018}. For slightly higher frequencies, where the Drude peak has only a Lorentzian tail we see some additional feature coming from vertex corrections, that are positive at all temperatures. This frequency range though is not very well represented by the approximate constant self-energy that we used and a closer look in this regime goes beyond our simplified study.

    For a half-empirical calculation of $\pi$-ton vertex corrections one could use the magnetic susceptibility, as was suggested already in the supplemental material to~Ref.~\onlinecite{pi_ton}. Here we have shown in a model calculation that one can replace the RPA ladder by a universal Ornstein-Zernike form. The aforementioned sharpening-broadening-sharpening upon lowering temperature originates from both, the enhancement of the vertex around $(\pi,\pi)$ due to an increase of the correlation length $\xi$ and the reduction of its relevant frequency window. The latter is because the two $\pi$-ton particle-hole excitations coupled to the incoming and outgoing light must be closer and closer to the Fermi energy with decreasing temperature in the low frequency range of the optical spectrum. 
    
    The final question that we have tried to answer in this work is whether the simplest ladder diagrams in the  $\overline{ph}$-channel are enough to at least qualitatively address the $\pi$-tons in the 2D Hubbard model. For the low frequency part of the optical conductivity, the answer seems to be positive, since the main effect, a non-monotonic temperature dependence resulting in broadening or sharpening of the Drude peak by the $\pi$-ton contributions is analogous in p-D$\Gamma$A and in the RPA-ladder approach.

    {\sl Note added:} During the completion of our work, we became aware of an independent calculation of $\pi$-ton contributions to optical conductivity  using RPA \cite{Simard2020} with a different physics focus, mainly on the 1D Hubbard model, and some differences in the implementation.

    {\sl Acknowledgments. } We would like to thank K. Astleithner,  J. Kune\v{s}, and P. Pudleiner for many stimulating discussions.
    This work was supported by the  
    Austrian Science Fund (FWF) through  projects P 30819 and P 30997, and the  Doctoral School W 1243 ``Building Solids for Function''. Calculations have been done on the Vienna Scientific Cluster (VSC).

\appendix
\begin{figure}
    \centering
    \scalebox{0.5}{\includegraphics[width=1\textwidth]{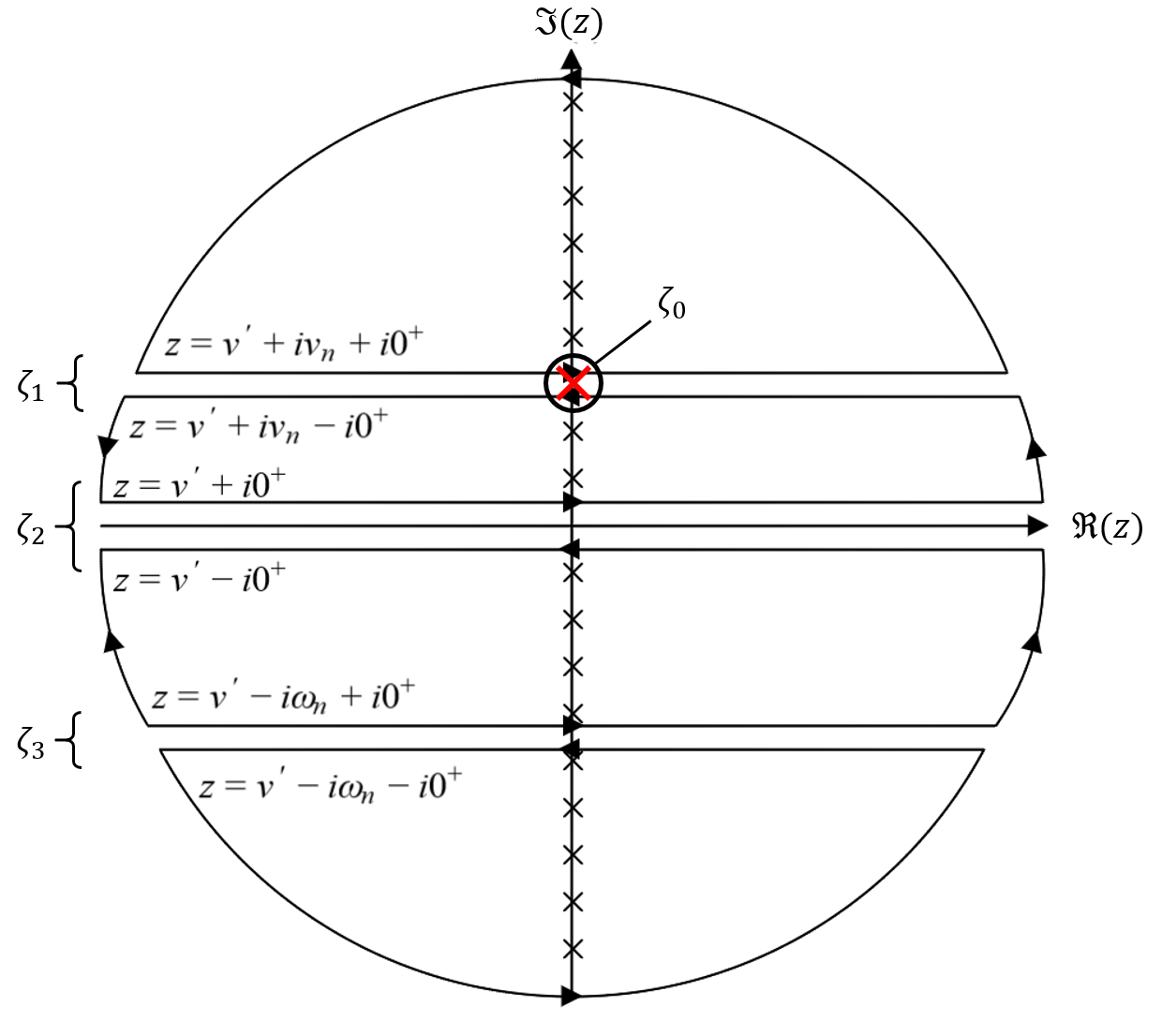}}
    \caption{The three branch-cuts for the analytical continuation ($\mi \nu_n' \to z$) of the first Matsubara sum in \eqref{eq:opt_cond_vert_1}. The red cross marks the Matsubara frequency $\mi \nu_n = \mi \nu_{n'}$ which has to be excluded from the sum and treated separately. The $\zeta_i$'s combine the line integral above and below the branch cut of two contour integrals (the arches of the contour integral vanish). In case of $\zeta_0$ the contour integral is around the single Matsubara frequency $i\nu'=i\nu$.
    }
    \label{fig:branchcuts_vert}
\end{figure}

\section{Derivation of the real frequency optical conductivity with vertex corrections}

\label{app:derivations}

In this Appendix we provide an explicit derivation of the real frequency expressions used for the numerical calculation of the current-current correlation function $\chi_{jj}$---the full one as well as its constituents, the bare bubble and vertex correction. The real part of the optical conductivity is then given by evaluating the expression $\sigma(\omega)=\Im \chi_{jj}(\omega,{\bf q}=0)/\omega$. To make the notation clearer we use in the following a different notation for Matsubara frequencies: i.e., we denote the imaginary fermionic frequencies by $\mi \nu_n = \mi \pi T (2n+1)$ and the bosonic by $\mi \omega_n = \mi 2\pi T n$ (with $n\in \mathbb{Z}$). The real frequencies are $\nu$ or $\omega$. We also keep the general bosonic momentum ${\bf q}$ dependence of the current-current correlation function and skip the $jj$ subscript in the following.

\subsection{Bubble contribution} 
To make the notation and assumptions clear we repeat a derivation of  \eqref{eq:bc_ana_cont_2} for the bare bubble contribution on the real frequency axis. Without vertex corrections the expression of interest is in Matsubara frequencies
\begin{equation}
\label{eq:current_current_bubble_matsubara}
\begin{split}
\chi_{\rm bub}(\mi \omega_n,\textbf{q}) = - &2 \sum_{\bf k} \gamma_{\alpha}^{\textbf{k} \textbf{q}} \gamma_{\alpha}^{\textbf{k} (-\textbf{q})} \, \frac{1}{\beta}\sum_{\mi \nu_n}   G^{\textbf{k}+\textbf{q}}_{\mi \nu_n + \mi \omega_n} \, G^{\textbf{k}}_{\mi \nu_n}  \\
\equiv & \; 2 \sum_{\bf k} \gamma_{\alpha}^{\textbf{k} \textbf{q}} \gamma_{\alpha}^{\textbf{k} (-\textbf{q})} \, \chi_0^{\textbf{k} \textbf{q}}(\mi \omega_n),
\end{split}
\end{equation}
where  the sign originates from the fermionic loop rule and the factor two is due to the spin.  The Matsubara sums in \eqref{eq:current_current_bubble_matsubara} can be translated into a real frequency integral by using the fact that the fermionic Matsubara frequencies are located at the poles of the Fermi-Dirac distribution $\eta_{F}(z)=\frac{1}{\me^{\beta z} + 1}$ and that $G^\textbf{k}(z)$ is analytical everywhere except for the real axis (where it has a branch cut). For $\Im z > 0 $ ($<0$)  the Green's function $G^\textbf{k}(z)$ is the analytic continuation of the retarded (advanced) Green's function $G^{R\:\textbf{k}}_{\omega}$ ($G^{A \: \textbf{k}}_{\omega}$). This leads to
\begin{equation}
\begin{array}{rcl}
\chi^{\textbf{k},\textbf{q}}_0(\mi \omega_n) &=& \frac{-1}{\beta}\sum_{\mi \nu_n}   G^{\textbf{k}+\textbf{q}}_{\mi \nu_n + \mi \omega_n} \, G^{\textbf{k}}_{\mi \nu_n} \\
&=& \frac{1}{2 \pi i} \sum_\mathcal{C} \oint_{\mathcal{C}} \md z \:  \eta_{F}(z) \, G^{\textbf{k} + \textbf{q}}_{z + \mi \omega_n} \, G^{\textbf{k}}_{z},
\label{eq:bc01}
\end{array}
\end{equation}
where the contours $\mathcal C$ encircle all the singularities of of $\eta_{F}(z)$. By Cauchy's theorem the contours can be deformed arbitrarily, unless they encounter a region of ambiguous analyticity. For the above expression these regions are located at $\Im z = 0$ and at $\Im z = - \mi \omega_n$. After stretching the contours in such a ways that they cover the whole complex plane, except  for stripes of infinitesimal height at $\Im z = 0$ and at $\Im z = - \mi \omega_n$, it can be shown that the stripes give the only nonzero contribution. The arcs of (infinite radius) that close the contour do not give a contribution because the integrand is decaying faster than $1/|z|$. One can therefore rewrite \eqref{eq:bc01} as 
\begin{equation}
\begin{split}
\chi^{\textbf{k},\textbf{q}}_0(\mi \omega_n) = \frac{1}{2 \pi i} \int_{-\infty}^{+\infty} \!\!\!\!\!\! \md \nu \: \eta_{F}(\nu) \bigg[ G_{\nu + i \omega_n}^{\textbf{k} + \textbf{q}} \big(G_{\nu + i 0^+}^{\textbf{k}}-G_{\nu-i 0^+}^{\textbf{k}} \big) & \\+
G_{\nu-i \omega_n }^{\textbf{k}} \big(G_{\nu + i 0^+}^{\textbf{k}+ \textbf{q}}-G_{\nu-i 0^+}^{\textbf{k}+ \textbf{q}} \big) \bigg].
\end{split}
\label{eq:bc02}
\end{equation}
Here, we used that $\eta_F(\nu + i \omega_n) = \eta_F(\nu )$ and $\eta_F(\nu) = \eta_F(\nu + \mi 0^+)$. We also assumed that $\mi \omega_n \neq 0$, which may be referred to as the dynamic limit. Using that the analytical continuation of the Green's function from above/below the branch cut gives $G^{\textbf{k}}_{\nu \pm \mi 0^+} = G^{R/A \: \textbf{k}}_{\nu}$ (which are in equilibrium related to one another by complex conjugation) and denoting the spectral function as $A^\textbf{k}_\nu = \frac{-1}{\pi } \Im\, G^{R \:  \textbf{k}}_{\nu} = \frac{1}{\pi } \Im\, G^{A \:  \textbf{k}}_{\nu} = \frac{-1}{2 \pi \mi } \left( G^{R \:  \textbf{k}}_{\nu} - G^{A \:  \textbf{k}}_{\nu}  \right) $ we get 
\begin{equation}
\begin{split}
\chi^{\textbf{k},\textbf{q}}_0(\mi \omega_n) = -\int_{-\infty}^{+\infty} \!\!\!\!\!\! \md \nu \: \eta_{F}(\nu) \left[ G_{\nu + i \omega_n}^{\textbf{k} + \textbf{q}} A_{\nu}^{\textbf{k}} +
G_{\nu-i \omega_n }^{\textbf{k}} A_{\nu}^{\textbf{k}+ \textbf{q}} \right].
\end{split}
\label{eq:bc03}
\end{equation}
After transforming \textit{all} Matsubara sums into integrals over real frequencies the remaining analytical continuation amounts to a simple substitution $\mi \omega_n \rightarrow \omega + \mi 0^+$. 

\begin{equation}
\chi^{R \: \textbf{k},\textbf{q}}_0(\omega) = -\int_{-\infty}^{+\infty} \!\!\!\!\!\! \md \nu \, \eta_{F}(\nu) \left[ G_{\nu + \omega}^{R \: \textbf{k} + \textbf{q}} A_{\nu}^{\textbf{k}} +
G_{\nu - \omega }^{A \: \textbf{k}} A_{\nu}^{\textbf{k}+ \textbf{q}} \right].
\label{eq:bc04}
\end{equation}
Inserting \eqref{eq:bc04} into \eqref{eq:current_current_bubble_matsubara} for $\textbf{q}=0$ leads to \eqref{eq:bc_ana_cont_2}. 

\subsection{Vertex Corrections} In this Section we  derive \eqref{eq:zeta_final}, starting again from the expression in Matsubara frequencies, but now for the vertex corrections:
\begin{widetext}
    \begin{equation}
    \chi_{\text{vert}}(i\omega_n,\textbf{q}) = -2  \sum_{\textbf{k}\textbf{k}'} \gamma_{\alpha}^{\textbf{k} \textbf{q}}\,  \gamma_{\alpha}^{\textbf{k}'(-\textbf{q})} \overbrace{ \frac{1}{\beta} \sum_{\mi\nu_n} G^\textbf{k}_{\mi \nu_n} \, G^{\textbf{k} + \textbf{q}}_{\mi \nu_n + \mi \omega_n} \,\underbrace{\frac{1}{\beta} \sum_{ \mi\nu_n'}  F^{\textbf{k}-\textbf{k}'}_{\mi \nu_n - \mi \nu_n'} \, G^{\textbf{k}'}_{\mi \nu_n'} \,  G^{\textbf{k}'+\textbf{q}}_{\mi \nu_n' + \mi \omega_n}}_{\equiv C^{\textbf{k},\textbf{k}',\textbf{q}}(\mi \nu_n, \mi \omega_n)}}^{\equiv C^{\textbf{k},\textbf{k}',\textbf{q}}(\mi \omega_n)}.
    \label{eq:opt_cond_vert_1}
    \end{equation}
\end{widetext}
The analytic continuation of \eqref{eq:opt_cond_vert_1} follows the same procedure as for the bubble contribution. We again assume $\mi \omega_n \neq 0$ (dynamic limit). The order in which the Matsubara sums are evaluated does not matter. Here we present the evaluation of the $\mi \nu_n'$-sum first [$C^{\textbf{k},\textbf{k}',\textbf{q}}(\mi \nu_n, \mi \omega_n)$ in \ref{eq:opt_cond_vert_1}]. The complex structure of the single frequency vertex must be the same as that of a physical magnetic susceptibility, i.e., $F^\textbf{k}_{z}$ is analytical in the whole complex plane except for $\Im z=0$ where it has a branch cut~\footnote{ as explained in \eqref{sec:methods_real_freq_vertex}}. The RPA-Bubble (\ref{eq:RPA_effective_vertex}) as well as the Ornstein-Zernike-susceptibility~(\ref{eq:OZ}) fulfill this assumption. Since branch-cuts appear in both the Green's function $G$ and the one-frequency vertex $F$, when the imaginary part of the frequency argument vanishes, there are three branch cuts when we analytically continue the argument of the Matsubara sum $i\nu'_n \rightarrow z$ into the whole complex plane, cf.~Fig.~\ref{fig:branchcuts_vert}:
\begin{equation}
z =
\begin{cases}
\nu' + \mi  \nu_n, \ &(\text{i})  \\
\nu',  \ &(\text{ii})\\
\nu'-\mi  \omega_n \ &(\text{iii}) 
\end{cases}
\label{eq:branchcut_vert}
\end{equation}
The first one, (i) in \eqref{eq:branchcut_vert}, originates from the vertex $F_{i\nu_n-z}^{\textbf{k}-\textbf{k}'} \, \stackrel{z \to \nu' + \mi \nu_n  \mp \mi 0^+}{=}\,F_{-\nu'}^{R/A\: \textbf{k}-\textbf{k}'}$ in \eqref{eq:opt_cond_vert_1}, which has a branch cut between retarded and advanced vertex at $z = \nu' + \mi \nu_n$ or $\Im z= + \mi \nu_n$.  Branch cuts \eqref{eq:branchcut_vert} (ii) and (iii) are from the two Green's functions $G_{z}^{\textbf{k}'}$ and $G_{z+i\omega_n}^{\textbf{k}'+\textbf{q}}$, respectively. 
Special care has to be taken additionally since the first branch-cut in \eqref{eq:branchcut_vert} is on top of a pole. Thus the point $\nu_n = \nu_{n}'$ has to be explicitly excluded from the second Matsubara sum and treated separately:
\begin{widetext}
    \begin{equation}
    \begin{array}{rcll}
    C^{\textbf{k},\textbf{k}',\textbf{q}}(\mi \nu_n, \mi \omega_n) &=&  \frac{1}{\beta} \sum_{\nu_n'}  F^{\textbf{k}-\textbf{k}'}_{\mi \nu_n - \mi \nu_n'} \, G^{\textbf{k}'}_{\mi \nu_n} \,  G^{\textbf{k}'+\textbf{q}}_{\mi \nu_n' + \mi \omega_n} &\\
    &=& \left. \frac{1}{\beta} F^{\textbf{k}-\textbf{k}'}_{\mi \nu_n - \mi \nu_n'} \, G^{\textbf{k}'}_{\mi \nu_n'} \,  G^{\textbf{k}'+\textbf{q}}_{\mi \nu_n' + \mi \omega_n} \right|_{\nu_n'=\nu_n} &+ \frac{1}{\beta} \sum_{ \nu_n' \neq \nu_n}  F^{\textbf{k}-\textbf{k}'}_{\mi \nu_n - \mi \nu_n'} \, G^{\textbf{k}'}_{\mi \nu_n'} \,  G^{\textbf{k}'+\textbf{q}}_{\mi \nu_n' + \mi \omega_n} \\
    &=& \frac{1}{\beta} F^{\textbf{k}-\textbf{k}'}_{0} \, G^{\textbf{k}'}_{\mi \nu_n} \,  G^{\textbf{k}'+\textbf{q}}_{\mi \nu_n + \mi \omega_n} &- \frac{1}{2 \pi \mi }\sum_\mathcal{C} \oint_{\mathcal{C}} \md z\: \eta_{F}(z) \, F^{\textbf{k}-\textbf{k}'}_{\mi \nu_n - z} \, G^{\textbf{k}'}_{z} \,  G^{\textbf{k}'+\textbf{q}}_{z + \mi \omega_n} 
    \label{eq:opt_cond_vert_1a}
    \end{array}
    \end{equation}
\end{widetext}
Here, we introduced $F_{i\omega_n=0}^{\textbf{q}} \equiv   F_0^{\textbf{q}} $. 
The contours $\mathcal{C}$ enclose all Matsubara frequencies of the $\nu_n'-$sum except for $\nu_n=\nu_n'$ and are visualized in  Fig.~\ref{fig:branchcuts_vert} as well. 
%

Again the arches (of infinite radius) do not contribute to the integral because the integrand decays faster than $1/|z|$. The only remaining contributions come from the three stripes $\zeta_i$ in Fig.~\ref{fig:branchcuts_vert} (leftward and rightward integral with infinitely small separation in-between), mathematically given by
\begin{widetext}
    \begin{equation}
    \begin{array}{rcll}
    C^{\textbf{k},\textbf{k}',\textbf{q}}(\mi \nu_n, \mi \omega_n) &=&
    \frac{F_{0}^{\textbf{k}-\textbf{k}'}}{\beta}   G_{i\nu_n}
    ^{\textbf{k}'} G_{i\nu_n+i\omega_n}^{\textbf{k}'+\textbf{q}} & \left( =\zeta_{0}\right) \\
    && +\int_{-\infty}^{+\infty} d\nu' \:\frac{-1}{ 2 \pi \mi } 
    \, G_{\nu'+i\nu_n}^{\textbf{k}'} G_{\nu' + i \nu_n +i\omega_n}^{\textbf{k}'+\textbf{q}} 
    \left( \eta_B(\nu' - \mi 0^+) \, F_{-\nu'}^{R \: \textbf{k}-\textbf{k}'} - F_{-\nu'}^{A \: \textbf{k}-\textbf{k}'} \, \eta_B(\nu' + \mi 0^+) \right) & \left( =\zeta_{1}\right) \\
    && +  \int_{-\infty}^{+\infty} d\nu' \: \eta_F(\nu') A_{\nu'}^{\textbf{k}'} 
    G_{\nu' + i\omega_n}^{\textbf{k}'+\textbf{q}} F_{i\nu_n-\nu'}^{\textbf{k}-\textbf{k}'}  & \left(=\zeta_{2} \right) \\
    && +  \int_{-\infty}^{+\infty} d\nu' \: \eta_F(\nu')  A_{\nu'}^{\textbf{k}' + \textbf{q}}
    G_{\nu'-i\omega_n}^{\textbf{k}'} F_{i\nu_n + i \omega_n -\nu'}^{\textbf{k}-\textbf{k}'} & \left( =\zeta_{3} \right)	
    \end{array}
    \label{eq:opt_cond_vert_3}
    \end{equation} 
\end{widetext}
For the second term ($\zeta_{1}$) we used $\eta_{F}(\nu' + \mi \nu_n \pm \mi 0^+  ) = - \eta_{B}(\nu'  \pm \mi 0^+)$. For the singularity of the Bose-Einstein function $n_B(x)=\frac{1}{\me^{\beta x} -1 } $ on can use a variation of the Plemelj-Sokhotskii theorem~\cite{Sokhotskii} that given a well behaved (holomorphic) function $g(x)$ reads
\begin{equation}
\begin{split}
&\int_{-\infty}^{\infty} \md x \: g(x) \; n_B(x-x_0\pm \mi 0^+ ) =  \\ &\mathcal{P} \int_{-\infty}^{\infty} \md x \: g(x) \; n_B(x-x_0) \mp \mi \frac{\pi}{\beta} \int_{-\infty}^{\infty} \md x \: \delta(x-x_0) \; g(x).
\end{split}
\end{equation}
The $\zeta_{1}$-term can be split up into two parts, where the contributions due to the delta-distribution cancels exactly the $\zeta_{0}$ term \footnote{We present here the solution for $F^{\textbf{q}}_{i \omega_n \rightarrow z \rightarrow 0} = F^{\textbf{q}}_{i \omega_n =0}$. While this is true for the systems considered here (RPA, Ornstein-Zernike) it is not correct for some special cases (e.g. non-ergodic systems \cite{Wilcox_1968, Kwok_1969, Suzuki_1971}). The general procedure does, however, not change. The only difference being that the $\zeta_0$ term does no longer cancel. Instead one gets the $\zeta_0$ as an additional contribution, but with $\left( F^{\textbf{q}}_{i \omega_n =0} - F^{\textbf{q}}_{i \omega_n \rightarrow z \rightarrow 0} \right)$ instead of $F^{\textbf{q}}_{0}$.}:
\begin{widetext}
    \begin{equation}
    \begin{array}{rcl}
    \zeta_{1}  &=& \int_{-\infty}^{+\infty} d\nu' \:\frac{-1}{ 2 \pi \mi } 
    \, G_{\nu'+i\nu_n}^{\textbf{k}'} G_{\nu' + i \nu_n +i\omega_n}^{\textbf{k}'+\textbf{q}} 
    \left( \eta_B(\nu' - \mi 0^+) F_{-\nu'}^{R \: \textbf{k}-\textbf{k}'} - F_{-\nu'}^{A \: \textbf{k}-\textbf{k}'} \eta_B(\nu' + \mi 0^+) \right) \\
    &=& \frac{-1}{ 2 \pi \mi }  \mathcal{P} \int_{-\infty}^{+\infty} d\nu' 
    \, G_{\nu'+i\nu_n}^{\textbf{k}'} G_{\nu' + i \nu_n +i\omega_n}^{\textbf{k}'+\textbf{q}} 
    \eta_B(\nu') \, \left( F_{-\nu'}^{R \:\textbf{k}-\textbf{k}'} - F_{-\nu'}^{A \: \textbf{k}-\textbf{k}'} \right)  - \frac{1}{2} G_{i\nu_n}^{\textbf{k}'} G_{i \nu_n +i\omega_n}^{\textbf{k}'+\textbf{q}} 
    \, \left( F_{0}^{R \: \textbf{k}-\textbf{k}'} + F_{0}^{A \: \textbf{k}-\textbf{k}'} \right) \\
    &=& \int_{-\infty}^{+\infty} d\nu' \:\frac{-1}{ 2 \pi \mi } 
    \, G_{\nu'+i\nu_n}^{\textbf{k}'} G_{\nu' + i \nu_n +i\omega_n}^{\textbf{k}'+\textbf{q}} 
    \eta_B(\nu') \, \left( F_{-\nu'}^{R\: \textbf{k}-\textbf{k}'} - F_{-\nu'}^{A \:\textbf{k}-\textbf{k}'} \right) - \zeta_{0} \\
    \end{array}
    \label{eq:opt_cond_vert_corrected}
    \end{equation} 
\end{widetext}

In the last line of \eqref{eq:opt_cond_vert_corrected} we used that $F^{R \: \textbf{k}-\textbf{k}'}_0 = F^{A \: \textbf{k}-\textbf{k}'}_0  = F^{\textbf{k}-\textbf{k}'}_0$ and that the singularity of $n_B(\nu')$ gets regularized by $ \left( F_{-\nu'}^{R \: \textbf{k}-\textbf{k}'} - F_{-\nu'}^{A \: \textbf{k}-\textbf{k}'} \right)$. The Cauchy-principle value $\mathcal{P}$ can therefore be replaced by a normal integral.
%
%
%
Inserting \eqref{eq:opt_cond_vert_3} into \eqref{eq:opt_cond_vert_1} and reordering in such a way that all terms containing $\mi \nu_{n}$ are together leads to

\begin{widetext}
    \begin{equation}
    \begin{array}{rcl}
    C^{\textbf{k},\textbf{k}',\textbf{q}}(\mi \omega_n) &=&  
    -  \frac{1}{2\pi \mi } \int_{-\infty}^\infty \md \nu' \: \eta_{B}(\nu') \, 
    \left( F^{R \: \textbf{k}-\textbf{k}'}_{-\nu'}  - F^{A \: \textbf{k}-\textbf{k}'}_{-\nu'}   \right) \, 
    \bar \zeta^{\textbf{k},\textbf{k}', \textbf{q}}_{1}(\mi \omega_n, \nu')      \\	 
    && +  \int_{-\infty}^\infty \md \nu' \: \eta_{F}(\nu') \, A^{\textbf{k}'}_{\nu'} \, G^{\textbf{k}'+\textbf{q}}_{\nu' + \mi \omega_n}  \,
    \bar \zeta^{\textbf{k},\textbf{k}', \textbf{q}}_{2}(\mi \omega_n, \nu')      \\	 
    && +  \int_{-\infty}^\infty \md \nu' \: \eta_{F}(\nu') \, A^{\textbf{k}'+\textbf{q}}_{\nu'} \, G^{\textbf{k}'+\textbf{q}}_{\nu' + \mi \omega_n} \, 
    \bar \zeta^{\textbf{k},\textbf{k}', \textbf{q}}_{3}(\mi \omega_n, \nu'),      
    \end{array}
    \end{equation}
    %

    \begin{equation}
    \begin{array}{rcl}
    \bar \zeta_{1}(\mi \omega_n, \nu') 
    &=& \frac{1}{\beta} \sum_{\nu_{n}} G^{\textbf{k}}_{\mi \nu_n} \, G^{\textbf{k}'}_{\mi \nu_n + \nu'} \, G^{\textbf{k} + \textbf{q}}_{\mi \nu_n + \mi \omega_n} \, G^{\textbf{k}' + \textbf{q}}_{\mi \nu_n + \nu' + \mi \omega_n} \\ 
    &=& \frac{-1}{2 \pi \mi } \sum_{\mathcal{C}} \oint_{\mathcal{C}} \md z \: \eta_{F}(z) \, 
    G^{\textbf{k}}_{z} \, G^{\textbf{k}'}_{z + \nu'} \, G^{\textbf{k} + \textbf{q}}_{z + \mi \omega_n} \, G^{\textbf{k}' + \textbf{q}}_{z + \nu' + \mi \omega_n}\\
    &=& \frac{-1}{2\pi \mi} \int_{-\infty}^\infty \md \nu \: \eta_{F}(\nu) \, 
    \left( 
    G^{R \: \textbf{k}}_{\nu} \, G^{R \: \textbf{k}'}_{\nu + \nu'} - 
    G^{A \: \textbf{k}}_{\nu} \, G^{A \: \textbf{k}'}_{\nu + \nu'}
    \right)
    G^{\textbf{k}+\textbf{q}}_{\nu + \mi \omega_n} \, 
    G^{\textbf{k}'+\textbf{q}}_{\nu + \nu' +  \mi \omega_n} + \\
    && \frac{-1}{2\pi \mi} \int_{-\infty}^\infty \md \nu \: \eta_{F}(\nu) \, 
    G^{\textbf{k}}_{\nu - \mi \omega_n} \, 
    G^{\textbf{k}'}_{\nu + \nu' -  \mi \omega_n}
    \left( 
    G^{R \: \textbf{k} + \textbf{q}}_{\nu} \, G^{R \: \textbf{k}' + \textbf{q}}_{\nu + \nu'} - 
    G^{A \: \textbf{k} + \textbf{q}}_{\nu} \, G^{A \: \textbf{k}' + \textbf{q}}_{\nu + \nu'}
    \right),\\
    \\
    \bar \zeta_{2}(\mi \omega_n, \nu') &=&
    \frac{1}{\beta} \sum_{\nu_{n}} G^{\textbf{k}}_{\mi \nu_n} \, G^{\textbf{k} + \textbf{q}}_{\mi \nu_n + \mi \omega_n} \, F^{\textbf{k}-\textbf{k}'}_{\mi \nu_n - \nu'} \\
    &=&  \frac{-1}{2\pi \mi } \sum_{\mathcal{C}} \oint_{\mathcal{C}} \md z \: \eta_{F}(z) \, G^{\textbf{k}}_{z} \, G^{\textbf{k} + \textbf{q}}_{z + \mi \omega_n} \, F^{\textbf{k}-\textbf{k}'}_{z - \nu'} \\
    
    &=& \frac{-1}{2 \pi \mi} \int_{-\infty}^{\infty} \md \nu \: \eta_{F}(\nu)
    \left(
    G^{R \: \textbf{k}}_{\nu} \, F^{R \: \textbf{k}-\textbf{k}'}_{\nu - \nu'} - 
    G^{A \: \textbf{k}}_{\nu} \, F^{A \: \textbf{k}-\textbf{k}'}_{\nu - \nu'}
    \right)
    G^{\textbf{k} + \textbf{q}}_{\nu +\mi \omega_n} +  \int_{-\infty}^{\infty} \md \nu \: \eta_{F}(\nu)
    G^{\textbf{k}}_{\nu - \mi \omega_n}
    F^{\textbf{k}-\textbf{k}'}_{\nu - \nu' - \mi \omega_n} 
    A^{\textbf{k} + \textbf{q}}_{\nu}\\
    \\
    \bar \zeta_{3}(\mi \omega_n, \nu') &=&
    \frac{1}{\beta} \sum_{\nu_{n}} G^{\textbf{k}}_{\mi \nu_n} \, G^{\textbf{k} + \textbf{q}}_{\mi \nu_n + \mi \omega_n} \, F^{\textbf{k}-\textbf{k}'}_{\mi \nu_n - \nu' + \mi \omega_n} \\
    &=&  \frac{-1}{2\pi \mi } \sum_{\mathcal{C}} \oint_{\mathcal{C}} \md z \: \eta_{F}(z) \, G^{\textbf{k}}_{z} \, G^{\textbf{k} + \textbf{q}}_{z + \mi \omega_n} \, F^{\textbf{k}-\textbf{k}'}_{z - \nu' + \mi \omega_n} \\
    &=& \int_{-\infty}^{\infty} \md \nu \: \eta_{F}(\nu) \,
    A^{\textbf{k}}_{\nu} \,
    G^{\textbf{k} + \textbf{q}}_{\nu + \mi \omega_n}
    F^{\textbf{k}-\textbf{k}'}_{\nu - \nu' + \mi \omega_n} + 
    \frac{-1}{2 \pi \mi} \int_{-\infty}^{\infty} \md \nu \: \eta_{F}(\nu) \,
    G^{\textbf{k}}_{\nu -\mi \omega_n}
    \left(
    G^{R \: \textbf{k} + \textbf{q}}_{\nu} \, F^{R \: \textbf{k}-\textbf{k}'}_{\nu - \nu'} - 
    G^{A \: \textbf{k} + \textbf{q}}_{\nu} \, F^{A \: \textbf{k}-\textbf{k}'}_{\nu - \nu'}
    \right). \\
    \end{array}
    \label{eq:zetaBars}
    \end{equation}
\end{widetext}

In \eqref{eq:zetaBars} we further  need to transform the Matusbara sums over the first index $\nu_{n}$ in \eqref{eq:opt_cond_vert_1} into contours integrals enclosing all poles of the Fermi-Dirac distribution function $\eta_{F}(z)$. All terms have branch-cuts at two additional lines regarding the analytically continued first summation index $\nu_n \rightarrow z$, namely at
\begin{equation}
z = 
\begin{cases}
\nu, \\
\nu-\mi  \omega_n.
\end{cases}
\label{eq:branchcut_vert_inu}
\end{equation}
Altogether we hence need to account for five branch cuts, when we analytically continue  the Matsubara frequency sums $\nu_n$ and $\nu'_n$ to real frequency integrations.

Inserting \eqref{eq:zetaBars} 
into \eqref{eq:opt_cond_vert_3} and performing the last analytical continuation of the external frequency by substituting $\mi \omega_n \rightarrow \omega + \mi 0^+$  gives 

\begin{equation}
\chi^R_{\text{vert}}(\omega, \textbf{q}) = 
2 \sum_{\textbf{k}\textbf{k}'} \, \gamma_{\alpha}^{\textbf{k}\textbf{q}}\,  \gamma_{\alpha}^{\textbf{k}'(-\textbf{q})} \, 
C^{R \: \textbf{k},\textbf{k}',\textbf{q}}( \omega),
\label{eq:chi_q_dependent}
\end{equation}
with 
\begin{equation} 
C^{R \: \textbf{k},\textbf{k}',\textbf{q}}( \omega) = 
\zeta^{R \: \textbf{k}\textbf{k}'\textbf{q}}_{1}(\omega) + 
\zeta^{R \: \textbf{k}\textbf{k}'\textbf{q}}_{2}(\omega) + 
\zeta^{R \: \textbf{k}\textbf{k}'\textbf{q}}_{3}(\omega),
\label{eq:C_q_dependent}
\end{equation}
and
%
%
%
%
\begin{widetext}
    \begin{subequations}
        \begin{equation}
        \begin{array}{rcll}
        \zeta^{R \: \textbf{k},\textbf{k}'\textbf{q}}_{1}(\omega) &=&    
        \multicolumn{2}{l}{
            \frac{-1}{(2\pi)^2} 
            \int_{-\infty}^{\infty} \md \nu' \: \eta_{B}(\nu') \,
            \left(
            F^{R \: \textbf{k} - \textbf{k}'}_{-\nu'} - 
            F^{A \: \textbf{k} - \textbf{k}'}_{-\nu'}
            \right)
        }
        \\
        && 
        \cdot \int_{-\infty}^{\infty} \md \nu  \: \eta_{F}(\nu) 
        \bigg[ &
        \left( 
        G^{R \: \textbf{k}}_{\nu} \, G^{R \: \textbf{k}'}_{\nu + \nu'} -   
        G^{A \: \textbf{k}}_{\nu} \, G^{A \: \textbf{k}'}_{\nu + \nu'}
        \right)
        G^{R \: \textbf{k}+\textbf{q} }_{\nu + \omega} \, 
        G^{R \: \textbf{k}'+\textbf{q}}_{\nu + \nu' +  \omega} \, +
        \\
        &&& 
        G^{A \: \textbf{k}}_{\nu - \omega} \, 
        G^{A \: \textbf{k}'}_{\nu + \nu' - \omega}
        \left( 
        G^{R \: \textbf{k}  + \textbf{q} }_{\nu} \, G^{R \: \textbf{k}' + \textbf{q} }_{\nu + \nu'} -   
        G^{A \: \textbf{k}  + \textbf{q} }_{\nu} \, G^{A \: \textbf{k}' + \textbf{q} }_{\nu + \nu'}
        \right)
        \bigg]
        \end{array}
        \label{eq:zetaQ2}
        \end{equation}

        \begin{equation}
        \begin{array}{rcl}
        \zeta^{R \: \textbf{k}\textbf{k}'\textbf{q}}_{2}(\omega) &=&    
        %
        \int_{-\infty}^{\infty} \md \nu' \: \eta_{F}(\nu') \,
        A^{ \textbf{k}' }_{\nu'} \, 
        G^{R \: \textbf{k'} + \textbf{q} }_{\nu' + \omega} 
        \,   \int_{-\infty}^{\infty} \md \nu \: \eta_{F}(\nu) \, 
        \\
        && 
        \cdot
        \left[
        \frac{-1}{2\pi \mi }
        \left(
        G^{R \: \textbf{k}}_{\nu} \, F^{R \: \textbf{k}-\textbf{k}'}_{\nu - \nu'} - 
        G^{A \: \textbf{k}}_{\nu} \, F^{A \: \textbf{k}-\textbf{k}'}_{\nu - \nu'}
        \right)
        G^{R \: \textbf{k} + \textbf{q}}_{\nu + \omega} + 
        %
        G^{A \: \textbf{k}}_{\nu - \omega} \, 
        F^{A \: \textbf{k}-\textbf{k}'}_{\nu - \nu' - \omega} \,  
        A^{\textbf{k} + \textbf{q}}_{\nu}
        \right]
        \end{array}
        \label{eq:zetaQ3}
        \end{equation}
        \begin{equation}
        \begin{array}{rcl}
        \zeta^{R \: \textbf{k}\textbf{k}'\textbf{q}}_{3}(\omega) &=&    
        %
        \int_{-\infty}^{\infty} \md \nu' \: \eta_{F}(\nu') \,
        G^{A \: \textbf{k'} }_{\nu' - \omega} \, 
        A^{ \textbf{k}' + \textbf{q} }_{\nu'} \,
        \int_{-\infty}^{\infty} \md \nu \: \eta_{F}(\nu) \, \cdot 
        \\
        && 
        \cdot
        \left[
        A^{\textbf{k}}_{\nu} \,
        G^{R \: \textbf{k} + \textbf{q}}_{\nu + \omega}
        F^{R \: \textbf{k}-\textbf{k}'}_{\nu - \nu' + \omega} + 
        \frac{-1}{2\pi \mi }
        G^{A \: \textbf{k}}_{\nu - \omega}
        \left(
        G^{R \: \textbf{k} + \textbf{q}}_{\nu} \, F^{R \: \textbf{k}-\textbf{k}'}_{\nu - \nu'} - 
        G^{A \: \textbf{k} + \textbf{q}}_{\nu} \, F^{A \: \textbf{k}-\textbf{k}'}_{\nu - \nu'}
        \right)
        \right].
        \end{array}
        \label{eq:zetaQ4}
        \end{equation}
    \end{subequations}
\end{widetext}
For the long-wavelength limit ($\textbf{q}=0$) \eqref{eq:C_q_dependent} may also be written as 
\begin{widetext}
    \begin{equation}
    \chi^{R}_{\text{vert}}(\omega,\textbf{q} = 0) = -2 \sum_{\textbf{k}\textbf{k}'}  \gamma_{\alpha}^\textbf{k} \gamma_{\alpha}^{\textbf{k}'} \left(
    \zeta_1^{\textbf{k}\textbf{k}'}(\omega) + 
    \zeta_{2p3}^{\textbf{k}\textbf{k}'}(\omega)
    \right),
    \label{eq:opt_cond_vert_final:appendix}
    \end{equation}
    %
    %
    \begin{subequations}
        \label{eq:zeta_final_appendix}
        \begin{flalign}
        \zeta_1(\omega)^{R \: \textbf{k}\textbf{k}'} & = 
        \begin{aligned}[t]
        \frac{-1}{4 \pi^2} \iint\limits_{\mathbb{R}^2}  \md\nu \, \md \nu'   \: \eta_F(\nu) \, \eta_B(\nu')\big[ F_{-\nu'}^{R \: \textbf{k}-\textbf{k}'}- F_{-\nu'}^{A \: \textbf{k}-\textbf{k}'}  \big]
        & \mathrlap{\big[G_{\nu + \omega}^{R \: \textbf{k}} G_{\nu+\nu'+\omega}^{R \: \textbf{k}'} + G_{\nu-\omega}^{A \: \textbf{k}} G_{\nu+\nu'-\omega}^{A \: \textbf{k}'} \big]} \\ 
        & \mathrlap{\big[G_{\nu}^{R \: \textbf{k}} G_{\nu'+\nu}^{R \: \textbf{k}'}-G_{\nu}^{A \: \textbf{k}} G_{\nu'+\nu}^{A \: \textbf{k}'} \big],}
        \label{eq:zeta_2}
        \end{aligned} \\
        \zeta_{2p3}(\omega)^{R \: \textbf{k}\textbf{k}'} & =  
        \begin{aligned}[t]
        \frac{i}{2 \pi} \iint\limits_{\mathbb{R}^2}  \md \nu \, \md \nu'  \: \eta_F(\nu) \, \eta_F(\nu') A_{\nu'}^{\textbf{k}'} \bigg[ & \big[ G_{\nu + \omega}^{R \: \textbf{k}} G_{\nu'+\omega}^{R \: \textbf{k}'} + G_{\nu-\omega}^{A \: \textbf{k}} G_{\nu'-\omega}^{A \: \textbf{k}'} \big] \big[G_{\nu}^{R \: \textbf{k}} F_{\nu-\nu'}^{R \: \textbf{k}-\textbf{k}'} - G_{\nu}^{A \: \textbf{k}} F_{\nu-\nu'}^{A \: \textbf{k}-\textbf{k}'} \big] + \\ 
        & (-2\pi i) A_{\nu}^{\textbf{k}} \big[ G_{\nu'+\omega}^{R \: \textbf{k}'} G_{\nu-\omega}^{A \: \textbf{k}} F_{\nu-\nu'-\omega}^{A \: \textbf{k}-\textbf{k}'} + G_{\nu'-\omega}^{A \: \textbf{k}'} G_{\nu+\omega}^{R \: \textbf{k}} F_{\nu-\nu'+\omega}^{R \: \textbf{k}-\textbf{k}'} 
        \big]
        \bigg].
        \label{eq:zeta_3p4}
        \end{aligned}
        \end{flalign}
    \end{subequations}
\end{widetext}
\begin{figure*}
    \centering
    \scalebox{1}{\includegraphics[width=1\textwidth]{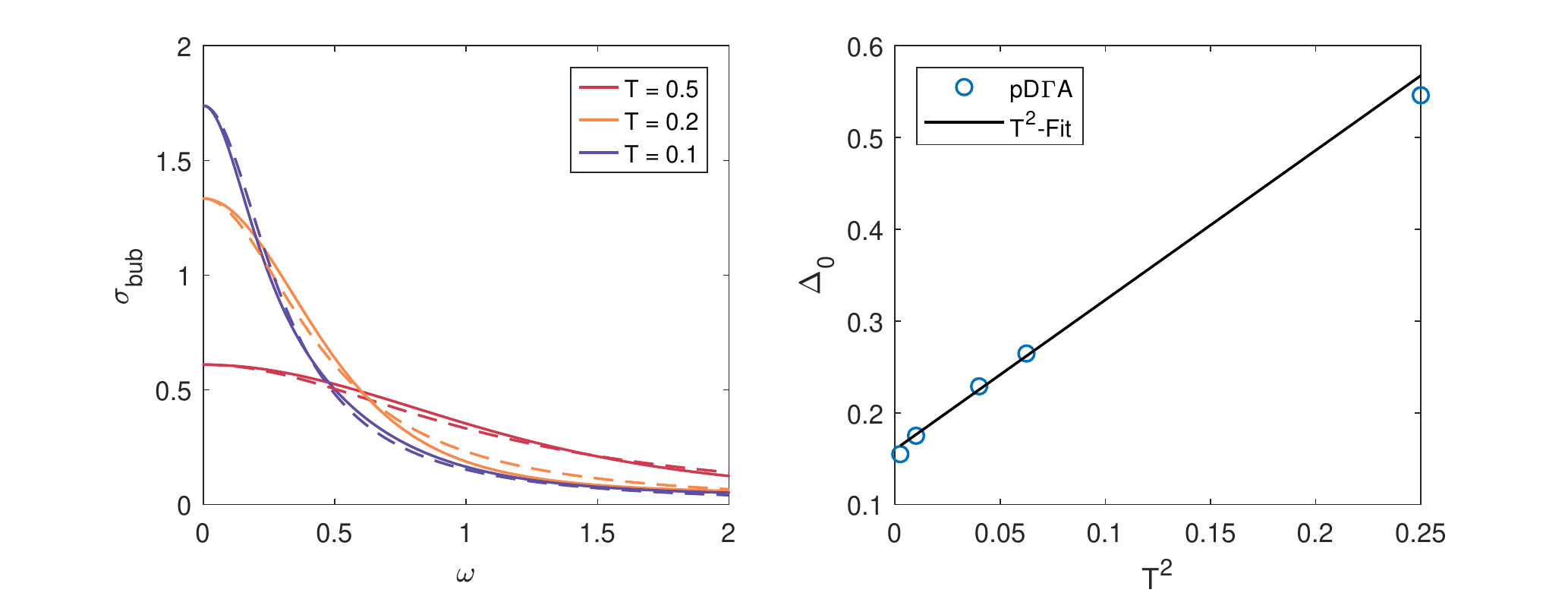}}
    \caption{Left: $\sigma_{\rm{bub}}$ from Ref.~\onlinecite{pi_ton} (solid line) and Drude fit (dashed). Right: $\Delta_0= \frac{1}{2 \tau}$ obtained from the Drude peak fit (blue circles) at various temperatures and the fitted temperature dependence $\Delta_0(T) =0.1547 + 1.637~T^2$  (black line). The parquet D$\Gamma$A calculations in Ref.~\onlinecite{pi_ton} were done  with a interaction value of $U_0=4$.
    }
    \label{fig:Drude_fit}
\end{figure*}

In \eqref{eq:zeta_3p4} we combined $\zeta_2^{R \: \textbf{k},\textbf{k}',\textbf{q}=0} $ plus  $\zeta_3^{R \: \textbf{k},\textbf{k}',\textbf{q}=0}$ into the sum $\zeta_{2p3}^{R \: \textbf{k},\textbf{k}'} $ so that each individual term  fulfills the symmetry $\zeta_{i}(-\omega)^* = \zeta_{i}(\omega)$ for $i\in\{1,\, 2p3 \}$. 

    \section{Computational Details}
    \label{app:numerical}
    \subsection{$k$ and $\omega$ grid sizes}
    
    For all calculations presented we used the  ${\bf k}$-grid of at least $30\times 30$ points. The numeric frequency integration was done using trapezoidal rule with at least $751$ points within the frequency range of $\omega \in [-45,45]$. 
    
    \subsection{Drude fit to pD$\Gamma$A bubble}
    \label{app:dga_fit}
    In order to obtain the temperature dependent scattering rate $\Delta_0(T)$, we used a $\chi^2$-fit of the Drude form  $\sigma(\omega) = \frac{\sigma_0}{1+\omega^2 \tau^2}\equiv \frac{\sigma_0}{1+\omega^2 /(2 \Delta_0)^2}$ to the analytically continued bubble contribution of the p-D$\Gamma$A optical conductivity obtained in Ref.~\onlinecite{pi_ton} for different temperatures and the bare interaction value of $U_0=4$. Both the Drude fit, as well as the original p-D$\Gamma$A bubble contribution are shown in the left panel of Fig.~\ref{fig:Drude_fit} for different temperatures. The extracted temperature dependence together with a $T^2$ fit is shown in the right panel of Fig.~\ref{fig:Drude_fit}.

    \begin{figure*}
        \centering
        \scalebox{1}{\includegraphics[width=1\textwidth]{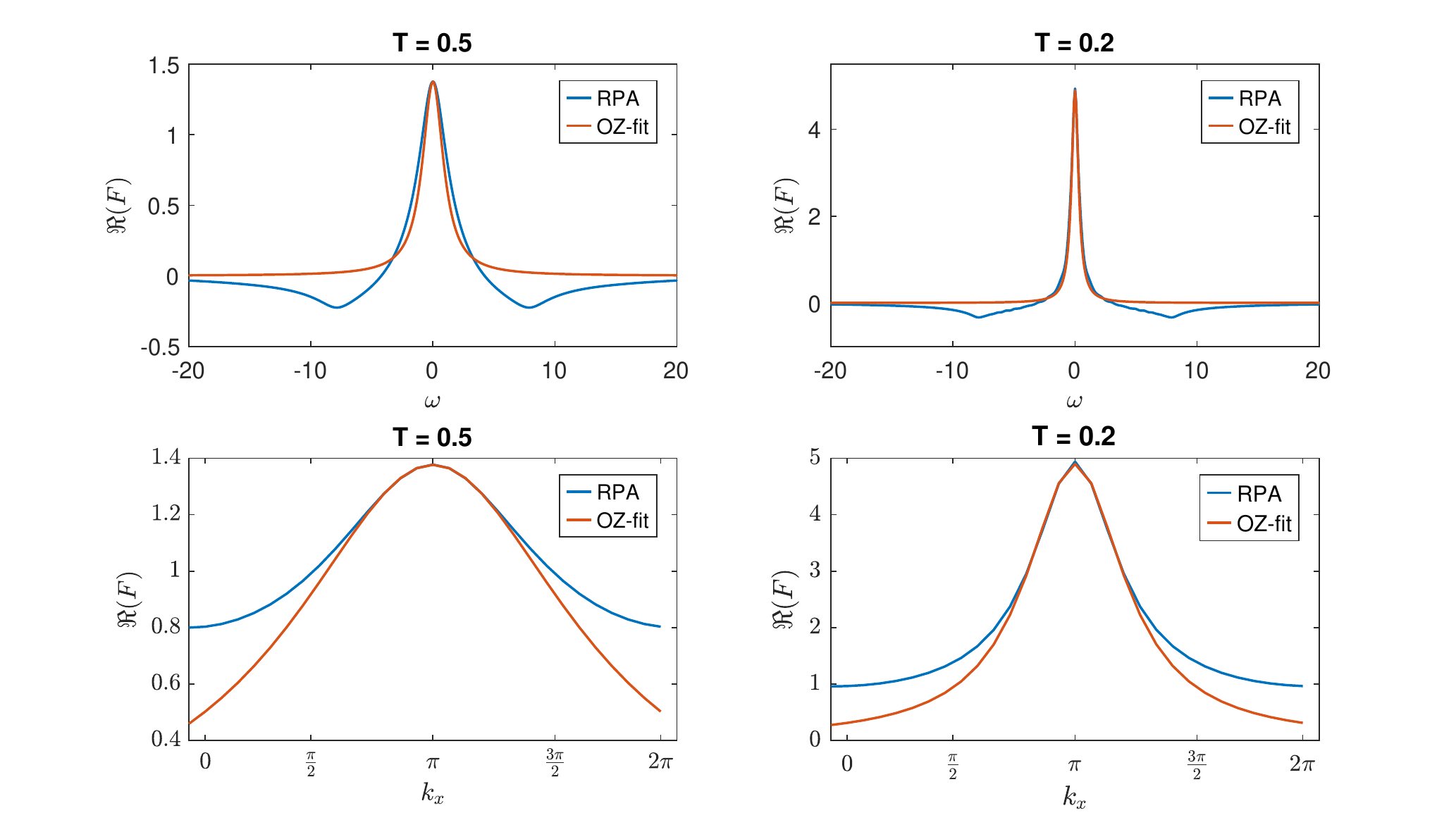}}
        \caption{Real part of the Ornstein-Zernike vertex (red) fitted to an RPA-ladder vertex (blue) for two different temperatures. Left: $T = 0.5$; right: $T = 0.2$, and  $U = 1.9$ for the RPA-ladder. Top row: as a function of $\omega$ for a fixed ${\bf q} =(\pi,\pi)$, yielding $\lambda$. Bottom row: as a function of $k_x$ for $k_y = \pi$ and $\omega=0$, determining $\xi$ and $A$.
        }
        \label{fig:OZ_fit_2_RPA}
    \end{figure*}

    \subsection{Fit of the Ornstein-Zernike vertex to the RPA ladder vertex}
    \label{app:OZ_fits}
    We extract the parameters of the Ornstein-Zernike vertex of \eqref{eq:OZ_vertex}: $A$, $\xi$, $\lambda$  through a  $\chi^2$-fit to the real part of the RPA-ladder vertex  \eqref{eq:RPA_effective_vertex}. All three parameters are, however, not extracted simultaneously. First $A$ and $\xi$ are obtained by fitting to the value at $\omega = 0$. Then, keeping $A$ and $\xi$ constant, $\lambda$ is obtained by a $\chi^2$-fit to the $\omega$ dependence. A comparison of the fitted OZ vertex to the RPA vertex is shown in Fig.\ref{fig:OZ_fit_2_RPA}.

    \begin{figure*}
        \centering
        \scalebox{1}{\includegraphics[width=1\textwidth]{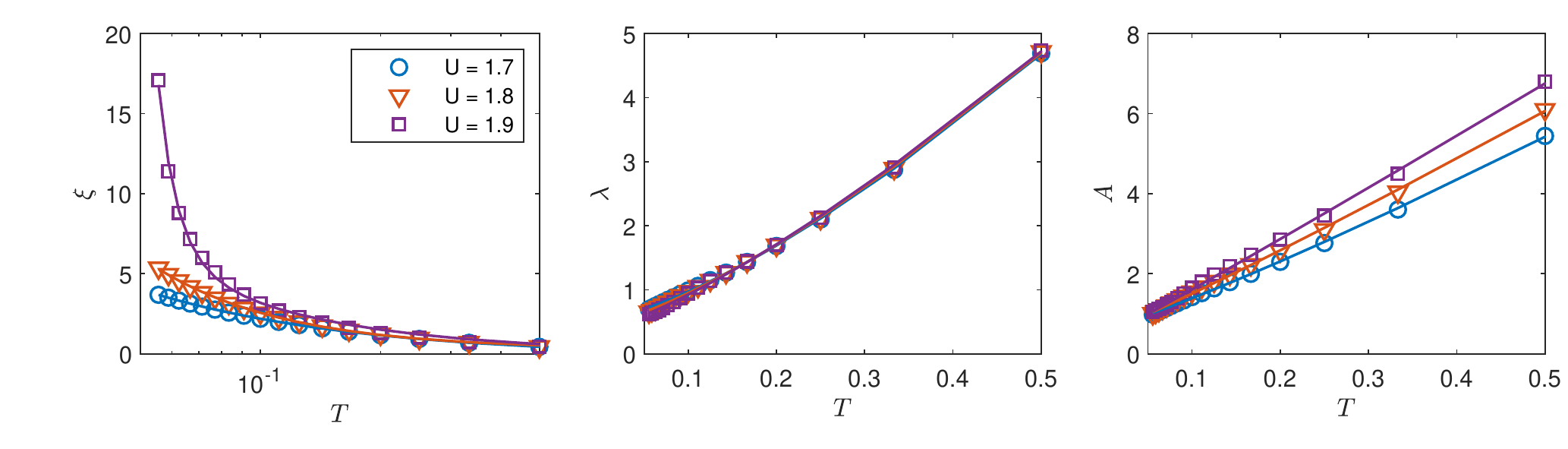}}
        \caption{Temperature dependence of the  Ornstein-Zernike parameters (symbols) $\xi$ (left), $\lambda$ (middle), and  $A$ (right)---together with the fit functions for the temperature dependence   (solid lines; as described in the text). Three different values of $U = 1.7,~1.8,~1.9$ are used to extract the OZ parameters  from the RPA ladders (blue circles, red triangles, purple squares). 
        }
        \label{fig:OZ_fit_2_RPA_parameter}
    \end{figure*}

    Fig.~\ref{fig:OZ_fit_2_RPA_parameter} in turn displays the extracted parameters of the Ornstein-Zernike vertex as a function of temperature (symbols). The solid lines represent a fitted temperature behavior of the form $a_{\rm{OZ}} + b_{\rm{OZ}} \cdot (1/T)^{C_{\rm{OZ}}} $,  except for fitting $\xi$ at $U = 1.9$, where an exponential fit of the form $ \frac{a_{\rm{OZ}}}{T}+ b_{\rm{OZ}} \cdot e^{\frac{C_{\rm{OZ}}}{T}}$ had to be employed to account for the strong increase $\xi$ at low temperatures, see purple line in Fig.~\ref{fig:OZ_parameter_scan_appendix} (left). The respective fit parameters $a_{\rm{OZ}},~b_{\rm{OZ}},~C_{\rm{OZ}}$  are given in Table~\ref{tab:OZ_fitparameter}.  
    
    \begin{table}
        \centering
        \caption{\label{tab:OZ_fitparameter} Parameters extracted from fitting the temperature dependence of the three Ornstein-Zernike parameters; see text for the fit function and note the different fit function for  $\xi$ at $U = 1.9$.}
        \begin{tabular}{cccccccccccc}  
            \toprule
            & \multicolumn{3}{c}{$a_{\rm{OZ}}$} && \multicolumn{3}{c}{$b_{\rm{OZ}}$} && \multicolumn{3}{c}{$C_{\rm{OZ}}$}    \\
            \cmidrule{2-4}                 \cmidrule{6-8}                  \cmidrule{10-12}
            $U$		   & 1.7     & 1.8    & 1.9            &&    1.7   &   1.8   & 1.9          && 1.7    &    1.8    & 1.9             \\
            \midrule
            $A$        & 0.63   & 0.53  & 0.41          &&  10.6   &  11.7  & 13.0        && -1.15  &  -1.08   & -1.03          \\
            $\xi$      & -0.19  & 0.26  & 0.30          &&  0.36   &  0.11  & 1e-3      &&  0.83  &  1.33   & 0.51           \\
            $\lambda$  & 0.53   & 0.46  & 0.38          &&  10.9   &  10.7  & 10.6        && -1.39  &  -1.37  & -1.29           \\	 	
            \bottomrule
        \end{tabular}
    \end{table}
    
   \section{Ornstein-Zernike vertex corrections to the optical conductivity for high temperature}
    \label{app:OZ_high_temp}
    
        \begin{figure*}
        \centering
        \scalebox{1}{\includegraphics[width=1\textwidth]{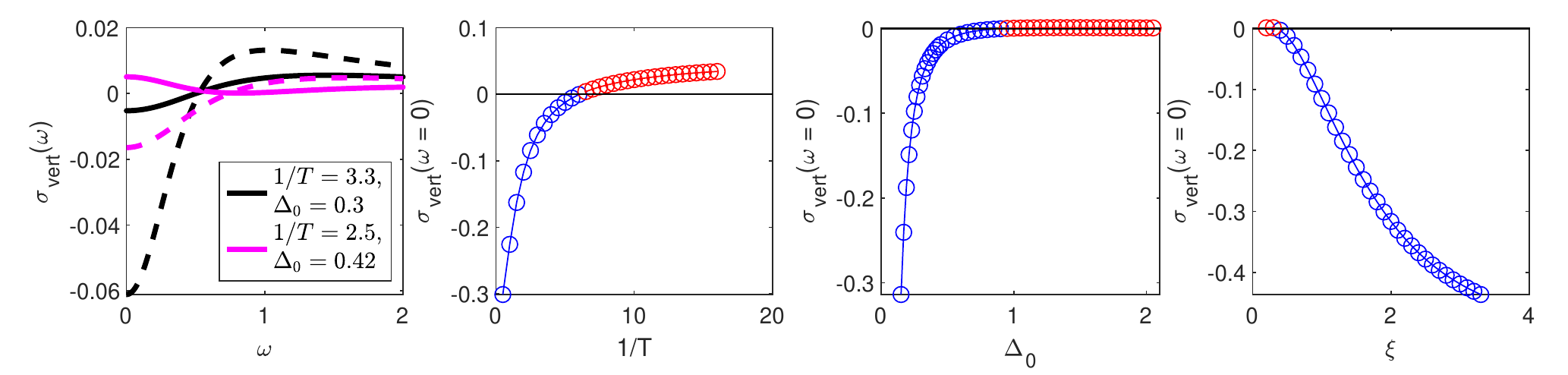}}
        \caption{Left: Vertex corrections calculated directly with the RPA vertex (solid lines) and from  OZ vertex with fitted parameters to the RPA (dashed lines) for parameters of the the black circle (black lines) and magenta triangle (magenta lines) in Fig.~\ref{fig:contour}. The other three plots show  $\sigma_{\vert}(\omega = 0)$ obtained with the OZ vertex with only one parameter changed: $1/T$ (middle left), $\Delta_0$ (middle right)  and $\xi$ (right),  all other parameters are kept constant. The starting point was the black circle in Fig.~\ref{fig:contour}, which corresponds to $1/T = 3.3, \Delta_0 = 0.3, \xi = 0.77$. To improve visibility, positive values are marked as red, negative ones as blue.}
        \label{fig:OZ_parameter_scan_appendix}
    \end{figure*}
    
In the high temperature regime we see a transition between broadening of the Drude peak and its sharpening when the temperature is increased in case we use the RPA vertex. This is illustrated in Fig.~\ref{fig:contour}, where for high temperatures or large $\Delta_0$ the vertex corrections are positive. We choose now two points on the green line, with $1/T=3.3$ and $\Delta_0=0.3$ (black circle in Fig.~\ref{fig:contour}) and with  $1/T=2.5$ and $\Delta_0=0.42$  (magenta triangle in Fig.~\ref{fig:contour}) and perform an OZ-fit to the RPA vertex for each of them. With the obtained OZ parameters,  we calculate the vertex corrections with the OZ vertex instead of the RPA vertex. The results for the two different ways of calculating  $\sigma_{\textrm{vert}}$, i.e. directly from the RPA vertex or alternatively from OZ vertex with parameters fitted on RPA, are shown in the leftmost plot in Fig.~\ref{fig:OZ_parameter_scan_appendix}. Solid lines denote the RPA vertex corrections that change sign when we move from the black circle (black in  Fig.~\ref{fig:OZ_parameter_scan_appendix}) to the magenta triangle (magenta). Dashed lines denote the OZ results for these points. We do not see a sign change for the OZ vertex, although its parameters were fitted to the RPA vertex. We neither see a sharpening of the Drude peak at high temperature with thus fitted OZ vertex.

In order to check if it is at all possible to obtain positive vertex corrections with the OZ vertex, we choose the black circle of  Fig.~\ref{fig:contour} as a starting point and change the temperature, correlation length $\xi$ and $\Delta_0$ treating them as independent parameters, as was done in Fig.~\ref{fig:OZ_opt_cond_xi_scan} for a lower temperature. We show in  Fig.~\ref{fig:OZ_parameter_scan_appendix}  only the value of  $\sigma_{\textrm{vert}}(\omega=0)$ as a function $1/T$, $\Delta_0$, and $\xi$, with other parameters kept fixed (left middle, right middle, and rightmost plot, respectively). With lowering the temperature we enter the sharpening regime with the OZ vertex as well, even if we keep the correlation length fixed. Staying at higher temperature ($1/T=3.3$) however, we get positive values of $\sigma_{\textrm{vert}}(\omega=0)$ only at $\Delta_0\gtrsim 1$ or with a very small correlation length $\xi< 0.3$. In the small correlation length or large scattering rate regime, the vertex corrections become positive at high temperature even with the OZ form of the vertex.

Let us note, however, that at high temperature the correlation length description of the system is not any more adequate, which also explains the differences between OZ and RPA results.  The correlation between next neighbors still changes with the correlation length if it is somewhat below 1. As a matter of course in this regime the description by means of a correlation length is not accurate any more, as also the quite substantial deviation from the Ornstein-Zernike form for $T=0.5$ in Fig.~\ref{fig:OZ_fit_2_RPA} (left) shows.
    
     \section{RPA ladders in $ph$ and $\overline{ph}$ channels}
     \label{app:rpa_ladders}
     Motivated by the results of Ref.~\onlinecite{pi_ton}, we state in Sec.~\ref{sec:conclusions}, that the dominant contribution to the vertex correction to optical conductivity comes from an RPA-type ladder in the $\overline{ph}$ channel. Here we discuss all possible contributions from ladder-type diagrams in the $ph$ and $\overline{ph}$ channels and explain why RPA-type ladders other than in the magnetic $\overline{ph}$ channel are not important.
     
     Assuming SU(2) spin-symmetry, the vertex entering Eq.~\ref{eq:current_current_correlator_F} is the full two-particle vertex $F_{\rm{d}}$ in the  \textit{density} spin channel~\cite{Tremblay2011}, which (in the notation of Ref.~\onlinecite{Rohringer_Diagrammatic_extensions}) is defined as 
     \begin{equation}
        F_{\text{d}} = F_{\uparrow\uparrow\uparrow\uparrow} + F_{\uparrow\uparrow\downarrow\downarrow} \equiv F_{\uparrow\uparrow} + F_{\uparrow\downarrow},
        \label{eq:density_vertex}
     \end{equation}
     where we used a shortened notation of ${\uparrow\uparrow}$ for ${\uparrow\uparrow\uparrow\uparrow}$ and ${\uparrow\downarrow}$ for ${\uparrow\uparrow\downarrow\downarrow}$. The remaining independent spin combination is ${\uparrow\downarrow\downarrow\uparrow}$, which we denote by $\overline{\uparrow\downarrow}$. For completeness we also introduce the \textit{magnetic} spin channel
          \begin{equation}
        F_{\text{m}} = F_{\uparrow\uparrow\uparrow\uparrow} - F_{\uparrow\uparrow\downarrow\downarrow} \equiv F_{\uparrow\uparrow} - F_{\uparrow\downarrow}.
        \label{eq:magnetic_vertex}
     \end{equation}
     
          \begin{figure*}
        \centering
        \scalebox{1}{\includegraphics[width=1\textwidth]{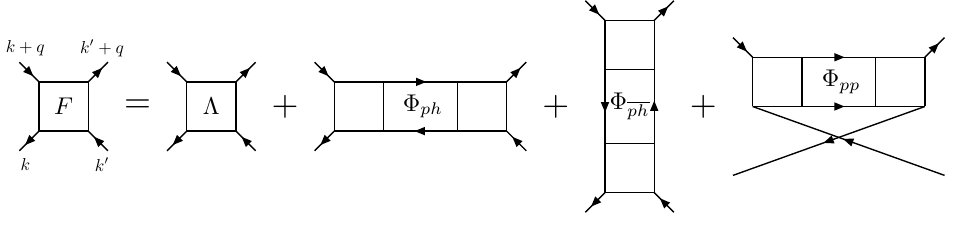}}
        \caption{Visual representation of the parquet decomposition in Eq.~\ref{eq:parquet_decomposition}. The full vertex $F$ consists of diagrams that are two-particle reducible in $ph$, $\overline{ph}$, or  $pp$ channel ($\Phi_{ph}$, $\Phi_{\overline{ph}}$, $\Phi_{pp}$) and diagrams that are not two-particle reducible (fully irreducible, $\Lambda$)}
        \label{fig:parquet_decomposition}
    \end{figure*}
    
     Using the parquet decomposition \cite{Rohringer_Diagrammatic_extensions} allows for unambiguous classification in terms of two-particle reducibility, see Fig.~\ref{fig:parquet_decomposition},
     \begin{equation}
     \label{eq:parquet_decomposition}
        F = \Lambda + \Phi_{ph} + \Phi_{\overline{ph}} + \Phi_{pp},
     \end{equation}
     with $\Lambda$ denoting the set of fully irreducible diagrams (diagrams that are not reducible in $ph$, nor $\overline{ph}$, nor $pp$ channels).
     
\begin{figure}
         \centering
         \begin{subfigure}[b]{0.4\textwidth}
             \centering
             \includegraphics[width=\textwidth]{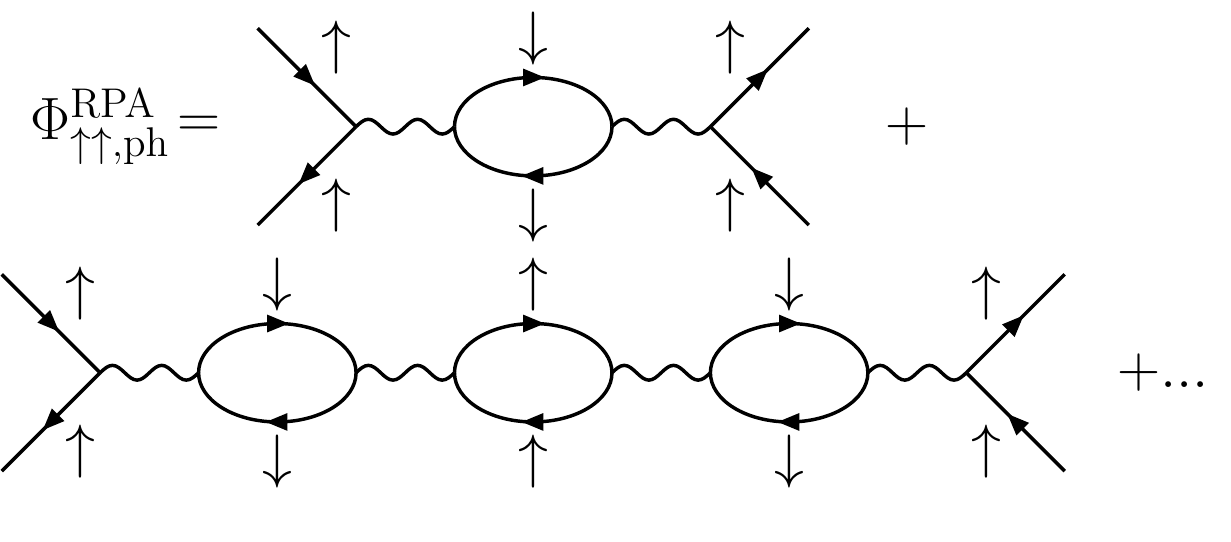}
             \label{fig:phupup}
         \end{subfigure}
         \begin{subfigure}[b]{0.4\textwidth}
             \centering
             \includegraphics[width=\textwidth]{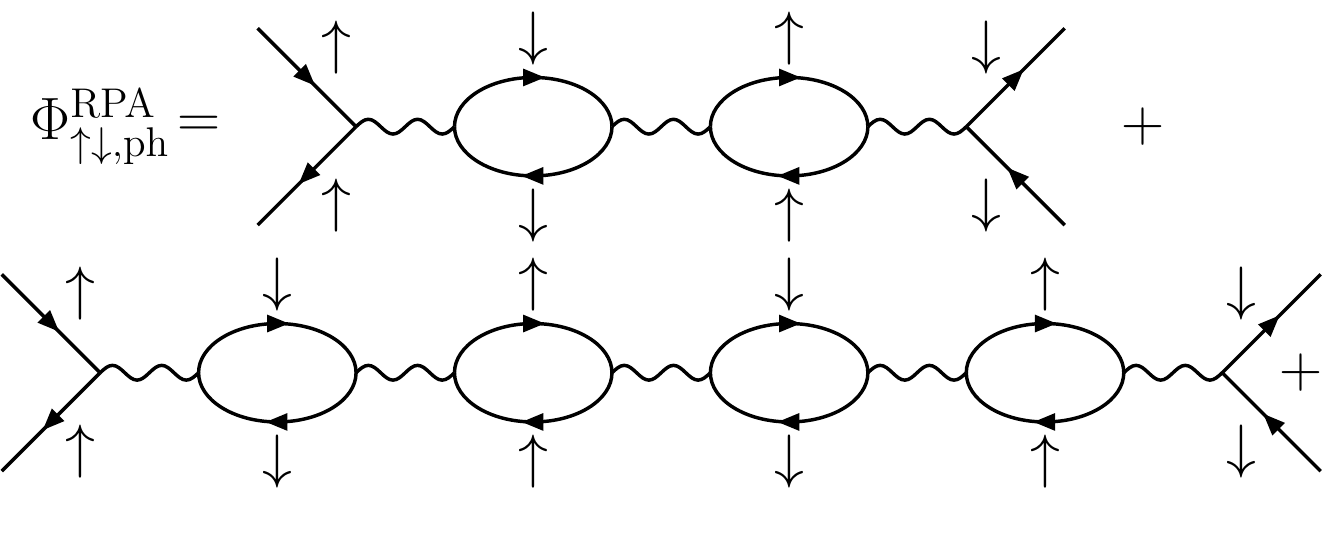}
             \label{fig:phupdo}
         \end{subfigure}
         \begin{subfigure}[b]{0.4\textwidth}
             \centering
             \includegraphics[width=\textwidth]{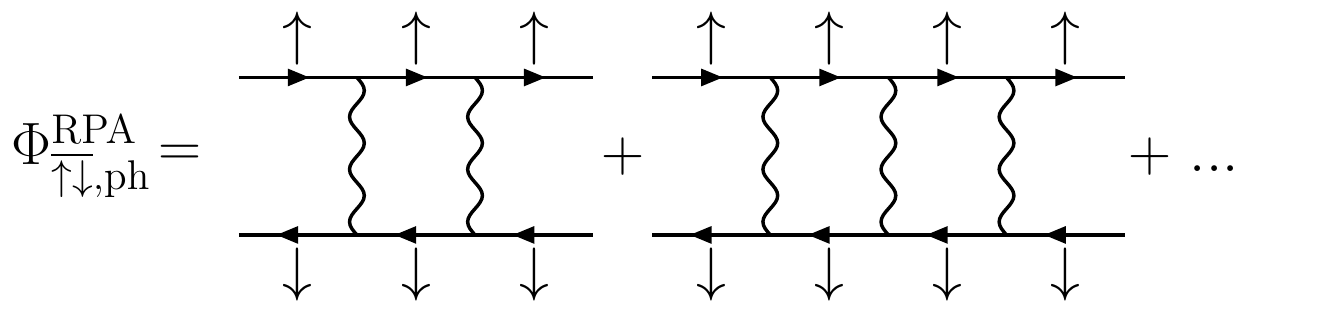}
             \label{fig:phupdobar}
         \end{subfigure}
            \caption{RPA-like ladders in the $ph$ channel.}
            \label{fig:RPAph}
    \end{figure}

\begin{figure}
         \centering
         \begin{subfigure}[b]{0.35\textwidth}
         \centering
             \includegraphics[width=\textwidth]{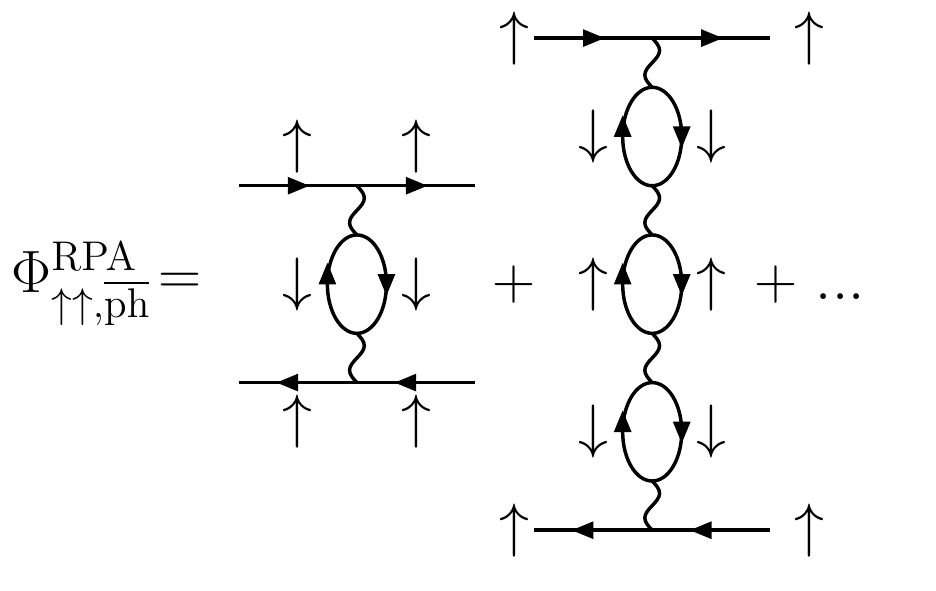}
             \label{fig:phbarupup}
         \end{subfigure}
         \begin{subfigure}[b]{0.4\textwidth}
         \centering
             \includegraphics[width=\textwidth]{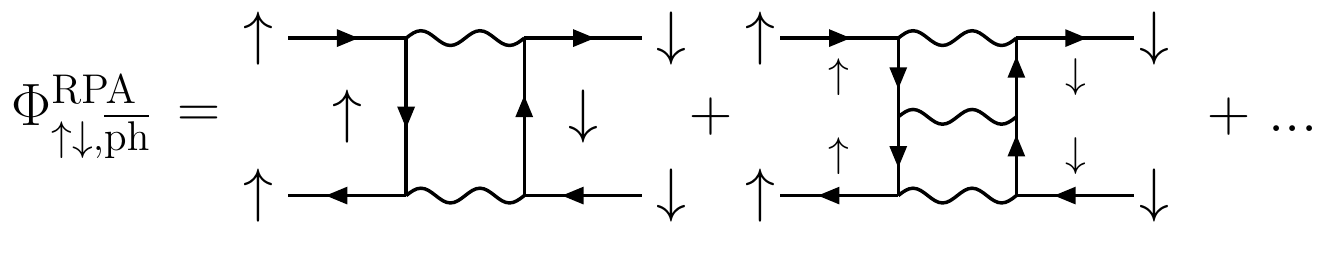}
             \label{fig:phbarupdo}
         \end{subfigure}
         \begin{subfigure}[b]{0.35\textwidth}
             \centering
             \includegraphics[width=\textwidth]{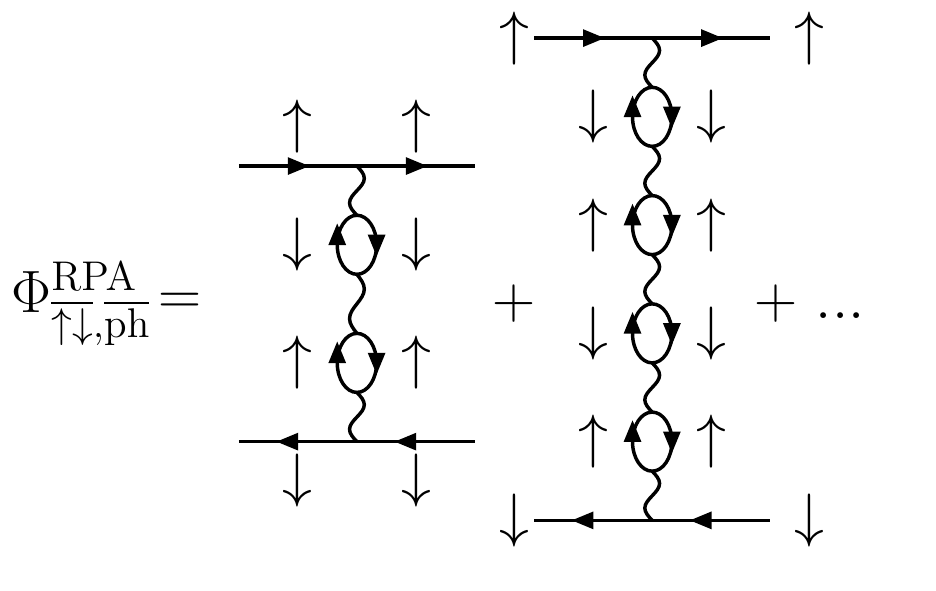}
             \label{fig:phbarupdobar}
         \end{subfigure}
            \caption{RPA-like ladders in the $\overline{ph}$ channel.}
            \label{fig:RPAphbar}
    \end{figure}

      All RPA-like ladders are by construction reducible in one of the three channels. Each channel contains three ladders, one for each of the three spin combinations ($\uparrow\uparrow$, $\uparrow\downarrow$, $\overline{\uparrow\downarrow}$). 
     For the $ph$ and $\overline{ph}$ channels these are given by:
     \begin{subequations}
        \label{eq:RPA_ladders}
        \begin{flalign}
                \Phi^{\text{RPA},kk'q}_{\uparrow\uparrow,ph} & = \frac{U^2\chi_0^{q}}{1-U^2(\chi_0^{q})^2}, \\ 
                \Phi^{\text{RPA},kk'q}_{\uparrow\downarrow,ph} & = \frac{-U^3(\chi_0^{q})^2}{1-U^2(\chi_0^{q})^2}, \\ 
                \Phi^{\text{RPA},kk'q}_{\overline{\uparrow\downarrow},ph} & = \frac{U^2\chi_0^{q}}{1-U\chi_0^{q}}, \\ 
                \Phi^{\text{RPA},kk'q}_{\uparrow\uparrow,\overline{ph}} & = \frac{-U^2\chi_0^{k-k'}}{1-U^2(\chi_0^{k-k'})^2}, \\ 
                \Phi^{\text{RPA},kk'q}_{\uparrow\downarrow,\overline{ph}} & = \frac{-U^2\chi_0^{k-k'}}{1-U\chi_0^{k-k'}}, \\ 
                \Phi^{\text{RPA},kk'q}_{\overline{\uparrow\downarrow},\overline{ph}} & = \frac{U^3(\chi_0^{k-k'})^2}{1-U^2(\chi_0^{k-k'})^2},
        \end{flalign}
     \end{subequations}
     with $\chi_{0}^q = -\frac{1}{\beta N}\sum_{k} G_k G_{k+q}$. The first two diagrams of those RPA-like ladders are also displayed in Fig.\ref{fig:RPAph} (in the $ph$ channel) and Fig.\ref{fig:RPAphbar} (in the $\overline{ph}$ channel). Note, that they have a reduced frequency-momentum dependence and depend only either on $q$  ($ph$ channel) or on $k-k'$  ($\overline{ph}$ channel).

     
     If we restrict ourselves only to such RPA-like ladders and assume the pp-channel to be small as observed in Ref.~\onlinecite{pi_ton} the vertex in Eq.~(\ref{eq:density_vertex}) can be written as,
     
    \begin{equation}
        F_{\text{d}}^{kk'q} = U + \Phi^{\text{RPA},kk'q}_{d,ph} + \Phi^{\text{RPA},kk'q}_{d,\overline{ph}}.
        \label{eq:density_vertex2}
    \end{equation}
    
    Using the so-called crossing symmetry that connects the $ph$ and $\overline{ph}$ channels (see e.g. Ref.~\cite{Rohringer_Diagrammatic_extensions}),
    \begin{equation}
        \Phi^{kk'q}_{d,\overline{ph}} = -\frac{1}{2} \Phi^{(k'+q)k'(k-k')}_{d,ph} - \frac{3}{2} \Phi^{(k'+q)k'(k-k')}_{m,ph}, 
    \end{equation}
    one can rewrite Eq.~\ref{eq:density_vertex2} in order to express everything in the $ph$-channel only
    \begin{equation}
        F_{\text{d}}^{kk'q} = U + \Phi^{\text{RPA},q}_{d,ph} -\frac{1}{2} \Phi^{\text{RPA},(k-k')}_{d,ph} - \frac{3}{2} \Phi^{\text{RPA},(k-k')}_{m,ph},
        \label{eq:density_vertex3}
    \end{equation}
     where we omitted in $\Phi$'s the first two (fermionic) frequency-momentum arguments, because in the RPA-ladder approximation in the $ph$-channel they depend only on the third, bosonic, argument (see Eqs.~(\ref{eq:RPA_ladders})). Note, that such constructed vertex $F$ obeys the crossing-symmetry, as opposed to a single RPA ladder in the $ph$ channel.
     
          \begin{figure*}
         \centering
        \includegraphics[width=1\textwidth]{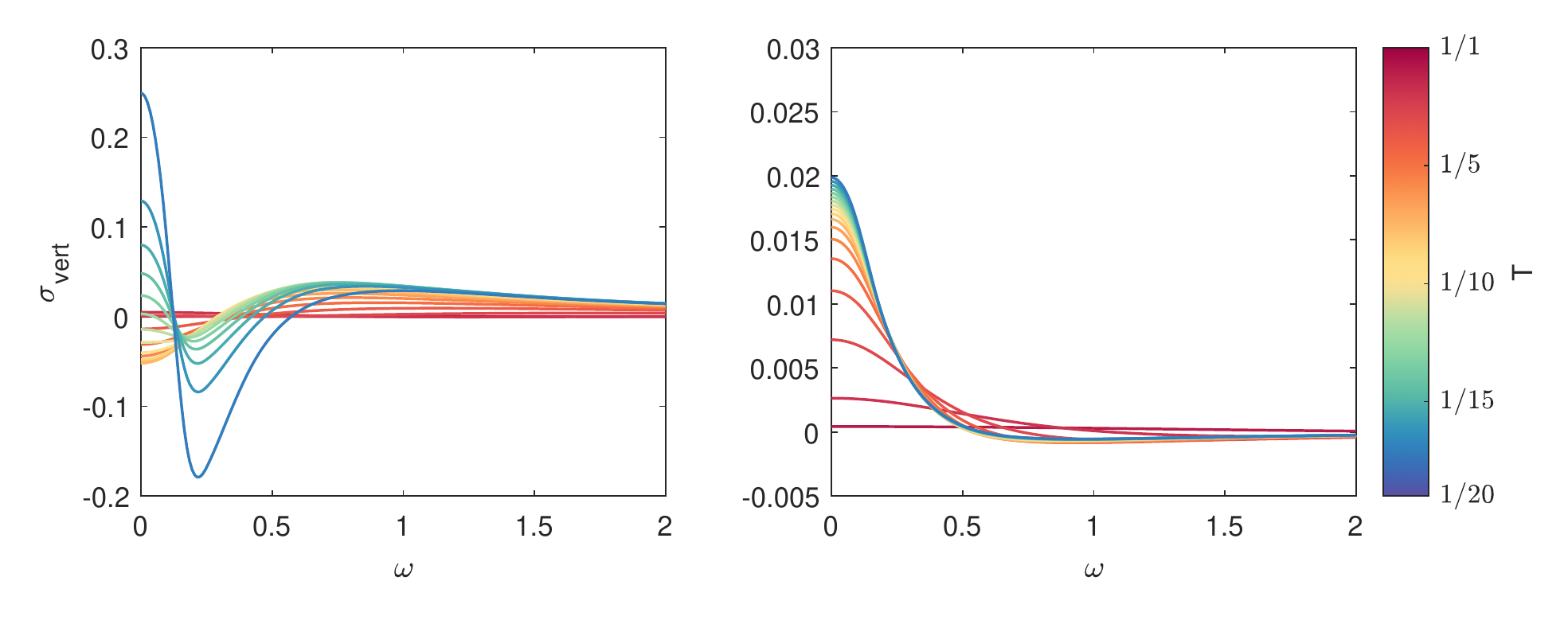}
        \caption{Left: Vertex corrections obtained by using both the magnetic as well as the density RPA-like ladder as vertex (Eq.~\ref{eq:density_vertex3}). Right: Vertex corrections stemming from the density type ladder only. The contribution from the density ladder is small compared to the magnetic ladder. As parameters we use $U = 1.9$ and a constant in frequency but temperature dependent scattering rate $\Delta(T)$ as described in the main text.}
        \label{fig:density_vertex_corrections}
 \end{figure*}
     
     The first two contributions to $F$ in Eq.~(\ref{eq:density_vertex3}) {do}  not give any vertex correction to the optical ($\mathbf{q}=0$) conductivity for a one-band model. This is because $\gamma_{\alpha}^{\textbf{k}}$ is anti-symmetric in $\textbf{k}$, while $G^{\textbf{k}}$ is symmetric and therefore any vertex $F$ that depends only on $q$ but not on $k,k'$ will give zero when inserted in ~\eqref{eq:current_current_correlator_F}. 
     
     For the remaining part ($-\frac{1}{2} \Phi^{\text{RPA},(k-k')}_{d,ph} - \frac{3}{2} \Phi^{\text{RPA},(k-k')}_{m,ph}$) in~\eqref{eq:density_vertex3} we can assume, that the magnetic channel will dominate because $\Phi^{\text{RPA},(k-k')}_{d,ph} = \frac{U}{1+U\chi_0^{k-k'}}-U$ and $\Phi^{\text{RPA},(k-k')}_{m,ph} = U-\frac{U}{1-U\chi_0^{k-k'}}$ and $U > 0$, $\chi_0^{k-k'} > 0$. In Fig.~\ref{fig:density_vertex_corrections} we show a comparison of vertex contributions from the density and magnetic channels. The density channel contribution is indeed much smaller, even though we take the same value of the effective $U$ in this calculation. In fact, taking a different value of the effective interaction in the density channel (less screened than in the magnetic channel) would make the contribution even smaller. Considering only the dominant magnetic channel, which corresponds to the $(\uparrow \downarrow,\overline{ph})$ diagram as drawn in Fig.~\ref{fig:rpa_ladder}, Sec.~\ref{sec:rpa_vertex}, allows us also avoiding the problem of choosing appropriate effective $U^{d/m}$~\cite{Tremblay2011}, leaving only one effective (magnetic) $U$-value that we treat as a free parameter in our model calculation.

    \section{Schwinger-Dyson equation in real frequencies for a  vertex depending on one frequency}
    \label{app:Schwinger_Dyson_real}
    The Schwinger-Dyson equation connects the two-particle vertex $F$ to the one-particle self-energy $\Sigma$ and reads for the Hubbard-Hamiltonian \cite{Rohringer_Diagrammatic_extensions}:
    
    \begin{equation}
    \label{eq:Schinger_Dyson_equation}
        \Sigma_{k} = \frac{{U}n}{2} - {U} \frac{1}{\beta^2 N^2} \sum_{k'q} F_{\uparrow\downarrow}^{kk'q}. G_{k'}G_{k'+q}G_{k+q}
    \end{equation}
  The first term in Eq.~(\ref{eq:Schinger_Dyson_equation}) is the Hartree-term, which is not important for the following discussion and will hence be omitted for simplicity. 
   
  To be specific we will proceed with the RPA-like ladder approximation of the vertex, as presented in Appendix~C, but the final equation can be used for any vertex that depends on only one frequency ($\widetilde{F}^q$ in \eqref{eq:Schwinger_Dyson_RPA2}). In RPA we will omit the contribution from the $pp$-channel and take only the  $ph$ and $\overline{ph}$ ladders in the Schwinger-Dyson equation. Note that while the  contribution of $ph$ ladder to the optical conductivity  can be neglected, as discussed in Appendix~\ref{app:rpa_ladders}, this is not the case for the contribution to the self-energy (as neither the restriction to ${\mathbf q}=0$ nor the antisymmetry of the $\gamma_\alpha^{\mathbf k}$ enters).
    Including  both,  $ph$ and $\overline{ph}$ ladders, in  $F$ yields
       \begin{equation}
    \begin{split}
        F_{\uparrow\downarrow}^{kk'q} &= U + \Phi^{\text{RPA},kk'q}_{\uparrow\downarrow,ph} +  \Phi^{\text{RPA},kk'q}_{\uparrow\downarrow,\overline{ph}} \\ 
        &= U + \Phi^{\text{RPA},kk'q}_{\uparrow\downarrow,ph} - \Phi^{\text{RPA},k(k+q)(k'-k)}_{\overline{\uparrow\downarrow},{ph}} \\ 
        &\equiv U + \Phi^{\text{RPA},q}_{\uparrow\downarrow,ph} - \Phi^{\text{RPA},k'-k}_{\overline{\uparrow\downarrow},{ph}},
        \label{eq:updo_ladder_vertex}
    \end{split}
    \end{equation}
    where we used the crossing symmetry~\cite{Rohringer_Diagrammatic_extensions}
    $\Phi^{\text{RPA},kk'q}_{\uparrow\downarrow,\overline{ph}}=-\Phi^{\text{RPA},(k'+q)k'(k-k')}_{\overline{\uparrow\downarrow},{ph}}=-\Phi^{\text{RPA},k(k+q)(k'-k)}_{\overline{\uparrow\downarrow},{ph}}$ in the second line and 
    in the notation of the third line we omitted the first two fermionic variables, since  the $ph$-ladder depends only on the last, bosonic variable, as shown in~\eqref{eq:RPA_ladders}. 
   
       \begin{figure*}
        \centering
        \includegraphics[width=1\textwidth]{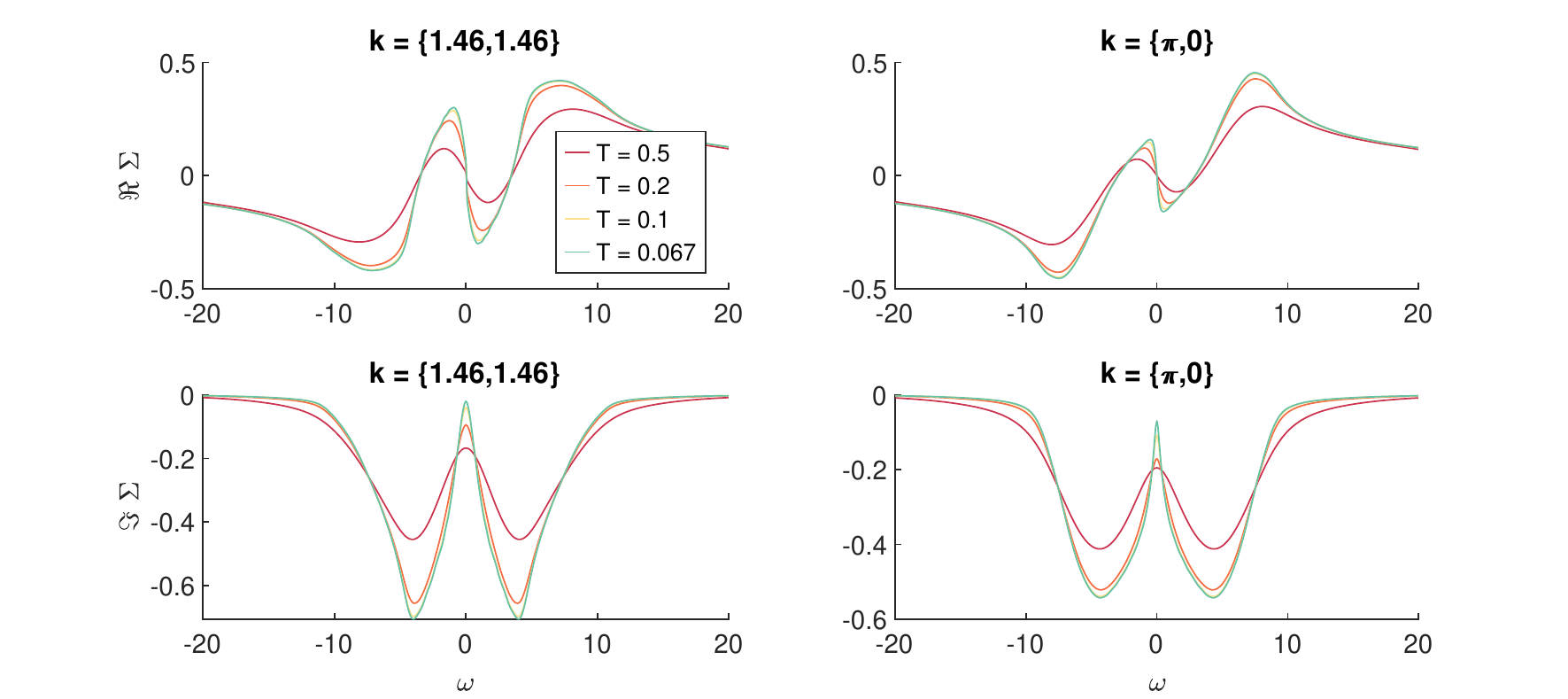}
        \caption{Real (upper row) and imaginary (lower row) part of the self-energy obtained using the Schwinger-Dyson equation  Eq.~(\ref{eq:Schwinger_Dyson_equation_real}) as a function of $\omega$ for two points in the Brillouin zone. Different colors denote different temperatures. The bare interaction value in the Schwinger-Dyson equation was taken as $U_0 = 4$ and the effective interaction for the vertex as $U = 1.8$.}
        \label{fig:SDE_Sigma_w}
    \end{figure*}
    
    \begin{figure*}
        \centering
        \includegraphics[width=1\textwidth]{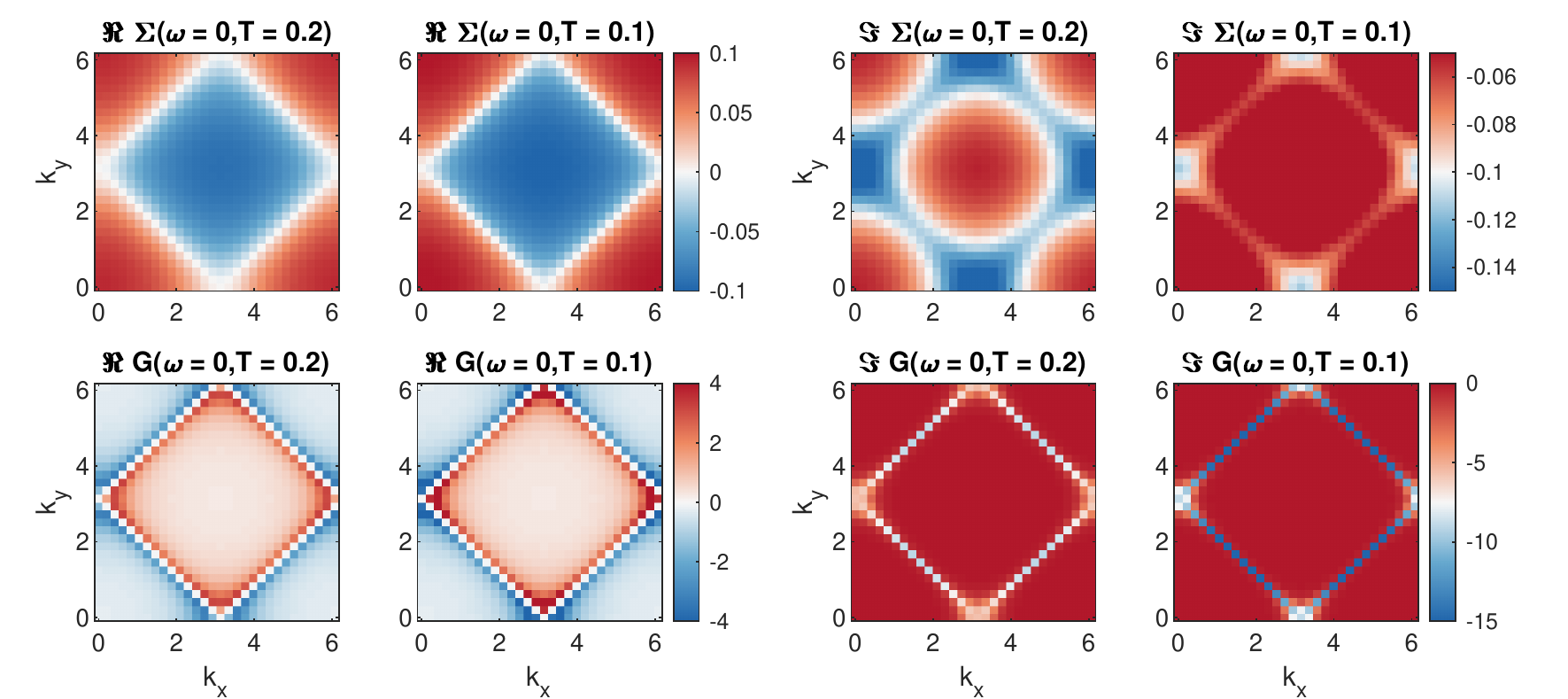}
        \caption{Real and imaginary part of the (retarded) self-energy obtained using the Schwinger-Dyson equation  Eq.~(\ref{eq:Schwinger_Dyson_equation_real}) (upper row) and Green's function (lower row) at $\omega = 0$, as a function of $k_x$ and $k_y$ for $T = 0.2$ and $T = 0.1$. The bare interaction value was taken as $U_0 = 4$ and the effective interaction for the vertex as $U = 1.8$.}
        \label{fig:SDE_Sigma_FS}
    \end{figure*}
   
   In this approximation Eq.~(\ref{eq:Schinger_Dyson_equation}) can be rewritten as:
   \begin{equation}
   \label{eq:Schwinger_Dyson_RPA1}
               \Sigma_{k} = {-U_0} \frac{1}{\beta^2 N^2} \sum_{q k'} ( U + \Phi^{\text{RPA},q}_{\uparrow\downarrow,ph} - \Phi^{\text{RPA},k'-k}_{\overline{\uparrow\downarrow},ph}) G_{k+q} G_{k'}G_{k'+q}. 
   \end{equation}
  From now on we will explicitly distinguish between the bare interaction value, which we now denote with $U_0$, and the effective (screened) $U$ that we take as the RPA vertex.
  
  Eq.~(\ref{eq:Schwinger_Dyson_RPA1}) can be further simplified by using again $\chi_{0}^q = -\frac{1}{\beta N}\sum_{k} G_k G_{k+q}$ and employing a variable change $k'' = k+q$ and $q' = k'-k$ for the third term.
   
      \begin{equation}
       \begin{split}
                   \Sigma_{k} &=  {U_0} \frac{1}{\beta N} \sum_{q} (U+\Phi^{\text{RPA},q}_{\uparrow\downarrow,ph}-\Phi^{\text{RPA},q}_{\overline{\uparrow\downarrow},{ph}}) G_{k+q} \chi_{0}^q \\
                   & \equiv U_0 \frac{1}{\beta N} \sum_{q} \widetilde{F}^{q} G_{k+q} \chi_{0}^q.
       \end{split}
        \label{eq:Schwinger_Dyson_RPA2}
   \end{equation}

    The real frequency version of \eqref{eq:Schwinger_Dyson_RPA2} can be obtained following along the same lines as in Section~\ref{app:derivations}. Branchcuts are at $z = \omega$ and $z = \omega -i \nu_n$. Rewriting the Matsubara sum in terms of integrals yields, 
    \begin{widetext}
        \begin{equation}
            \label{eq:Schwinger_Dyson_equation_real}
            \Sigma_{\nu}^{\text{R} \: \textbf{k}} = U_0 \frac{i}{2\pi} \sum_{\textbf{q}} \int_{-\infty}^{\infty}~d\omega \big[ - 2 i \pi  \eta_{\text{F}}(\omega) \tilde{F}^{\text{A} \: \textbf{q}}_{\omega} \chi_{0,\omega}^{\text{A} \: \textbf{q}} A^{\textbf{k}+\textbf{q}}_{\omega+\nu} +\eta_{\text{B}}(\omega) G_{\nu+\omega}^{\text{R} \: \textbf{k}+\textbf{q}} (\tilde{F}^{\text{R} \: \textbf{q}}_{\omega} \chi_{0,\omega}^{\text{R} \: \textbf{q}} - \tilde{F}^{\text{A} \: \textbf{q}}_{\omega} \chi_{0,\omega}^{\text{A} \: \textbf{q}}) \big].
        \end{equation}
        \label{eq:Schinger_Dyson_equation_real}
    \end{widetext}
    The above expression can be used for calculating the self-energy for a general $\widetilde{F}^q$, provided the reduced frequency dependence. 

       \begin{figure*}
        \centering
        \includegraphics[width=1\textwidth]{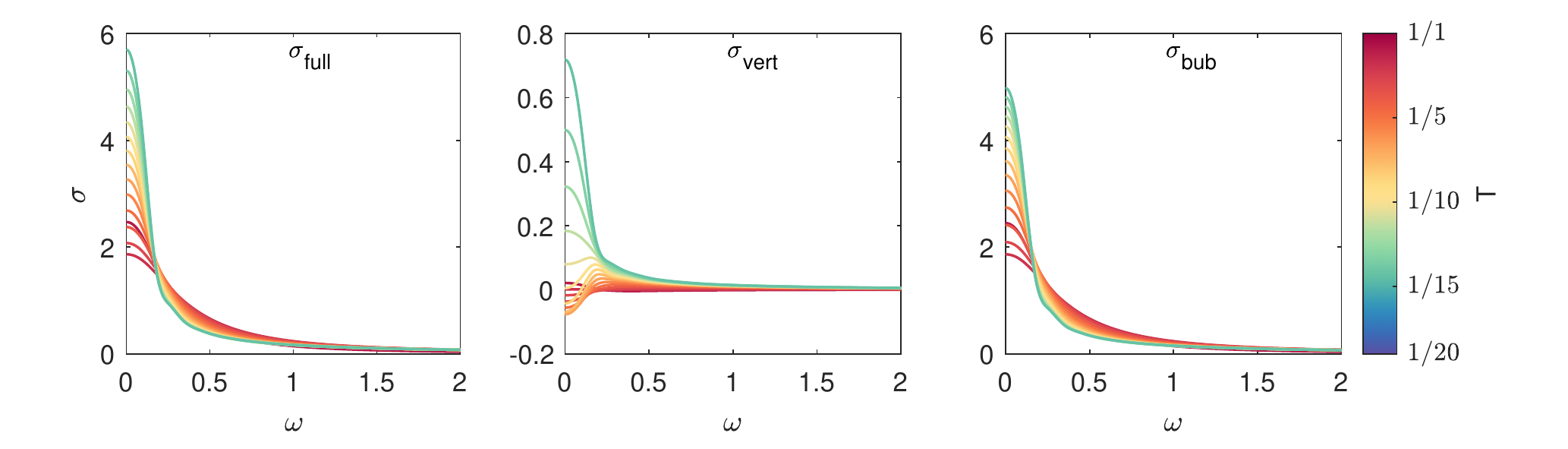}
        \caption{Full optical conductivity $\sigma_{\rm full}$ (left), the vertex corrections in the RPA ladder $\sigma_{\rm vert}$  (middle) and the bare particle-hole bubble $\sigma_{\rm bub}$ (right) at  effective interaction $U=1.8$ and various temperatures $T$. This figure is the same as Fig.~\ref{fig:RPA_temp}, but the one-particle Green's function was calculated with a one-shot vertex correction obtained  as described in the text.}
        \label{fig:opt_cond_vs_T_SDE}
    \end{figure*}

       In Fig.~\ref{fig:SDE_Sigma_w} we show the  real and imaginary part of the self-energy as a function of frequency $\omega$ for two points in the Brillouin zone, for several temperatures, as obtained from Eq.~(\ref{eq:Schwinger_Dyson_equation_real}) with the RPA vertex as in  Eq.~(\ref{eq:Schwinger_Dyson_RPA2}).  The input self-energy to calculate the RPA vertex as well as the one-particle Green's functions  was the constant but temperature dependent self-energy we used in the main text (see Sec.~\ref{sec:self_energy}), with $\Delta_0(T)= 0.1547 + 1.637\,T^2$. We performed here only a one-shot calculation. Further improvement could be achieved (albeit at a significant numerical cost) by using e.g. the two-particle self-consistent approach as in~Ref.~\onlinecite{Tremblay2011}.  We used here the bare interaction value of $U_0 = 4$, which corresponds to the value from the parquet D$\Gamma$A calculation of Ref.~\cite{pi_ton} (as the bare $U_0$ enters the Schwinger-Dyson equation) and the effective interaction $U=1.8$ in the RPA ladders (as beyond RPA ladder contributions reduce the magnetic fluctuations).

       The self-energy  in Fig.~\ref{fig:SDE_Sigma_w} shows the typical Fermi liquid behavior with a linear $\omega$-dependence for the real part of the self-energy and a quadratic behavior for the imaginary part \footnote{Particular strong scattering at the van Hove singularity will be cut-off by the finite scattering rate of the the starting self-energy}. On top of this quadratic $\omega$-dependence, there is a finite $\omega=0$ scattering rate that is suppressed upon decreasing temperature.  This low energy scattering is more pronounced for the antinodal point ${\mathbf k}=(\pi,0)$, as one can also see in  momentum dependence of the self-energy at the Fermi level ($\omega = 0$)  shown in Fig.~\ref{fig:SDE_Sigma_FS}.  Despite this momentum differentiation there is no sign for the formation of a pseudogap, which would be signaled by a $1/\omega$ pole in the real part of the self-energy. Please note that, while showing a reasonable Fermi liquid frequency dependence, this RPA recalculated self-energy cannot be considered (in the frequency range relevant for the Drude peak) to be superior   to the constant $\Sigma(\omega)=-i\Delta_0(T)$ fitted to the more precise numerical parquet data.
Indeed due to anti-ferromagnetic fluctuations the low frequency scattering rate will not vanish but after a plateau increase again at low temperatures~\cite{Rohringer2016}. In the next Section, we will see that the difference between the fitted constant self-energy and the RPA recalculated self-energy in any case does not alter our conclusions.

     \begin{figure}
        \centering
        \includegraphics[width=1.0\linewidth]{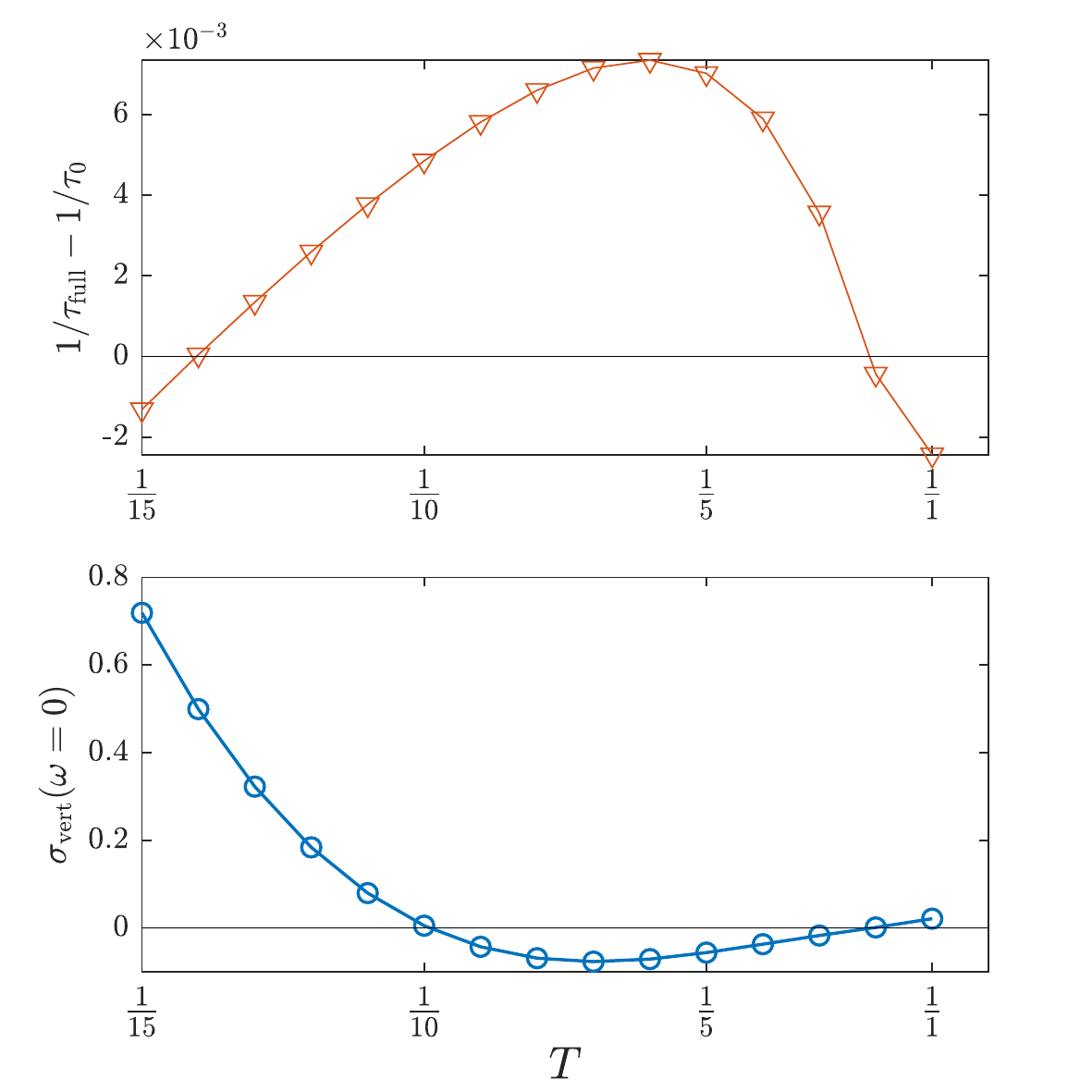}
        \caption{Top: Difference between the width of the Drude peak including vertex corrections ($1/\tau_{\rm{full}}$) and   the width  of the Drude peak using the bubble contribution only ($1/\tau_{\rm{0}}$)  as a function of temperature (on a scale linear in $1/T=\beta$) for $U=1.8$. Bottom: vertex contribution $\sigma_{\rm vert}(\omega=0)$ to the DC conductivity as a function of temperature for the same parameters. The figure is the same as Fig.~\ref{fig:brodening_vs_T} but the bubble contribution and the external Green's functions  in the vertex corrections were calculated with the self-energy from Eq.~\ref{eq:Schinger_Dyson_equation_real} as described in the text.
        }
        \label{fig:brodening_vs_T_SDE}
    \end{figure}
    
      \section{The effect of RPA-ladder vertex corrections to  self-energy on the optical conductivity}
      \label{app:sde_selfenergy}
    Throughout the main paper we used a model, featureless, self-energy as described in Sec.~\ref{sec:self_energy}. This approach does not capture the interplay between vertex corrections to one- and two-particle quantities. It is however not feasible to apply a fully one- and two-particle consistent approach, as e.g.\ the parquet approach of Ref.~\onlinecite{pi_ton} does, and to keep the same momentum resolution or do calculations on the real frequency axis. In the following we only try to assess if and how including the frequency dependence and self-consistency effects for the self-energy alters our  conclusions of the main text. To this aim we use the self-energy with a vertex correction from Eq.~(\ref{eq:Schwinger_Dyson_equation_real}) in the calculation of the optical conductivity. We make only a one-shot correction, which means the self-energy from  Eq.~(\ref{eq:Schwinger_Dyson_equation_real}) is used only in the bubble contribution and in the outer lines in Eq.~(\ref{eq:opt_cond}), but the vertex entering Eq.~(\ref{eq:opt_cond_vert_final}) is calculated with Green's functions for which we use the constant scattering rate as described in the main text. This way the vertex used in Eq.~(\ref{eq:Schwinger_Dyson_equation_real}) and Eq.~(\ref{eq:opt_cond_vert_final}) is the same.

     The results are shown in Fig.~\ref{fig:opt_cond_vs_T_SDE}, which is constructed analogously to Fig.~\ref{fig:RPA_temp}. We also see the three temperature regimes: (i) a sharpening for high temperatures; (ii) broadening in the intermediate regime; (iii) and again sharpening in the low temperature regime. Quantitatively now both the bubble contribution as well as the vertex corrections are larger at $\omega=0$ at low temperatures, which is caused by the reduced scattering rate around $\omega = 0$ (compared to the constant we took in the model Green's function in the main text) as can be seen in the imaginary part of the self-energy in Fig.~\ref{fig:SDE_Sigma_w}. The qualitative behavior in the lowest frequency regime is however unchanged. In particular the lowest frequency regime of both the full optical conductivity and the bubble contribution can still be well fitted with a Drude peak. For intermediate frequencies, $\omega\approx0.3$, we see a different behavior already in the bubble contribution. Also the vertex corrections do not become negative in this frequency regime as it was visible in  Fig.~\ref{fig:RPA_temp} for a frequency independent self-energy. These differences come from the strong frequency dependence of the self-energy for the intermediate frequency range (cf. Fig.~\ref{fig:SDE_Sigma_w}). 
     
     In Fig.~\ref{fig:brodening_vs_T_SDE} we present, in analogy to Fig.~\ref{fig:brodening_vs_T}, the difference between the width of the Drude peak fitted to full optical conductivity $1/\tau_{\textrm{full}}$ and fitted to only the bubble contribution  $1/\tau_{\textrm{bub}}$. Also the vertex corrections to the DC conductivity are shown. We again see the sharpening--broadening--sharpening behavior with temperature. Overall the corrections to $\sigma_{\textrm{vert}}(\omega=0)$ are larger as already seen in Fig.~\ref{fig:opt_cond_vs_T_SDE}. The corrections to the width of the Drude peak on the other hand are smaller than the ones in the case of constant self-energy (cf. Fig.~\ref{fig:brodening_vs_T}).
     
      We could alternatively do a different calculation and also use the vertex corrected self-energy to obtain an updated RPA vertex and then calculate the vertex correction with Eq.~(\ref{eq:opt_cond_vert_final}). The outer and inner Green's function lines in the calculation of optical conductivity from Eq.~(\ref{eq:opt_cond}) would then be the same (where by 'inner lines' are the Green's functions used to calculate the RPA vertex). We have done so but this additional self-energy feedback  to the vertex has only minor influence on the vertex corrections to optical conductivity and is hence not shown here.


\begin{thebibliography}{86}%
    \makeatletter
    \providecommand \@ifxundefined [1]{%
        \@ifx{#1\undefined}
    }%
    \providecommand \@ifnum [1]{%
        \ifnum #1\expandafter \@firstoftwo
        \else \expandafter \@secondoftwo
        \fi
    }%
    \providecommand \@ifx [1]{%
        \ifx #1\expandafter \@firstoftwo
        \else \expandafter \@secondoftwo
        \fi
    }%
    \providecommand \natexlab [1]{#1}%
    \providecommand \enquote  [1]{``#1''}%
    \providecommand \bibnamefont  [1]{#1}%
    \providecommand \bibfnamefont [1]{#1}%
    \providecommand \citenamefont [1]{#1}%
    \providecommand \href@noop [0]{\@secondoftwo}%
    \providecommand \href [0]{\begingroup \@sanitize@url \@href}%
    \providecommand \@href[1]{\@@startlink{#1}\@@href}%
    \providecommand \@@href[1]{\endgroup#1\@@endlink}%
    \providecommand \@sanitize@url [0]{\catcode `\\12\catcode `\$12\catcode
        `\&12\catcode `\#12\catcode `\^12\catcode `\_12\catcode `\%12\relax}%
    \providecommand \@@startlink[1]{}%
    \providecommand \@@endlink[0]{}%
    \providecommand \url  [0]{\begingroup\@sanitize@url \@url }%
    \providecommand \@url [1]{\endgroup\@href {#1}{\urlprefix }}%
    \providecommand \urlprefix  [0]{URL }%
    \providecommand \Eprint [0]{\href }%
    \providecommand \doibase [0]{https://doi.org/}%
    \providecommand \selectlanguage [0]{\@gobble}%
    \providecommand \bibinfo  [0]{\@secondoftwo}%
    \providecommand \bibfield  [0]{\@secondoftwo}%
    \providecommand \translation [1]{[#1]}%
    \providecommand \BibitemOpen [0]{}%
    \providecommand \bibitemStop [0]{}%
    \providecommand \bibitemNoStop [0]{.\EOS\space}%
    \providecommand \EOS [0]{\spacefactor3000\relax}%
    \providecommand \BibitemShut  [1]{\csname bibitem#1\endcsname}%
    \let\auto@bib@innerbib\@empty
    \bibitem [{\citenamefont {Drude}(1900{\natexlab{a}})}]{Drude1900a}%
    \BibitemOpen
    \bibfield  {author} {\bibinfo {author} {\bibfnamefont {P.}~\bibnamefont
            {Drude}},\ }\bibfield  {title} {\bibinfo {title} {Zur elektronentheorie der
            metalle},\ }\href@noop {} {\bibfield  {journal} {\bibinfo  {journal} {Annalen
                der Physik}\ }\textbf {\bibinfo {volume} {306}},\ \bibinfo {pages} {566}
        (\bibinfo {year} {1900}{\natexlab{a}})}\BibitemShut {NoStop}%
    \bibitem [{\citenamefont {Drude}(1900{\natexlab{b}})}]{Drude1900b}%
    \BibitemOpen
    \bibfield  {author} {\bibinfo {author} {\bibfnamefont {P.}~\bibnamefont
            {Drude}},\ }\bibfield  {title} {\bibinfo {title} {Zur elektronentheorie der
            metalle; ii. teil. galvanomagnetische und thermomagnetische effecte},\
    }\href@noop {} {\bibfield  {journal} {\bibinfo  {journal} {Annalen der
                Physik}\ }\textbf {\bibinfo {volume} {308}},\ \bibinfo {pages} {369}
        (\bibinfo {year} {1900}{\natexlab{b}})}\BibitemShut {NoStop}%
    \bibitem [{\citenamefont {{Sommerfeld}}(1928)}]{Sommerfeld1928}%
    \BibitemOpen
    \bibfield  {author} {\bibinfo {author} {\bibfnamefont {A.}~\bibnamefont
            {{Sommerfeld}}},\ }\bibfield  {title} {\bibinfo {title} {{Zur
                Elektronentheorie der Metalle auf Grund der Fermischen Statistik}},\ }\href
    {https://doi.org/10.1007/BF01391052} {\bibfield  {journal} {\bibinfo
            {journal} {Z. Physik}\ }\textbf {\bibinfo {volume} {47}},\ \bibinfo {pages}
        {1} (\bibinfo {year} {1928})}\BibitemShut {NoStop}%
    \bibitem [{\citenamefont {Frenkel}(1931)}]{Frenkel1931}%
    \BibitemOpen
    \bibfield  {author} {\bibinfo {author} {\bibfnamefont {J.}~\bibnamefont
            {Frenkel}},\ }\bibfield  {title} {\bibinfo {title} {On the transformation of
            light into heat in solids. i},\ }\href
    {https://doi.org/10.1103/PhysRev.37.17} {\bibfield  {journal} {\bibinfo
            {journal} {Phys. Rev.}\ }\textbf {\bibinfo {volume} {37}},\ \bibinfo {pages}
        {17} (\bibinfo {year} {1931})}\BibitemShut {NoStop}%
    \bibitem [{\citenamefont {Wannier}(1937)}]{Wannier37}%
    \BibitemOpen
    \bibfield  {author} {\bibinfo {author} {\bibfnamefont {G.~H.}\ \bibnamefont
            {Wannier}},\ }\bibfield  {title} {\bibinfo {title} {The structure of
            electronic excitation levels in insulating crystals},\ }\href
    {https://doi.org/10.1103/PhysRev.52.191} {\bibfield  {journal} {\bibinfo
            {journal} {Phys. Rev.}\ }\textbf {\bibinfo {volume} {52}},\ \bibinfo {pages}
        {191} (\bibinfo {year} {1937})}\BibitemShut {NoStop}%
    \bibitem [{\citenamefont {Maebashi}\ and\ \citenamefont
        {Fukuyama}(1997)}]{Fukuyama1997}%
    \BibitemOpen
    \bibfield  {author} {\bibinfo {author} {\bibfnamefont {H.}~\bibnamefont
            {Maebashi}}\ and\ \bibinfo {author} {\bibfnamefont {H.}~\bibnamefont
            {Fukuyama}},\ }\bibfield  {title} {\bibinfo {title} {Electrical conductivity
            of interacting fermions. i. general formulation},\ }\href
    {https://doi.org/10.1143/JPSJ.66.3577} {\bibfield  {journal} {\bibinfo
            {journal} {Journal of the Physical Society of Japan}\ }\textbf {\bibinfo
            {volume} {66}},\ \bibinfo {pages} {3577} (\bibinfo {year}
        {1997})}\BibitemShut {NoStop}%
    \bibitem [{\citenamefont {Peierls}(1930)}]{Peierls1930}%
    \BibitemOpen
    \bibfield  {author} {\bibinfo {author} {\bibfnamefont {R.}~\bibnamefont
            {Peierls}},\ }\bibfield  {title} {\bibinfo {title} {Zur theorie der
            elektrischen und thermischen leitf{\"a}higkeit von metallen},\ }\href
    {https://doi.org/https://doi.org/10.1002/andp.19303960202} {\bibfield
        {journal} {\bibinfo  {journal} {Annalen der Physik}\ }\textbf {\bibinfo
            {volume} {396}},\ \bibinfo {pages} {121} (\bibinfo {year}
        {1930})}\BibitemShut {NoStop}%
    \bibitem [{\citenamefont {Kontani}(2008)}]{Kontani2008}%
    \BibitemOpen
    \bibfield  {author} {\bibinfo {author} {\bibfnamefont {H.}~\bibnamefont
            {Kontani}},\ }\bibfield  {title} {\bibinfo {title} {Anomalous transport
            phenomena in fermi liquids with strong magnetic fluctuations},\ }\href
    {https://doi.org/10.1088/0034-4885/71/2/026501} {\bibfield  {journal}
        {\bibinfo  {journal} {Reports on Progress in Physics}\ }\textbf {\bibinfo
            {volume} {71}},\ \bibinfo {pages} {026501} (\bibinfo {year}
        {2008})}\BibitemShut {NoStop}%
    \bibitem [{\citenamefont {Maslov}\ and\ \citenamefont
        {Chubukov}(2017)}]{Maslov2017}%
    \BibitemOpen
    \bibfield  {author} {\bibinfo {author} {\bibfnamefont {D.~L.}\ \bibnamefont
            {Maslov}}\ and\ \bibinfo {author} {\bibfnamefont {A.~V.}\ \bibnamefont
            {Chubukov}},\ }\bibfield  {title} {\bibinfo {title} {Optical response of
            correlated electron systems},\ }\href
    {https://doi.org/10.1088/1361-6633/80/2/026503} {\bibfield  {journal}
        {\bibinfo  {journal} {Rep. Prog. Phys.}\ }\textbf {\bibinfo {volume} {80}},\
        \bibinfo {pages} {026503} (\bibinfo {year} {2017})}\BibitemShut {NoStop}%
    \bibitem [{\citenamefont {Keimer}(2013)}]{Keimer2013}%
    \BibitemOpen
    \bibfield  {author} {\bibinfo {author} {\bibfnamefont {B.}~\bibnamefont
            {Keimer}},\ }\bibinfo {title} {Autumn {S}chool on {C}orrelated
        {E}lectrons.emergent phenomena in correlated matter}\ (\bibinfo  {publisher}
    {Forschungszentrum J{\"u}lich},\ \bibinfo {year} {2013})\ Chap.\ \bibinfo
    {chapter} {Recent Advances in Experimental Research on High-Temperature
        Superconductivity}\BibitemShut {NoStop}%
    \bibitem [{\citenamefont {Aeppli}(2004)}]{Aeppli2014}%
    \BibitemOpen
    \bibfield  {author} {\bibinfo {author} {\bibfnamefont {G.}~\bibnamefont
            {Aeppli}},\ }\bibfield  {title} {\bibinfo {title} {The ubiquity of
            antiferromagnetism in strange solids},\ }\href
    {https://doi.org/10.1016/j.jmmm.2003.12.962} {\bibfield  {journal} {\bibinfo
            {journal} {Journal of Magnetism and Magnetic Materials}\ }\textbf {\bibinfo
            {volume} {272-276}},\ \bibinfo {pages} {7} (\bibinfo {year}
        {2004})}\BibitemShut {NoStop}%
    \bibitem [{\citenamefont {Kauch}\ \emph {et~al.}(2020)\citenamefont {Kauch},
        \citenamefont {Pudleiner}, \citenamefont {Astleithner}, \citenamefont
        {Thunstr\"om}, \citenamefont {Ribic},\ and\ \citenamefont {Held}}]{pi_ton}%
    \BibitemOpen
    \bibfield  {author} {\bibinfo {author} {\bibfnamefont {A.}~\bibnamefont
            {Kauch}}, \bibinfo {author} {\bibfnamefont {P.}~\bibnamefont {Pudleiner}},
        \bibinfo {author} {\bibfnamefont {K.}~\bibnamefont {Astleithner}}, \bibinfo
        {author} {\bibfnamefont {P.}~\bibnamefont {Thunstr\"om}}, \bibinfo {author}
        {\bibfnamefont {T.}~\bibnamefont {Ribic}},\ and\ \bibinfo {author}
        {\bibfnamefont {K.}~\bibnamefont {Held}},\ }\bibfield  {title} {\bibinfo
        {title} {Generic optical excitations of correlated systems:
            $\ensuremath{\pi}$-tons},\ }\href
    {https://doi.org/10.1103/PhysRevLett.124.047401} {\bibfield  {journal}
        {\bibinfo  {journal} {Phys. Rev. Lett.}\ }\textbf {\bibinfo {volume} {124}},\
        \bibinfo {pages} {047401} (\bibinfo {year} {2020})}\BibitemShut {NoStop}%
    \bibitem [{\citenamefont {Clarke}(1993)}]{Clarke1993}%
    \BibitemOpen
    \bibfield  {author} {\bibinfo {author} {\bibfnamefont {D.~G.}\ \bibnamefont
            {Clarke}},\ }\bibfield  {title} {\bibinfo {title} {Particle-hole bound states
            in mott-hubbard insulators},\ }\href
    {https://doi.org/10.1103/PhysRevB.48.7520} {\bibfield  {journal} {\bibinfo
            {journal} {Phys. Rev. B}\ }\textbf {\bibinfo {volume} {48}},\ \bibinfo
        {pages} {7520} (\bibinfo {year} {1993})}\BibitemShut {NoStop}%
    \bibitem [{\citenamefont {Wr\'obel}\ and\ \citenamefont
        {Eder}(2002)}]{Wrobel2001}%
    \BibitemOpen
    \bibfield  {author} {\bibinfo {author} {\bibfnamefont {P.}~\bibnamefont
            {Wr\'obel}}\ and\ \bibinfo {author} {\bibfnamefont {R.}~\bibnamefont
            {Eder}},\ }\bibfield  {title} {\bibinfo {title} {Excitons in mott
            insulators},\ }\href {https://doi.org/10.1103/PhysRevB.66.035111} {\bibfield
        {journal} {\bibinfo  {journal} {Phys. Rev. B}\ }\textbf {\bibinfo {volume}
            {66}},\ \bibinfo {pages} {035111} (\bibinfo {year} {2002})}\BibitemShut
    {NoStop}%
    \bibitem [{\citenamefont {Essler}\ \emph {et~al.}(2001)\citenamefont {Essler},
        \citenamefont {Gebhard},\ and\ \citenamefont {Jeckelmann}}]{Essler2001}%
    \BibitemOpen
    \bibfield  {author} {\bibinfo {author} {\bibfnamefont {F.~H.~L.}\
            \bibnamefont {Essler}}, \bibinfo {author} {\bibfnamefont {F.}~\bibnamefont
            {Gebhard}},\ and\ \bibinfo {author} {\bibfnamefont {E.}~\bibnamefont
            {Jeckelmann}},\ }\bibfield  {title} {\bibinfo {title} {Excitons in
            one-dimensional mott insulators},\ }\href
    {https://doi.org/10.1103/PhysRevB.64.125119} {\bibfield  {journal} {\bibinfo
            {journal} {Phys. Rev. B}\ }\textbf {\bibinfo {volume} {64}},\ \bibinfo
        {pages} {125119} (\bibinfo {year} {2001})}\BibitemShut {NoStop}%
    \bibitem [{\citenamefont {Jeckelmann}(2003)}]{Jeckelmann2003}%
    \BibitemOpen
    \bibfield  {author} {\bibinfo {author} {\bibfnamefont {E.}~\bibnamefont
            {Jeckelmann}},\ }\bibfield  {title} {\bibinfo {title} {Optical excitations in
            a one-dimensional mott insulator},\ }\href
    {https://doi.org/10.1103/PhysRevB.67.075106} {\bibfield  {journal} {\bibinfo
            {journal} {Phys. Rev. B}\ }\textbf {\bibinfo {volume} {67}},\ \bibinfo
        {pages} {075106} (\bibinfo {year} {2003})}\BibitemShut {NoStop}%
    \bibitem [{\citenamefont {Kontani}(2006)}]{Kontani2006}%
    \BibitemOpen
    \bibfield  {author} {\bibinfo {author} {\bibfnamefont {H.}~\bibnamefont
            {Kontani}},\ }\bibfield  {title} {\bibinfo {title} {Optical conductivity and
            hall coefficient in high-tc superconductors: Significant role of current
            vertex corrections},\ }\href {https://doi.org/10.1143/JPSJ.75.013703}
    {\bibfield  {journal} {\bibinfo  {journal} {Journal of the Physical Society
                of Japan}\ }\textbf {\bibinfo {volume} {75}},\ \bibinfo {pages} {013703}
        (\bibinfo {year} {2006})}\BibitemShut {NoStop}%
    \bibitem [{\citenamefont {Lin}\ \emph {et~al.}(2009)\citenamefont {Lin},
        \citenamefont {Gull},\ and\ \citenamefont {Millis}}]{Gull2009a}%
    \BibitemOpen
    \bibfield  {author} {\bibinfo {author} {\bibfnamefont {N.}~\bibnamefont
            {Lin}}, \bibinfo {author} {\bibfnamefont {E.}~\bibnamefont {Gull}},\ and\
        \bibinfo {author} {\bibfnamefont {A.~J.}\ \bibnamefont {Millis}},\ }\bibfield
    {title} {\bibinfo {title} {Optical conductivity from cluster dynamical
            mean-field theory: Formalism and application to high-temperature
            superconductors},\ }\href {https://doi.org/10.1103/PhysRevB.80.161105}
    {\bibfield  {journal} {\bibinfo  {journal} {Phys. Rev. B}\ }\textbf {\bibinfo
            {volume} {80}},\ \bibinfo {pages} {161105} (\bibinfo {year}
        {2009})}\BibitemShut {NoStop}%
    \bibitem [{\citenamefont {Chubukov}\ \emph {et~al.}(2014)\citenamefont
        {Chubukov}, \citenamefont {Maslov},\ and\ \citenamefont
        {Yudson}}]{Chubukov2014}%
    \BibitemOpen
    \bibfield  {author} {\bibinfo {author} {\bibfnamefont {A.~V.}\ \bibnamefont
            {Chubukov}}, \bibinfo {author} {\bibfnamefont {D.~L.}\ \bibnamefont
            {Maslov}},\ and\ \bibinfo {author} {\bibfnamefont {V.~I.}\ \bibnamefont
            {Yudson}},\ }\bibfield  {title} {\bibinfo {title} {Optical conductivity of a
            two-dimensional metal at the onset of spin-density-wave order},\ }\href
    {https://doi.org/10.1103/PhysRevB.89.155126} {\bibfield  {journal} {\bibinfo
            {journal} {Phys. Rev. B}\ }\textbf {\bibinfo {volume} {89}},\ \bibinfo
        {pages} {155126} (\bibinfo {year} {2014})}\BibitemShut {NoStop}%
    \bibitem [{\citenamefont {Bergeron}\ \emph {et~al.}(2011)\citenamefont
        {Bergeron}, \citenamefont {Hankevych}, \citenamefont {Kyung},\ and\
        \citenamefont {Tremblay}}]{Tremblay2011}%
    \BibitemOpen
    \bibfield  {author} {\bibinfo {author} {\bibfnamefont {D.}~\bibnamefont
            {Bergeron}}, \bibinfo {author} {\bibfnamefont {V.}~\bibnamefont {Hankevych}},
        \bibinfo {author} {\bibfnamefont {B.}~\bibnamefont {Kyung}},\ and\ \bibinfo
        {author} {\bibfnamefont {A.-M.~S.}\ \bibnamefont {Tremblay}},\ }\bibfield
    {title} {\bibinfo {title} {Optical and dc conductivity of the two-dimensional
            hubbard model in the pseudogap regime and across the antiferromagnetic
            quantum critical point including vertex corrections},\ }\href
    {https://doi.org/10.1103/PhysRevB.84.085128} {\bibfield  {journal} {\bibinfo
            {journal} {Phys. Rev. B}\ }\textbf {\bibinfo {volume} {84}},\ \bibinfo
        {pages} {085128} (\bibinfo {year} {2011})}\BibitemShut {NoStop}%
    \bibitem [{\citenamefont {Vu\ifmmode \check{c}\else \v{c}\fi{}i\ifmmode
            \check{c}\else \v{c}\fi{}evi\ifmmode~\acute{c}\else \'{c}\fi{}}\ \emph
        {et~al.}(2019)\citenamefont {Vu\ifmmode \check{c}\else \v{c}\fi{}i\ifmmode
            \check{c}\else \v{c}\fi{}evi\ifmmode~\acute{c}\else \'{c}\fi{}},
        \citenamefont {Kokalj}, \citenamefont {\ifmmode~\check{Z}\else
            \v{Z}\fi{}itko}, \citenamefont {Wentzell}, \citenamefont
        {Tanaskovi\ifmmode~\acute{c}\else \'{c}\fi{}},\ and\ \citenamefont
        {Mravlje}}]{Mravlje2018}%
    \BibitemOpen
    \bibfield  {author} {\bibinfo {author} {\bibfnamefont {J.}~\bibnamefont
            {Vu\ifmmode \check{c}\else \v{c}\fi{}i\ifmmode \check{c}\else
                \v{c}\fi{}evi\ifmmode~\acute{c}\else \'{c}\fi{}}}, \bibinfo {author}
        {\bibfnamefont {J.}~\bibnamefont {Kokalj}}, \bibinfo {author} {\bibfnamefont
            {R.}~\bibnamefont {\ifmmode~\check{Z}\else \v{Z}\fi{}itko}}, \bibinfo
        {author} {\bibfnamefont {N.}~\bibnamefont {Wentzell}}, \bibinfo {author}
        {\bibfnamefont {D.}~\bibnamefont {Tanaskovi\ifmmode~\acute{c}\else
                \'{c}\fi{}}},\ and\ \bibinfo {author} {\bibfnamefont {J.}~\bibnamefont
            {Mravlje}},\ }\bibfield  {title} {\bibinfo {title} {Conductivity in the
            square lattice hubbard model at high temperatures: Importance of vertex
            corrections},\ }\href {https://doi.org/10.1103/PhysRevLett.123.036601}
    {\bibfield  {journal} {\bibinfo  {journal} {Phys. Rev. Lett.}\ }\textbf
        {\bibinfo {volume} {123}},\ \bibinfo {pages} {036601} (\bibinfo {year}
        {2019})}\BibitemShut {NoStop}%
    \bibitem [{\citenamefont {Pudleiner}\ \emph {et~al.}(2019)\citenamefont
        {Pudleiner}, \citenamefont {Thunstr\"om}, \citenamefont {Valli},
        \citenamefont {Kauch}, \citenamefont {Li},\ and\ \citenamefont
        {Held}}]{Pudleiner2019}%
    \BibitemOpen
    \bibfield  {author} {\bibinfo {author} {\bibfnamefont {P.}~\bibnamefont
            {Pudleiner}}, \bibinfo {author} {\bibfnamefont {P.}~\bibnamefont
            {Thunstr\"om}}, \bibinfo {author} {\bibfnamefont {A.}~\bibnamefont {Valli}},
        \bibinfo {author} {\bibfnamefont {A.}~\bibnamefont {Kauch}}, \bibinfo
        {author} {\bibfnamefont {G.}~\bibnamefont {Li}},\ and\ \bibinfo {author}
        {\bibfnamefont {K.}~\bibnamefont {Held}},\ }\bibfield  {title} {\bibinfo
        {title} {Parquet approximation for molecules: Spectrum and optical
            conductivity of the pariser-parr-pople model},\ }\href
    {https://doi.org/10.1103/PhysRevB.99.125111} {\bibfield  {journal} {\bibinfo
            {journal} {Phys. Rev. B}\ }\textbf {\bibinfo {volume} {99}},\ \bibinfo
        {pages} {125111} (\bibinfo {year} {2019})}\BibitemShut {NoStop}%
    \bibitem [{\citenamefont {Astleithner}\ \emph {et~al.}(2020)\citenamefont
        {Astleithner}, \citenamefont {Kauch}, \citenamefont {Ribic},\ and\
        \citenamefont {Held}}]{Astleithner2020}%
    \BibitemOpen
    \bibfield  {author} {\bibinfo {author} {\bibfnamefont {K.}~\bibnamefont
            {Astleithner}}, \bibinfo {author} {\bibfnamefont {A.}~\bibnamefont {Kauch}},
        \bibinfo {author} {\bibfnamefont {T.}~\bibnamefont {Ribic}},\ and\ \bibinfo
        {author} {\bibfnamefont {K.}~\bibnamefont {Held}},\ }\bibfield  {title}
    {\bibinfo {title} {Parquet dual fermion approach for the falicov-kimball
            model},\ }\href {https://doi.org/10.1103/PhysRevB.101.165101} {\bibfield
        {journal} {\bibinfo  {journal} {Phys. Rev. B}\ }\textbf {\bibinfo {volume}
            {101}},\ \bibinfo {pages} {165101} (\bibinfo {year} {2020})}\BibitemShut
    {NoStop}%
    \bibitem [{\citenamefont {Bickers}(2004)}]{Bickers2004}%
    \BibitemOpen
    \bibfield  {author} {\bibinfo {author} {\bibfnamefont {N.~E.}\ \bibnamefont
            {Bickers}},\ }\bibinfo {title} {Theoretical methods for strongly correlated
        electrons}\ (\bibinfo  {publisher} {Springer-Verlag New York Berlin
        Heidelbert},\ \bibinfo {year} {2004})\ Chap.~\bibinfo {chapter} {6}, pp.\
    \bibinfo {pages} {237--296}\BibitemShut {NoStop}%
    \bibitem [{\citenamefont {Yang}\ \emph {et~al.}(2009)\citenamefont {Yang},
        \citenamefont {Fotso}, \citenamefont {Liu}, \citenamefont {Maier},
        \citenamefont {Tomko}, \citenamefont {D'Azevedo}, \citenamefont {Scalettar},
        \citenamefont {Pruschke},\ and\ \citenamefont {Jarrell}}]{Yang2009}%
    \BibitemOpen
    \bibfield  {author} {\bibinfo {author} {\bibfnamefont {S.~X.}\ \bibnamefont
            {Yang}}, \bibinfo {author} {\bibfnamefont {H.}~\bibnamefont {Fotso}},
        \bibinfo {author} {\bibfnamefont {J.}~\bibnamefont {Liu}}, \bibinfo {author}
        {\bibfnamefont {T.~A.}\ \bibnamefont {Maier}}, \bibinfo {author}
        {\bibfnamefont {K.}~\bibnamefont {Tomko}}, \bibinfo {author} {\bibfnamefont
            {E.~F.}\ \bibnamefont {D'Azevedo}}, \bibinfo {author} {\bibfnamefont {R.~T.}\
            \bibnamefont {Scalettar}}, \bibinfo {author} {\bibfnamefont {T.}~\bibnamefont
            {Pruschke}},\ and\ \bibinfo {author} {\bibfnamefont {M.}~\bibnamefont
            {Jarrell}},\ }\bibfield  {title} {\bibinfo {title} {Parquet approximation for
            the $4\ifmmode\times\else\texttimes\fi{}4$ hubbard cluster},\ }\href
    {https://doi.org/10.1103/PhysRevE.80.046706} {\bibfield  {journal} {\bibinfo
            {journal} {Phys. Rev. E}\ }\textbf {\bibinfo {volume} {80}},\ \bibinfo
        {pages} {046706} (\bibinfo {year} {2009})}\BibitemShut {NoStop}%
    \bibitem [{\citenamefont {Li}\ \emph {et~al.}(2016)\citenamefont {Li},
        \citenamefont {Wentzell}, \citenamefont {Pudleiner}, \citenamefont
        {Thunstr\"om},\ and\ \citenamefont {Held}}]{Li2016}%
    \BibitemOpen
    \bibfield  {author} {\bibinfo {author} {\bibfnamefont {G.}~\bibnamefont
            {Li}}, \bibinfo {author} {\bibfnamefont {N.}~\bibnamefont {Wentzell}},
        \bibinfo {author} {\bibfnamefont {P.}~\bibnamefont {Pudleiner}}, \bibinfo
        {author} {\bibfnamefont {P.}~\bibnamefont {Thunstr\"om}},\ and\ \bibinfo
        {author} {\bibfnamefont {K.}~\bibnamefont {Held}},\ }\bibfield  {title}
    {\bibinfo {title} {Efficient implementation of the parquet equations: Role of
            the reducible vertex function and its kernel approximation},\ }\href
    {https://doi.org/10.1103/PhysRevB.93.165103} {\bibfield  {journal} {\bibinfo
            {journal} {Phys. Rev. B}\ }\textbf {\bibinfo {volume} {93}},\ \bibinfo
        {pages} {165103} (\bibinfo {year} {2016})}\BibitemShut {NoStop}%
    \bibitem [{\citenamefont {Li}\ \emph {et~al.}(2017)\citenamefont {Li},
        \citenamefont {Kauch}, \citenamefont {Pudleiner},\ and\ \citenamefont
        {Held}}]{victory}%
    \BibitemOpen
    \bibfield  {author} {\bibinfo {author} {\bibfnamefont {G.}~\bibnamefont
            {Li}}, \bibinfo {author} {\bibfnamefont {A.}~\bibnamefont {Kauch}}, \bibinfo
        {author} {\bibfnamefont {P.}~\bibnamefont {Pudleiner}},\ and\ \bibinfo
        {author} {\bibfnamefont {K.}~\bibnamefont {Held}},\ }\bibfield  {title}
    {\bibinfo {title} {The victory project v1.0: an efficient parquet equations
            solver},\ }\href {https://doi.org/10.1016/j.cpc.2019.03.008} {\bibfield
        {journal} {\bibinfo  {journal} {Comput. Phys. Commun.}\ ,\ \bibinfo {pages}
            {146}} (\bibinfo {year} {2017})}\BibitemShut {NoStop}%
    \bibitem [{\citenamefont {Valli}\ \emph {et~al.}(2015)\citenamefont {Valli},
        \citenamefont {Sch\"afer}, \citenamefont {Thunstr\"om}, \citenamefont
        {Rohringer}, \citenamefont {Andergassen}, \citenamefont {Sangiovanni},
        \citenamefont {Held},\ and\ \citenamefont {Toschi}}]{Valli2015}%
    \BibitemOpen
    \bibfield  {author} {\bibinfo {author} {\bibfnamefont {A.}~\bibnamefont
            {Valli}}, \bibinfo {author} {\bibfnamefont {T.}~\bibnamefont {Sch\"afer}},
        \bibinfo {author} {\bibfnamefont {P.}~\bibnamefont {Thunstr\"om}}, \bibinfo
        {author} {\bibfnamefont {G.}~\bibnamefont {Rohringer}}, \bibinfo {author}
        {\bibfnamefont {S.}~\bibnamefont {Andergassen}}, \bibinfo {author}
        {\bibfnamefont {G.}~\bibnamefont {Sangiovanni}}, \bibinfo {author}
        {\bibfnamefont {K.}~\bibnamefont {Held}},\ and\ \bibinfo {author}
        {\bibfnamefont {A.}~\bibnamefont {Toschi}},\ }\bibfield  {title} {\bibinfo
        {title} {Dynamical vertex approximation in its parquet implementation:
            Application to hubbard nanorings},\ }\href
    {https://doi.org/10.1103/PhysRevB.91.115115} {\bibfield  {journal} {\bibinfo
            {journal} {Phys. Rev. B}\ }\textbf {\bibinfo {volume} {91}},\ \bibinfo
        {pages} {115115} (\bibinfo {year} {2015})}\BibitemShut {NoStop}%
    \bibitem [{\citenamefont {Rohringer}\ \emph {et~al.}(2018)\citenamefont
        {Rohringer}, \citenamefont {Hafermann}, \citenamefont {Toschi}, \citenamefont
        {Katanin}, \citenamefont {Antipov}, \citenamefont {Katsnelson}, \citenamefont
        {Lichtenstein}, \citenamefont {Rubtsov},\ and\ \citenamefont
        {Held}}]{Rohringer_Diagrammatic_extensions}%
    \BibitemOpen
    \bibfield  {author} {\bibinfo {author} {\bibfnamefont {G.}~\bibnamefont
            {Rohringer}}, \bibinfo {author} {\bibfnamefont {H.}~\bibnamefont
            {Hafermann}}, \bibinfo {author} {\bibfnamefont {A.}~\bibnamefont {Toschi}},
        \bibinfo {author} {\bibfnamefont {A.~A.}\ \bibnamefont {Katanin}}, \bibinfo
        {author} {\bibfnamefont {A.~E.}\ \bibnamefont {Antipov}}, \bibinfo {author}
        {\bibfnamefont {M.~I.}\ \bibnamefont {Katsnelson}}, \bibinfo {author}
        {\bibfnamefont {A.~I.}\ \bibnamefont {Lichtenstein}}, \bibinfo {author}
        {\bibfnamefont {A.~N.}\ \bibnamefont {Rubtsov}},\ and\ \bibinfo {author}
        {\bibfnamefont {K.}~\bibnamefont {Held}},\ }\bibfield  {title} {\bibinfo
        {title} {Diagrammatic routes to nonlocal correlations beyond dynamical mean
            field theory},\ }\href {https://doi.org/10.1103/RevModPhys.90.025003}
    {\bibfield  {journal} {\bibinfo  {journal} {Rev. Mod. Phys.}\ }\textbf
        {\bibinfo {volume} {90}},\ \bibinfo {pages} {025003} (\bibinfo {year}
        {2018})}\BibitemShut {NoStop}%
    \bibitem [{Note1()}]{Note1}%
    \BibitemOpen
    \bibinfo {note} {The numerical bottleneck is in particle the memory
        consumption of a general vertex depending on three frequencies and three
        momenta, which might be mitigated in the future by an efficient sparse
        parameterization thereof~\cite
        {Honerkamp2018,Shinaoka2020,Li2020,Eckhardt2020,Krien2020}}\BibitemShut
    {NoStop}%
    \bibitem [{\citenamefont {Altshuler}\ and\ \citenamefont
        {Aronov}(1985)}]{Altshuler1985}%
    \BibitemOpen
    \bibfield  {author} {\bibinfo {author} {\bibfnamefont {B.~L.}\ \bibnamefont
            {Altshuler}}\ and\ \bibinfo {author} {\bibfnamefont {A.~G.}\ \bibnamefont
            {Aronov}},\ }\href@noop {} {\emph {\bibinfo {title} {Electron-Electron
                interaction in disordered conductors}}},\ edited by\ \bibinfo {editor}
    {\bibfnamefont {A.~I.}\ \bibnamefont {Efros}}\ and\ \bibinfo {editor}
    {\bibfnamefont {M.}~\bibnamefont {Pollak}}\ (\bibinfo  {publisher} {Elsevier
        Science Publisher},\ \bibinfo {year} {1985})\BibitemShut {NoStop}%
    \bibitem [{\citenamefont {Rammer}(1998)}]{Rammer1998QuantumTransportTheory}%
    \BibitemOpen
    \bibfield  {author} {\bibinfo {author} {\bibfnamefont {J.}~\bibnamefont
            {Rammer}},\ }\href {https://doi.org/10.1201/9780429502835} {\emph {\bibinfo
            {title} {Quantum Transport Theory}}}\ (\bibinfo  {publisher} {{CRC} Press},\
    \bibinfo {year} {1998})\BibitemShut {NoStop}%
    \bibitem [{\citenamefont {Peierls}(1933)}]{Peierls1933}%
    \BibitemOpen
    \bibfield  {author} {\bibinfo {author} {\bibfnamefont {R.}~\bibnamefont
            {Peierls}},\ }\bibfield  {title} {\bibinfo {title} {Zur {T}heorie des
            {D}iamagnetismus von {L}eitungselektronen},\ }\href@noop {} {\bibfield
        {journal} {\bibinfo  {journal} {Z. Phys.}\ }\textbf {\bibinfo {volume}
            {80}},\ \bibinfo {pages} {763} (\bibinfo {year} {1933})}\BibitemShut
    {NoStop}%
    \bibitem [{\citenamefont {Altland}\ and\ \citenamefont
        {Simons}(2010)}]{altland_simons_2010}%
    \BibitemOpen
    \bibfield  {author} {\bibinfo {author} {\bibfnamefont {A.}~\bibnamefont
            {Altland}}\ and\ \bibinfo {author} {\bibfnamefont {B.~D.}\ \bibnamefont
            {Simons}},\ }\href {https://doi.org/10.1017/CBO9780511789984} {\emph
        {\bibinfo {title} {Condensed Matter Field Theory}}},\ \bibinfo {edition}
    {2nd}\ ed.\ (\bibinfo  {publisher} {Cambridge University Press},\ \bibinfo
    {year} {2010})\BibitemShut {NoStop}%
    \bibitem [{\citenamefont {Jarrell}\ and\ \citenamefont
        {Gubernatis}(1996)}]{Max_Ent_Jarrell}%
    \BibitemOpen
    \bibfield  {author} {\bibinfo {author} {\bibfnamefont {M.}~\bibnamefont
            {Jarrell}}\ and\ \bibinfo {author} {\bibfnamefont {J.~E.}\ \bibnamefont
            {Gubernatis}},\ }\bibfield  {title} {\bibinfo {title} {Bayesian inference and
            the analytic continuation of imaginary-time quantum monte carlo data},\
    }\href {https://doi.org/10.1016/0370-1573(95)00074-7} {\bibfield  {journal}
        {\bibinfo  {journal} {Phys. Rep.}\ }\textbf {\bibinfo {volume} {269}},\
        \bibinfo {pages} {133} (\bibinfo {year} {1996})}\BibitemShut {NoStop}%
    \bibitem [{\citenamefont {Mishchenko}\ \emph {et~al.}(2000)\citenamefont
        {Mishchenko}, \citenamefont {Prokof'ev}, \citenamefont {Sakamoto},\ and\
        \citenamefont {Svistunov}}]{SOM1}%
    \BibitemOpen
    \bibfield  {author} {\bibinfo {author} {\bibfnamefont {A.~S.}\ \bibnamefont
            {Mishchenko}}, \bibinfo {author} {\bibfnamefont {N.~V.}\ \bibnamefont
            {Prokof'ev}}, \bibinfo {author} {\bibfnamefont {A.}~\bibnamefont
            {Sakamoto}},\ and\ \bibinfo {author} {\bibfnamefont {B.~V.}\ \bibnamefont
            {Svistunov}},\ }\bibfield  {title} {\bibinfo {title} {Diagrammatic quantum
            monte carlo study of the fr\"ohlich polaron},\ }\href
    {https://doi.org/10.1103/PhysRevB.62.6317} {\bibfield  {journal} {\bibinfo
            {journal} {Phys. Rev. B}\ }\textbf {\bibinfo {volume} {62}},\ \bibinfo
        {pages} {6317} (\bibinfo {year} {2000})}\BibitemShut {NoStop}%
    \bibitem [{\citenamefont {Otsuki}\ \emph {et~al.}(2017)\citenamefont {Otsuki},
        \citenamefont {Ohzeki}, \citenamefont {Shinaoka},\ and\ \citenamefont
        {Yoshimi}}]{SpM}%
    \BibitemOpen
    \bibfield  {author} {\bibinfo {author} {\bibfnamefont {J.}~\bibnamefont
            {Otsuki}}, \bibinfo {author} {\bibfnamefont {M.}~\bibnamefont {Ohzeki}},
        \bibinfo {author} {\bibfnamefont {H.}~\bibnamefont {Shinaoka}},\ and\
        \bibinfo {author} {\bibfnamefont {K.}~\bibnamefont {Yoshimi}},\ }\bibfield
    {title} {\bibinfo {title} {Sparse modeling approach to analytical
            continuation of imaginary-time quantum monte carlo data},\ }\href
    {https://doi.org/10.1103/PhysRevE.95.061302} {\bibfield  {journal} {\bibinfo
            {journal} {Phys. Rev. E}\ }\textbf {\bibinfo {volume} {95}},\ \bibinfo
        {pages} {061302} (\bibinfo {year} {2017})}\BibitemShut {NoStop}%
    \bibitem [{\citenamefont {Yoshimi}\ \emph {et~al.}(2019)\citenamefont
        {Yoshimi}, \citenamefont {Otsuki}, \citenamefont {Motoyama}, \citenamefont
        {Ohzeki},\ and\ \citenamefont {Shinaoka}}]{SpM_YOSHIMI2019}%
    \BibitemOpen
    \bibfield  {author} {\bibinfo {author} {\bibfnamefont {K.}~\bibnamefont
            {Yoshimi}}, \bibinfo {author} {\bibfnamefont {J.}~\bibnamefont {Otsuki}},
        \bibinfo {author} {\bibfnamefont {Y.}~\bibnamefont {Motoyama}}, \bibinfo
        {author} {\bibfnamefont {M.}~\bibnamefont {Ohzeki}},\ and\ \bibinfo {author}
        {\bibfnamefont {H.}~\bibnamefont {Shinaoka}},\ }\bibfield  {title} {\bibinfo
        {title} {Spm: Sparse modeling tool for analytic continuation of
            imaginary-time green’s function},\ }\href
    {https://doi.org/https://doi.org/10.1016/j.cpc.2019.07.001} {\bibfield
        {journal} {\bibinfo  {journal} {Computer Physics Communications}\ }\textbf
        {\bibinfo {volume} {244}},\ \bibinfo {pages} {319 } (\bibinfo {year}
        {2019})}\BibitemShut {NoStop}%
    \bibitem [{\citenamefont {Tikhonov}\ and\ \citenamefont
        {Arsenin}(1977)}]{Tikhonov_Regularization}%
    \BibitemOpen
    \bibfield  {author} {\bibinfo {author} {\bibfnamefont {A.~N.}\ \bibnamefont
            {Tikhonov}}\ and\ \bibinfo {author} {\bibfnamefont {V.~I.}\ \bibnamefont
            {Arsenin}},\ }\href {https://nla.gov.au/nla.cat-vn720231} {\emph {\bibinfo
            {title} {Solutions to ill-posed problems}}}\ (\bibinfo  {publisher} {Winston
        ; distributed solely by Halsted Press Washington : New York},\ \bibinfo
    {year} {1977})\ pp.\ \bibinfo {pages} {xiii, 258 p. :}\BibitemShut {NoStop}%
    \bibitem [{\citenamefont {Kappl}\ \emph {et~al.}(2020)\citenamefont {Kappl},
        \citenamefont {Wallerberger}, \citenamefont {Kaufmann}, \citenamefont
        {Pickem},\ and\ \citenamefont {Held}}]{Kappl_2020}%
    \BibitemOpen
    \bibfield  {author} {\bibinfo {author} {\bibfnamefont {P.}~\bibnamefont
            {Kappl}}, \bibinfo {author} {\bibfnamefont {M.}~\bibnamefont {Wallerberger}},
        \bibinfo {author} {\bibfnamefont {J.}~\bibnamefont {Kaufmann}}, \bibinfo
        {author} {\bibfnamefont {M.}~\bibnamefont {Pickem}},\ and\ \bibinfo {author}
        {\bibfnamefont {K.}~\bibnamefont {Held}},\ }\bibfield  {title} {\bibinfo
        {title} {Statistical error estimates in dynamical mean-field theory and
            extensions thereof},\ }\href {https://doi.org/10.1103/PhysRevB.102.085124}
    {\bibfield  {journal} {\bibinfo  {journal} {Phys. Rev. B}\ }\textbf {\bibinfo
            {volume} {102}},\ \bibinfo {pages} {085124} (\bibinfo {year}
        {2020})}\BibitemShut {NoStop}%
    \bibitem [{\citenamefont {Galler}\ \emph {et~al.}(2017)\citenamefont {Galler},
        \citenamefont {Thunstr\"om}, \citenamefont {Gunacker}, \citenamefont
        {Tomczak},\ and\ \citenamefont {Held}}]{ladder_DGA}%
    \BibitemOpen
    \bibfield  {author} {\bibinfo {author} {\bibfnamefont {A.}~\bibnamefont
            {Galler}}, \bibinfo {author} {\bibfnamefont {P.}~\bibnamefont {Thunstr\"om}},
        \bibinfo {author} {\bibfnamefont {P.}~\bibnamefont {Gunacker}}, \bibinfo
        {author} {\bibfnamefont {J.~M.}\ \bibnamefont {Tomczak}},\ and\ \bibinfo
        {author} {\bibfnamefont {K.}~\bibnamefont {Held}},\ }\bibfield  {title}
    {\bibinfo {title} {Ab initio dynamical vertex approximation},\ }\href
    {https://doi.org/10.1103/PhysRevB.95.115107} {\bibfield  {journal} {\bibinfo
            {journal} {Phys. Rev. B}\ }\textbf {\bibinfo {volume} {95}},\ \bibinfo
        {pages} {115107} (\bibinfo {year} {2017})}\BibitemShut {NoStop}%
    \bibitem [{\citenamefont {Georges}\ \emph {et~al.}(1996)\citenamefont
        {Georges}, \citenamefont {Kotliar}, \citenamefont {Krauth},\ and\
        \citenamefont {Rozenberg}}]{kotliar_dmft}%
    \BibitemOpen
    \bibfield  {author} {\bibinfo {author} {\bibfnamefont {A.}~\bibnamefont
            {Georges}}, \bibinfo {author} {\bibfnamefont {G.}~\bibnamefont {Kotliar}},
        \bibinfo {author} {\bibfnamefont {W.}~\bibnamefont {Krauth}},\ and\ \bibinfo
        {author} {\bibfnamefont {M.~J.}\ \bibnamefont {Rozenberg}},\ }\bibfield
    {title} {\bibinfo {title} {Dynamical mean-field theory of strongly correlated
            fermion systems and the limit of infinite dimensions},\ }\href
    {https://doi.org/10.1103/RevModPhys.68.13} {\bibfield  {journal} {\bibinfo
            {journal} {Rev. Mod. Phys.}\ }\textbf {\bibinfo {volume} {68}},\ \bibinfo
        {pages} {13} (\bibinfo {year} {1996})}\BibitemShut {NoStop}%
    \bibitem [{\citenamefont {Kune\v{s}}(2011)}]{Kunes2011}%
    \BibitemOpen
    \bibfield  {author} {\bibinfo {author} {\bibfnamefont {J.}~\bibnamefont
            {Kune\v{s}}},\ }\bibfield  {title} {\bibinfo {title} {Efficient treatment of
            two-particle vertices in dynamical mean-field theory},\ }\href
    {https://doi.org/10.1103/PhysRevB.83.085102} {\bibfield  {journal} {\bibinfo
            {journal} {Phys. Rev. B}\ }\textbf {\bibinfo {volume} {83}},\ \bibinfo
        {pages} {085102} (\bibinfo {year} {2011})}\BibitemShut {NoStop}%
    \bibitem [{\citenamefont {L\"ohneysen}\ \emph {et~al.}(2007)\citenamefont
        {L\"ohneysen}, \citenamefont {Rosch}, \citenamefont {Vojta},\ and\
        \citenamefont {W\"olfle}}]{RevModPhys.79.1015}%
    \BibitemOpen
    \bibfield  {author} {\bibinfo {author} {\bibfnamefont {H.~v.}\ \bibnamefont
            {L\"ohneysen}}, \bibinfo {author} {\bibfnamefont {A.}~\bibnamefont {Rosch}},
        \bibinfo {author} {\bibfnamefont {M.}~\bibnamefont {Vojta}},\ and\ \bibinfo
        {author} {\bibfnamefont {P.}~\bibnamefont {W\"olfle}},\ }\bibfield  {title}
    {\bibinfo {title} {Fermi-liquid instabilities at magnetic quantum phase
            transitions},\ }\href {https://doi.org/10.1103/RevModPhys.79.1015} {\bibfield
        {journal} {\bibinfo  {journal} {Rev. Mod. Phys.}\ }\textbf {\bibinfo
            {volume} {79}},\ \bibinfo {pages} {1015} (\bibinfo {year}
        {2007})}\BibitemShut {NoStop}%
    \bibitem [{\citenamefont {Hertz}(1976)}]{Hertz}%
    \BibitemOpen
    \bibfield  {author} {\bibinfo {author} {\bibfnamefont {J.~A.}\ \bibnamefont
            {Hertz}},\ }\bibfield  {title} {\bibinfo {title} {Quantum critical
            phenomena},\ }\href {https://doi.org/10.1103/PhysRevB.14.1165} {\bibfield
        {journal} {\bibinfo  {journal} {Phys. Rev. B}\ }\textbf {\bibinfo {volume}
            {14}},\ \bibinfo {pages} {1165} (\bibinfo {year} {1976})}\BibitemShut
    {NoStop}%
    \bibitem [{\citenamefont {Millis}(1993)}]{Millis}%
    \BibitemOpen
    \bibfield  {author} {\bibinfo {author} {\bibfnamefont {A.~J.}\ \bibnamefont
            {Millis}},\ }\bibfield  {title} {\bibinfo {title} {Effect of a nonzero
            temperature on quantum critical points in itinerant fermion systems},\ }\href
    {https://doi.org/10.1103/PhysRevB.48.7183} {\bibfield  {journal} {\bibinfo
            {journal} {Phys. Rev. B}\ }\textbf {\bibinfo {volume} {48}},\ \bibinfo
        {pages} {7183} (\bibinfo {year} {1993})}\BibitemShut {NoStop}%
    \bibitem [{\citenamefont {Moriya}\ and\ \citenamefont
        {Kawabata}(1973)}]{Moriya}%
    \BibitemOpen
    \bibfield  {author} {\bibinfo {author} {\bibfnamefont {T.}~\bibnamefont
            {Moriya}}\ and\ \bibinfo {author} {\bibfnamefont {A.}~\bibnamefont
            {Kawabata}},\ }\bibfield  {title} {\bibinfo {title} {Effect of spin
            fluctuations on itinerant electron ferromagnetism},\ }\href
    {https://doi.org/10.1143/JPSJ.34.639} {\bibfield  {journal} {\bibinfo
            {journal} {J. Phys. Soc. Jpn.}\ }\textbf {\bibinfo {volume} {34}},\ \bibinfo
        {pages} {639} (\bibinfo {year} {1973})}\BibitemShut {NoStop}%
    \bibitem [{\citenamefont {Mahan}(2000)}]{Mahan2000}%
    \BibitemOpen
    \bibfield  {author} {\bibinfo {author} {\bibfnamefont {G.~D.}\ \bibnamefont
            {Mahan}},\ }\href@noop {} {\emph {\bibinfo {title} {Many-Particle Physics}}}\
    (\bibinfo  {publisher} {Kluwer Academic/Plenum Publishers, New York},\
    \bibinfo {year} {2000})\BibitemShut {NoStop}%
    \bibitem [{\citenamefont {Mahan}(1966)}]{Mahan1966}%
    \BibitemOpen
    \bibfield  {author} {\bibinfo {author} {\bibfnamefont {G.~D.}\ \bibnamefont
            {Mahan}},\ }\bibfield  {title} {\bibinfo {title} {Mobility of polarons},\
    }\href {https://doi.org/10.1103/PhysRev.142.366} {\bibfield  {journal}
        {\bibinfo  {journal} {Phys. Rev.}\ }\textbf {\bibinfo {volume} {142}},\
        \bibinfo {pages} {366} (\bibinfo {year} {1966})}\BibitemShut {NoStop}%
    \bibitem [{\citenamefont {Simard}\ \emph {et~al.}(2021)\citenamefont {Simard},
        \citenamefont {Takayoshi},\ and\ \citenamefont {Werner}}]{Simard2020}%
    \BibitemOpen
    \bibfield  {author} {\bibinfo {author} {\bibfnamefont {O.}~\bibnamefont
            {Simard}}, \bibinfo {author} {\bibfnamefont {S.}~\bibnamefont {Takayoshi}},\
        and\ \bibinfo {author} {\bibfnamefont {P.}~\bibnamefont {Werner}},\
    }\bibfield  {title} {\bibinfo {title} {Diagrammatic study of optical
            excitations in correlated systems},\ }\href
    {https://doi.org/10.1103/PhysRevB.103.104415} {\bibfield  {journal} {\bibinfo
            {journal} {Phys. Rev. B}\ }\textbf {\bibinfo {volume} {103}},\ \bibinfo
        {pages} {104415} (\bibinfo {year} {2021})}\BibitemShut {NoStop}%
    \bibitem [{Note2()}]{Note2}%
    \BibitemOpen
    \bibinfo {note} {The value of the interaction in p-D$\Gamma $A is $U_0=4$,
        whereas here we use different values of $U=1.7-1.9$. The reason is that the
        RPA series is a good approximation only for much smaller interaction values,
        and it grossly overestimates the AFM fluctuations. To represent in our
        approximate approach the intermediate coupling regime of $U_0=4$, where $\pi
        $-tons were observed, we hence need to use an effective value of $U<4$ in the
        RPA ladder, as it is done e.g. in the TPSC approach~\cite
        {Vilk1994}.}\BibitemShut {Stop}%
    \bibitem [{Note3()}]{Note3}%
    \BibitemOpen
    \bibinfo {note} {The fit we use will no longer be valid for the low
        temperature regime, $T\lesssim 0.05$, where we expect the pseudogap behavior
        in the Hubbard model~\cite {Schaefer2015-2,schaefer2020tracking}
        .}\BibitemShut {Stop}%
    \bibitem [{\citenamefont {Berthod}\ \emph {et~al.}(2013)\citenamefont
        {Berthod}, \citenamefont {Mravlje}, \citenamefont {Deng}, \citenamefont
        {\ifmmode~\check{Z}\else \v{Z}\fi{}itko}, \citenamefont {van~der Marel},\
        and\ \citenamefont {Georges}}]{Berthod2013}%
    \BibitemOpen
    \bibfield  {author} {\bibinfo {author} {\bibfnamefont {C.}~\bibnamefont
            {Berthod}}, \bibinfo {author} {\bibfnamefont {J.}~\bibnamefont {Mravlje}},
        \bibinfo {author} {\bibfnamefont {X.}~\bibnamefont {Deng}}, \bibinfo {author}
        {\bibfnamefont {R.}~\bibnamefont {\ifmmode~\check{Z}\else \v{Z}\fi{}itko}},
        \bibinfo {author} {\bibfnamefont {D.}~\bibnamefont {van~der Marel}},\ and\
        \bibinfo {author} {\bibfnamefont {A.}~\bibnamefont {Georges}},\ }\bibfield
    {title} {\bibinfo {title} {Non-drude universal scaling laws for the optical
            response of local fermi liquids},\ }\href
    {https://doi.org/10.1103/PhysRevB.87.115109} {\bibfield  {journal} {\bibinfo
            {journal} {Phys. Rev. B}\ }\textbf {\bibinfo {volume} {87}},\ \bibinfo
        {pages} {115109} (\bibinfo {year} {2013})}\BibitemShut {NoStop}%
    \bibitem [{Note4()}]{Note4}%
    \BibitemOpen
    \bibinfo {note} {This relation holds for higher temperatures only
        approximately.}\BibitemShut {Stop}%
    \bibitem [{\citenamefont {Vrani\ifmmode~\acute{c}\else \'{c}\fi{}}\ \emph
        {et~al.}(2020)\citenamefont {Vrani\ifmmode~\acute{c}\else \'{c}\fi{}},
        \citenamefont {Vu\ifmmode \check{c}\else \v{c}\fi{}i\ifmmode \check{c}\else
            \v{c}\fi{}evi\ifmmode~\acute{c}\else \'{c}\fi{}}, \citenamefont {Kokalj},
        \citenamefont {Skolimowski}, \citenamefont {\ifmmode~\check{Z}\else
            \v{Z}\fi{}itko}, \citenamefont {Mravlje},\ and\ \citenamefont
        {Tanaskovi\ifmmode~\acute{c}\else \'{c}\fi{}}}]{Vranic2020}%
    \BibitemOpen
    \bibfield  {author} {\bibinfo {author} {\bibfnamefont {A.}~\bibnamefont
            {Vrani\ifmmode~\acute{c}\else \'{c}\fi{}}}, \bibinfo {author} {\bibfnamefont
            {J.}~\bibnamefont {Vu\ifmmode \check{c}\else \v{c}\fi{}i\ifmmode
                \check{c}\else \v{c}\fi{}evi\ifmmode~\acute{c}\else \'{c}\fi{}}}, \bibinfo
        {author} {\bibfnamefont {J.}~\bibnamefont {Kokalj}}, \bibinfo {author}
        {\bibfnamefont {J.}~\bibnamefont {Skolimowski}}, \bibinfo {author}
        {\bibfnamefont {R.}~\bibnamefont {\ifmmode~\check{Z}\else \v{Z}\fi{}itko}},
        \bibinfo {author} {\bibfnamefont {J.}~\bibnamefont {Mravlje}},\ and\ \bibinfo
        {author} {\bibfnamefont {D.}~\bibnamefont {Tanaskovi\ifmmode~\acute{c}\else
                \'{c}\fi{}}},\ }\bibfield  {title} {\bibinfo {title} {Charge transport in the
            hubbard model at high temperatures: Triangular versus square lattice},\
    }\href {https://doi.org/10.1103/PhysRevB.102.115142} {\bibfield  {journal}
        {\bibinfo  {journal} {Phys. Rev. B}\ }\textbf {\bibinfo {volume} {102}},\
        \bibinfo {pages} {115142} (\bibinfo {year} {2020})}\BibitemShut {NoStop}%
    \bibitem [{Note5()}]{Note5}%
    \BibitemOpen
    \bibinfo {note} {For clarity: the temperature enters in the RPA-ladder
        expression for the vertex,~Eq.~(\ref {eq:RPA_effective_vertex}), through the
        Fermi distribution function present in $\chi _0$.}\BibitemShut {Stop}%
    \bibitem [{\citenamefont {Greger}\ \emph {et~al.}(2013)\citenamefont {Greger},
        \citenamefont {Kollar},\ and\ \citenamefont {Vollhardt}}]{Greger2013}%
    \BibitemOpen
    \bibfield  {author} {\bibinfo {author} {\bibfnamefont {M.}~\bibnamefont
            {Greger}}, \bibinfo {author} {\bibfnamefont {M.}~\bibnamefont {Kollar}},\
        and\ \bibinfo {author} {\bibfnamefont {D.}~\bibnamefont {Vollhardt}},\
    }\bibfield  {title} {\bibinfo {title} {Isosbestic points: How a narrow
            crossing region of curves determines their leading parameter dependence},\
    }\href {https://doi.org/10.1103/PhysRevB.87.195140} {\bibfield  {journal}
        {\bibinfo  {journal} {Phys. Rev. B}\ }\textbf {\bibinfo {volume} {87}},\
        \bibinfo {pages} {195140} (\bibinfo {year} {2013})}\BibitemShut {NoStop}%
    \bibitem [{\citenamefont {Maki}(1968)}]{Maki1968}%
    \BibitemOpen
    \bibfield  {author} {\bibinfo {author} {\bibfnamefont {K.}~\bibnamefont
            {Maki}},\ }\bibfield  {title} {\bibinfo {title} {{Critical Fluctuation of the
                Order Parameter in a Superconductor. I}},\ }\href
    {https://doi.org/10.1143/PTP.40.193} {\bibfield  {journal} {\bibinfo
            {journal} {Progress of Theoretical Physics}\ }\textbf {\bibinfo {volume}
            {40}},\ \bibinfo {pages} {193} (\bibinfo {year} {1968})}\BibitemShut
    {NoStop}%
    \bibitem [{\citenamefont {Thompson}(1970)}]{Thompson1970}%
    \BibitemOpen
    \bibfield  {author} {\bibinfo {author} {\bibfnamefont {R.~S.}\ \bibnamefont
            {Thompson}},\ }\bibfield  {title} {\bibinfo {title} {Microwave, flux flow,
            and fluctuation resistance of dirty type-ii superconductors},\ }\href
    {https://doi.org/10.1103/PhysRevB.1.327} {\bibfield  {journal} {\bibinfo
            {journal} {Phys. Rev. B}\ }\textbf {\bibinfo {volume} {1}},\ \bibinfo {pages}
        {327} (\bibinfo {year} {1970})}\BibitemShut {NoStop}%
    \bibitem [{\citenamefont {Aslamazov}\ and\ \citenamefont
        {Larkin}(1968)}]{Aslamazov1968}%
    \BibitemOpen
    \bibfield  {author} {\bibinfo {author} {\bibfnamefont {L.~G.}\ \bibnamefont
            {Aslamazov}}\ and\ \bibinfo {author} {\bibfnamefont {A.~I.}\ \bibnamefont
            {Larkin}},\ }\bibfield  {title} {\bibinfo {title} {Effect of fluctuations on
            the properties of a superconductor above the critical temperature},\
    }\href@noop {} {\bibfield  {journal} {\bibinfo  {journal} {Sov. Phys. Solid
                State}\ }\textbf {\bibinfo {volume} {10}},\ \bibinfo {pages} {875} (\bibinfo
        {year} {1968})}\BibitemShut {NoStop}%
    \bibitem [{\citenamefont {Kontani}\ \emph {et~al.}(1999)\citenamefont
        {Kontani}, \citenamefont {Kanki},\ and\ \citenamefont {Ueda}}]{Kontani1999}%
    \BibitemOpen
    \bibfield  {author} {\bibinfo {author} {\bibfnamefont {H.}~\bibnamefont
            {Kontani}}, \bibinfo {author} {\bibfnamefont {K.}~\bibnamefont {Kanki}},\
        and\ \bibinfo {author} {\bibfnamefont {K.}~\bibnamefont {Ueda}},\ }\bibfield
    {title} {\bibinfo {title} {Hall effect and resistivity in high-${T}_{c}$
            superconductors: The conserving approximation},\ }\href
    {https://doi.org/10.1103/PhysRevB.59.14723} {\bibfield  {journal} {\bibinfo
            {journal} {Phys. Rev. B}\ }\textbf {\bibinfo {volume} {59}},\ \bibinfo
        {pages} {14723} (\bibinfo {year} {1999})}\BibitemShut {NoStop}%
    \bibitem [{\citenamefont {Kampf}\ and\ \citenamefont
        {Brenig}(1992)}]{Kampf1992}%
    \BibitemOpen
    \bibfield  {author} {\bibinfo {author} {\bibfnamefont {A.~P.}\ \bibnamefont
            {Kampf}}\ and\ \bibinfo {author} {\bibfnamefont {W.}~\bibnamefont {Brenig}},\
    }\bibfield  {title} {\bibinfo {title} {Raman scattering from
            antiferromagnetic spin fluctuations},\ }\href
    {https://doi.org/10.1007/BF01318162} {\bibfield  {journal} {\bibinfo
            {journal} {Zeitschrift f{\"u}r Physik B Condensed Matter}\ }\textbf {\bibinfo
            {volume} {89}},\ \bibinfo {pages} {313} (\bibinfo {year} {1992})}\BibitemShut
    {NoStop}%
    \bibitem [{Note6()}]{Note6}%
    \BibitemOpen
    \bibinfo {note} {The second order in $U$ diagram of the RPA ladder is of AL,
        the others are of MT type; there is no first order diagram.}\BibitemShut
    {Stop}%
    \bibitem [{Note7()}]{Note7}%
    \BibitemOpen
    \bibinfo {note} {For the functional renormalization group \cite
        {Metzner2012}, which is related to the parquet through its multiloop
        extensions~\cite {Kugler2017,Hille2020} a simplified quasiparticle
        description directly on the real axis has been recently proposed~\cite
        {Rohe2020}. For a p-D$\Gamma $A additionally a local vertex for real
        frequencies is needed, which one might obtain, e.g., through matrix product
        states~\cite {Evertz2017}}\BibitemShut {NoStop}%
    \bibitem [{Note8()}]{Note8}%
    \BibitemOpen
    \bibinfo {note} {Please note that within the RPA-approach using an effective
        interaction is necessary, and hence we cannot reliably treat different
        channels on an equal footing.}\BibitemShut {Stop}%
    \bibitem [{Note9()}]{Note9}%
    \BibitemOpen
    \bibinfo {note} {As explained in Eq.~(\ref
        {sec:methods_real_freq_vertex})}\BibitemShut {NoStop}%
    \bibitem [{\citenamefont {Sokhotskii}(1873)}]{Sokhotskii}%
    \BibitemOpen
    \bibfield  {author} {\bibinfo {author} {\bibfnamefont {Y.~W.}\ \bibnamefont
            {Sokhotskii}},\ }\emph {\bibinfo {title} {On definite integrals and functions
            used in series expansions}},\ \href@noop {} {Ph.D. thesis},\ \bibinfo
    {school} {St. Petersburg} (\bibinfo {year} {1873})\BibitemShut {NoStop}%
    \bibitem [{Note10()}]{Note10}%
    \BibitemOpen
    \bibinfo {note} {We present here the solution for $F^{\protect \textbf
            {q}}_{i \omega _n \rightarrow z \rightarrow 0} = F^{\protect \textbf {q}}_{i
            \omega _n =0}$. While this is true for the systems considered here (RPA,
        Ornstein-Zernike) it is not correct for some special cases (e.g. non-ergodic
        systems \cite {Wilcox_1968, Kwok_1969, Suzuki_1971}). The general procedure
        does, however, not change. The only difference being that the $\zeta _0$ term
        does no longer cancel. Instead one gets the $\zeta _0$ as an additional
        contribution, but with $\left ( F^{\protect \textbf {q}}_{i \omega _n =0} -
        F^{\protect \textbf {q}}_{i \omega _n \rightarrow z \rightarrow 0} \right )$
        instead of $F^{\protect \textbf {q}}_{0}$.}\BibitemShut {Stop}%
    \bibitem [{Note11()}]{Note11}%
    \BibitemOpen
    \bibinfo {note} {Particular strong scattering at the van Hove singularity
        will be cut-off by the finite scattering rate of the the starting
        self-energy}\BibitemShut {NoStop}%
    \bibitem [{\citenamefont {Rohringer}\ and\ \citenamefont
        {Toschi}(2016)}]{Rohringer2016}%
    \BibitemOpen
    \bibfield  {author} {\bibinfo {author} {\bibfnamefont {G.}~\bibnamefont
            {Rohringer}}\ and\ \bibinfo {author} {\bibfnamefont {A.}~\bibnamefont
            {Toschi}},\ }\bibfield  {title} {\bibinfo {title} {Impact of non-local
            correlations over different energy scales: A dynamical vertex approximation
            study},\ }\href {https://doi.org/10.1103/PhysRevB.94.125144} {\bibfield
        {journal} {\bibinfo  {journal} {Phys. Rev. B}\ }\textbf {\bibinfo {volume}
            {94}},\ \bibinfo {pages} {125144} (\bibinfo {year} {2016})}\BibitemShut
    {NoStop}%
    \bibitem [{\citenamefont {Honerkamp}(2018)}]{Honerkamp2018}%
    \BibitemOpen
    \bibfield  {author} {\bibinfo {author} {\bibfnamefont {C.}~\bibnamefont
            {Honerkamp}},\ }\bibfield  {title} {\bibinfo {title} {Efficient vertex
            parametrization for the constrained functional renormalization group for
            effective low-energy interactions in multiband systems},\ }\href
    {https://doi.org/10.1103/PhysRevB.98.155132} {\bibfield  {journal} {\bibinfo
            {journal} {Phys. Rev. B}\ }\textbf {\bibinfo {volume} {98}},\ \bibinfo
        {pages} {155132} (\bibinfo {year} {2018})}\BibitemShut {NoStop}%
    \bibitem [{\citenamefont {Shinaoka}\ \emph {et~al.}(2020)\citenamefont
        {Shinaoka}, \citenamefont {Geffroy}, \citenamefont {Wallerberger},
        \citenamefont {Otsuki}, \citenamefont {Yoshimi}, \citenamefont {Gull},\ and\
        \citenamefont {Kuneš}}]{Shinaoka2020}%
    \BibitemOpen
    \bibfield  {author} {\bibinfo {author} {\bibfnamefont {H.}~\bibnamefont
            {Shinaoka}}, \bibinfo {author} {\bibfnamefont {D.}~\bibnamefont {Geffroy}},
        \bibinfo {author} {\bibfnamefont {M.}~\bibnamefont {Wallerberger}}, \bibinfo
        {author} {\bibfnamefont {J.}~\bibnamefont {Otsuki}}, \bibinfo {author}
        {\bibfnamefont {K.}~\bibnamefont {Yoshimi}}, \bibinfo {author} {\bibfnamefont
            {E.}~\bibnamefont {Gull}},\ and\ \bibinfo {author} {\bibfnamefont
            {J.}~\bibnamefont {Kuneš}},\ }\bibfield  {title} {\bibinfo {title} {{Sparse
                sampling and tensor network representation of two-particle Green's
                functions}},\ }\href {https://doi.org/10.21468/SciPostPhys.8.1.012}
    {\bibfield  {journal} {\bibinfo  {journal} {SciPost Phys.}\ }\textbf
        {\bibinfo {volume} {8}},\ \bibinfo {pages} {12} (\bibinfo {year}
        {2020})}\BibitemShut {NoStop}%
    \bibitem [{\citenamefont {Li}\ \emph {et~al.}(2020)\citenamefont {Li},
        \citenamefont {Wallerberger}, \citenamefont {Chikano}, \citenamefont {Yeh},
        \citenamefont {Gull},\ and\ \citenamefont {Shinaoka}}]{Li2020}%
    \BibitemOpen
    \bibfield  {author} {\bibinfo {author} {\bibfnamefont {J.}~\bibnamefont
            {Li}}, \bibinfo {author} {\bibfnamefont {M.}~\bibnamefont {Wallerberger}},
        \bibinfo {author} {\bibfnamefont {N.}~\bibnamefont {Chikano}}, \bibinfo
        {author} {\bibfnamefont {C.-N.}\ \bibnamefont {Yeh}}, \bibinfo {author}
        {\bibfnamefont {E.}~\bibnamefont {Gull}},\ and\ \bibinfo {author}
        {\bibfnamefont {H.}~\bibnamefont {Shinaoka}},\ }\bibfield  {title} {\bibinfo
        {title} {Sparse sampling approach to efficient ab initio calculations at
            finite temperature},\ }\href {https://doi.org/10.1103/PhysRevB.101.035144}
    {\bibfield  {journal} {\bibinfo  {journal} {Phys. Rev. B}\ }\textbf {\bibinfo
            {volume} {101}},\ \bibinfo {pages} {035144} (\bibinfo {year}
        {2020})}\BibitemShut {NoStop}%
    \bibitem [{\citenamefont {Eckhardt}\ \emph {et~al.}(2020)\citenamefont
        {Eckhardt}, \citenamefont {Honerkamp}, \citenamefont {Held},\ and\
        \citenamefont {Kauch}}]{Eckhardt2020}%
    \BibitemOpen
    \bibfield  {author} {\bibinfo {author} {\bibfnamefont {C.~J.}\ \bibnamefont
            {Eckhardt}}, \bibinfo {author} {\bibfnamefont {C.}~\bibnamefont {Honerkamp}},
        \bibinfo {author} {\bibfnamefont {K.}~\bibnamefont {Held}},\ and\ \bibinfo
        {author} {\bibfnamefont {A.}~\bibnamefont {Kauch}},\ }\bibfield  {title}
    {\bibinfo {title} {Truncated unity parquet solver},\ }\href
    {https://doi.org/10.1103/PhysRevB.101.155104} {\bibfield  {journal} {\bibinfo
            {journal} {Phys. Rev. B}\ }\textbf {\bibinfo {volume} {101}},\ \bibinfo
        {pages} {155104} (\bibinfo {year} {2020})}\BibitemShut {NoStop}%
    \bibitem [{\citenamefont {Krien}\ \emph {et~al.}(2021)\citenamefont {Krien},
        \citenamefont {Kauch},\ and\ \citenamefont {Held}}]{Krien2020}%
    \BibitemOpen
    \bibfield  {author} {\bibinfo {author} {\bibfnamefont {F.}~\bibnamefont
            {Krien}}, \bibinfo {author} {\bibfnamefont {A.}~\bibnamefont {Kauch}},\ and\
        \bibinfo {author} {\bibfnamefont {K.}~\bibnamefont {Held}},\ }\bibfield
    {title} {\bibinfo {title} {Tiling with triangles: parquet and
            $gw\ensuremath{\gamma}$ methods unified},\ }\href
    {https://doi.org/10.1103/PhysRevResearch.3.013149} {\bibfield  {journal}
        {\bibinfo  {journal} {Phys. Rev. Research}\ }\textbf {\bibinfo {volume}
            {3}},\ \bibinfo {pages} {013149} (\bibinfo {year} {2021})}\BibitemShut
    {NoStop}%
    \bibitem [{\citenamefont {Vilk}\ \emph {et~al.}(1994)\citenamefont {Vilk},
        \citenamefont {Chen},\ and\ \citenamefont {Tremblay}}]{Vilk1994}%
    \BibitemOpen
    \bibfield  {author} {\bibinfo {author} {\bibfnamefont {Y.~M.}\ \bibnamefont
            {Vilk}}, \bibinfo {author} {\bibfnamefont {L.}~\bibnamefont {Chen}},\ and\
        \bibinfo {author} {\bibfnamefont {A.-M.~S.}\ \bibnamefont {Tremblay}},\
    }\bibfield  {title} {\bibinfo {title} {Theory of spin and charge fluctuations
            in the hubbard model},\ }\href {https://doi.org/10.1103/PhysRevB.49.13267}
    {\bibfield  {journal} {\bibinfo  {journal} {Phys. Rev. B}\ }\textbf {\bibinfo
            {volume} {49}},\ \bibinfo {pages} {13267} (\bibinfo {year}
        {1994})}\BibitemShut {NoStop}%
    \bibitem [{\citenamefont {Sch\"afer}\ \emph {et~al.}(2015)\citenamefont
        {Sch\"afer}, \citenamefont {Geles}, \citenamefont {Rost}, \citenamefont
        {Rohringer}, \citenamefont {Arrigoni}, \citenamefont {Held}, \citenamefont
        {Bl\"umer}, \citenamefont {Aichhorn},\ and\ \citenamefont
        {Toschi}}]{Schaefer2015-2}%
    \BibitemOpen
    \bibfield  {author} {\bibinfo {author} {\bibfnamefont {T.}~\bibnamefont
            {Sch\"afer}}, \bibinfo {author} {\bibfnamefont {F.}~\bibnamefont {Geles}},
        \bibinfo {author} {\bibfnamefont {D.}~\bibnamefont {Rost}}, \bibinfo {author}
        {\bibfnamefont {G.}~\bibnamefont {Rohringer}}, \bibinfo {author}
        {\bibfnamefont {E.}~\bibnamefont {Arrigoni}}, \bibinfo {author}
        {\bibfnamefont {K.}~\bibnamefont {Held}}, \bibinfo {author} {\bibfnamefont
            {N.}~\bibnamefont {Bl\"umer}}, \bibinfo {author} {\bibfnamefont
            {M.}~\bibnamefont {Aichhorn}},\ and\ \bibinfo {author} {\bibfnamefont
            {A.}~\bibnamefont {Toschi}},\ }\bibfield  {title} {\bibinfo {title} {Fate of
            the false mott-hubbard transition in two dimensions},\ }\href
    {https://doi.org/10.1103/PhysRevB.91.125109} {\bibfield  {journal} {\bibinfo
            {journal} {Phys. Rev. B}\ }\textbf {\bibinfo {volume} {91}},\ \bibinfo
        {pages} {125109} (\bibinfo {year} {2015})}\BibitemShut {NoStop}%
    \bibitem [{\citenamefont {Sch\"afer}\ \emph {et~al.}(2021)\citenamefont
        {Sch\"afer}, \citenamefont {Wentzell}, \citenamefont {\ifmmode~\check{S}\else
            \v{S}\fi{}imkovic}, \citenamefont {He}, \citenamefont {Hille}, \citenamefont
        {Klett}, \citenamefont {Eckhardt}, \citenamefont {Arzhang}, \citenamefont
        {Harkov}, \citenamefont {Le~R\'egent}, \citenamefont {Kirsch}, \citenamefont
        {Wang}, \citenamefont {Kim}, \citenamefont {Kozik}, \citenamefont {Stepanov},
        \citenamefont {Kauch}, \citenamefont {Andergassen}, \citenamefont {Hansmann},
        \citenamefont {Rohe}, \citenamefont {Vilk}, \citenamefont {LeBlanc},
        \citenamefont {Zhang}, \citenamefont {Tremblay}, \citenamefont {Ferrero},
        \citenamefont {Parcollet},\ and\ \citenamefont
        {Georges}}]{schaefer2020tracking}%
    \BibitemOpen
    \bibfield  {author} {\bibinfo {author} {\bibfnamefont {T.}~\bibnamefont
            {Sch\"afer}}, \bibinfo {author} {\bibfnamefont {N.}~\bibnamefont {Wentzell}},
        \bibinfo {author} {\bibfnamefont {F.}~\bibnamefont {\ifmmode~\check{S}\else
                \v{S}\fi{}imkovic}}, \bibinfo {author} {\bibfnamefont {Y.-Y.}\ \bibnamefont
            {He}}, \bibinfo {author} {\bibfnamefont {C.}~\bibnamefont {Hille}}, \bibinfo
        {author} {\bibfnamefont {M.}~\bibnamefont {Klett}}, \bibinfo {author}
        {\bibfnamefont {C.~J.}\ \bibnamefont {Eckhardt}}, \bibinfo {author}
        {\bibfnamefont {B.}~\bibnamefont {Arzhang}}, \bibinfo {author} {\bibfnamefont
            {V.}~\bibnamefont {Harkov}}, \bibinfo {author} {\bibfnamefont {F.~m. c.-M.}\
            \bibnamefont {Le~R\'egent}}, \bibinfo {author} {\bibfnamefont
            {A.}~\bibnamefont {Kirsch}}, \bibinfo {author} {\bibfnamefont
            {Y.}~\bibnamefont {Wang}}, \bibinfo {author} {\bibfnamefont {A.~J.}\
            \bibnamefont {Kim}}, \bibinfo {author} {\bibfnamefont {E.}~\bibnamefont
            {Kozik}}, \bibinfo {author} {\bibfnamefont {E.~A.}\ \bibnamefont {Stepanov}},
        \bibinfo {author} {\bibfnamefont {A.}~\bibnamefont {Kauch}}, \bibinfo
        {author} {\bibfnamefont {S.}~\bibnamefont {Andergassen}}, \bibinfo {author}
        {\bibfnamefont {P.}~\bibnamefont {Hansmann}}, \bibinfo {author}
        {\bibfnamefont {D.}~\bibnamefont {Rohe}}, \bibinfo {author} {\bibfnamefont
            {Y.~M.}\ \bibnamefont {Vilk}}, \bibinfo {author} {\bibfnamefont {J.~P.~F.}\
            \bibnamefont {LeBlanc}}, \bibinfo {author} {\bibfnamefont {S.}~\bibnamefont
            {Zhang}}, \bibinfo {author} {\bibfnamefont {A.-M.~S.}\ \bibnamefont
            {Tremblay}}, \bibinfo {author} {\bibfnamefont {M.}~\bibnamefont {Ferrero}},
        \bibinfo {author} {\bibfnamefont {O.}~\bibnamefont {Parcollet}},\ and\
        \bibinfo {author} {\bibfnamefont {A.}~\bibnamefont {Georges}},\ }\bibfield
    {title} {\bibinfo {title} {Tracking the footprints of spin fluctuations: A
            multimethod, multimessenger study of the two-dimensional hubbard model},\
    }\href {https://doi.org/10.1103/PhysRevX.11.011058} {\bibfield  {journal}
        {\bibinfo  {journal} {Phys. Rev. X}\ }\textbf {\bibinfo {volume} {11}},\
        \bibinfo {pages} {011058} (\bibinfo {year} {2021})}\BibitemShut {NoStop}%
    \bibitem [{\citenamefont {Metzner}\ \emph {et~al.}(2012)\citenamefont
        {Metzner}, \citenamefont {Salmhofer}, \citenamefont {Honerkamp},
        \citenamefont {Meden},\ and\ \citenamefont {Sch\"onhammer}}]{Metzner2012}%
    \BibitemOpen
    \bibfield  {author} {\bibinfo {author} {\bibfnamefont {W.}~\bibnamefont
            {Metzner}}, \bibinfo {author} {\bibfnamefont {M.}~\bibnamefont {Salmhofer}},
        \bibinfo {author} {\bibfnamefont {C.}~\bibnamefont {Honerkamp}}, \bibinfo
        {author} {\bibfnamefont {V.}~\bibnamefont {Meden}},\ and\ \bibinfo {author}
        {\bibfnamefont {K.}~\bibnamefont {Sch\"onhammer}},\ }\bibfield  {title}
    {\bibinfo {title} {Functional renormalization group approach to correlated
            fermion systems},\ }\href {https://doi.org/10.1103/RevModPhys.84.299}
    {\bibfield  {journal} {\bibinfo  {journal} {Rev. Mod. Phys.}\ }\textbf
        {\bibinfo {volume} {84}},\ \bibinfo {pages} {299} (\bibinfo {year}
        {2012})}\BibitemShut {NoStop}%
    \bibitem [{\citenamefont {Kugler}\ and\ \citenamefont {von
            Delft}(2018)}]{Kugler2017}%
    \BibitemOpen
    \bibfield  {author} {\bibinfo {author} {\bibfnamefont {F.~B.}\ \bibnamefont
            {Kugler}}\ and\ \bibinfo {author} {\bibfnamefont {J.}~\bibnamefont {von
                Delft}},\ }\bibfield  {title} {\bibinfo {title} {Multiloop functional
            renormalization group that sums up all parquet diagrams},\ }\href
    {https://doi.org/10.1103/PhysRevLett.120.057403} {\bibfield  {journal}
        {\bibinfo  {journal} {Phys. Rev. Lett.}\ }\textbf {\bibinfo {volume} {120}},\
        \bibinfo {pages} {057403} (\bibinfo {year} {2018})}\BibitemShut {NoStop}%
    \bibitem [{\citenamefont {Hille}\ \emph {et~al.}(2020)\citenamefont {Hille},
        \citenamefont {Kugler}, \citenamefont {Eckhardt}, \citenamefont {He},
        \citenamefont {Kauch}, \citenamefont {Honerkamp}, \citenamefont {Toschi},\
        and\ \citenamefont {Andergassen}}]{Hille2020}%
    \BibitemOpen
    \bibfield  {author} {\bibinfo {author} {\bibfnamefont {C.}~\bibnamefont
            {Hille}}, \bibinfo {author} {\bibfnamefont {F.~B.}\ \bibnamefont {Kugler}},
        \bibinfo {author} {\bibfnamefont {C.~J.}\ \bibnamefont {Eckhardt}}, \bibinfo
        {author} {\bibfnamefont {Y.-Y.}\ \bibnamefont {He}}, \bibinfo {author}
        {\bibfnamefont {A.}~\bibnamefont {Kauch}}, \bibinfo {author} {\bibfnamefont
            {C.}~\bibnamefont {Honerkamp}}, \bibinfo {author} {\bibfnamefont
            {A.}~\bibnamefont {Toschi}},\ and\ \bibinfo {author} {\bibfnamefont
            {S.}~\bibnamefont {Andergassen}},\ }\bibfield  {title} {\bibinfo {title}
        {Quantitative functional renormalization group description of the
            two-dimensional hubbard model},\ }\href
    {https://doi.org/10.1103/PhysRevResearch.2.033372} {\bibfield  {journal}
        {\bibinfo  {journal} {Phys. Rev. Research}\ }\textbf {\bibinfo {volume}
            {2}},\ \bibinfo {pages} {033372} (\bibinfo {year} {2020})}\BibitemShut
    {NoStop}%
    \bibitem [{\citenamefont {Rohe}\ and\ \citenamefont
        {Honerkamp}(2020)}]{Rohe2020}%
    \BibitemOpen
    \bibfield  {author} {\bibinfo {author} {\bibfnamefont {D.}~\bibnamefont
            {Rohe}}\ and\ \bibinfo {author} {\bibfnamefont {C.}~\bibnamefont
            {Honerkamp}},\ }\bibfield  {title} {\bibinfo {title} {{Quasi-particle
                functional Renormalisation Group calculations in the two-dimensional
                half-filled Hubbard model at finite temperatures}},\ }\href
    {https://doi.org/10.21468/SciPostPhys.9.6.084} {\bibfield  {journal}
        {\bibinfo  {journal} {SciPost Phys.}\ }\textbf {\bibinfo {volume} {9}},\
        \bibinfo {pages} {84} (\bibinfo {year} {2020})}\BibitemShut {NoStop}%
    \bibitem [{\citenamefont {Bauernfeind}\ \emph {et~al.}(2017)\citenamefont
        {Bauernfeind}, \citenamefont {Zingl}, \citenamefont {Triebl}, \citenamefont
        {Aichhorn},\ and\ \citenamefont {Evertz}}]{Evertz2017}%
    \BibitemOpen
    \bibfield  {author} {\bibinfo {author} {\bibfnamefont {D.}~\bibnamefont
            {Bauernfeind}}, \bibinfo {author} {\bibfnamefont {M.}~\bibnamefont {Zingl}},
        \bibinfo {author} {\bibfnamefont {R.}~\bibnamefont {Triebl}}, \bibinfo
        {author} {\bibfnamefont {M.}~\bibnamefont {Aichhorn}},\ and\ \bibinfo
        {author} {\bibfnamefont {H.~G.}\ \bibnamefont {Evertz}},\ }\bibfield  {title}
    {\bibinfo {title} {Fork tensor-product states: Efficient multiorbital
            real-time dmft solver},\ }\href {https://doi.org/10.1103/PhysRevX.7.031013}
    {\bibfield  {journal} {\bibinfo  {journal} {Phys. Rev. X}\ }\textbf {\bibinfo
            {volume} {7}},\ \bibinfo {pages} {031013} (\bibinfo {year}
        {2017})}\BibitemShut {NoStop}%
    \bibitem [{\citenamefont {Wilcox}(1968)}]{Wilcox_1968}%
    \BibitemOpen
    \bibfield  {author} {\bibinfo {author} {\bibfnamefont {R.~M.}\ \bibnamefont
            {Wilcox}},\ }\bibfield  {title} {\bibinfo {title} {Bounds for the isothermal,
            adiabatic, and isolated static susceptibility tensors},\ }\href
    {https://doi.org/10.1103/PhysRev.174.624} {\bibfield  {journal} {\bibinfo
            {journal} {Phys. Rev.}\ }\textbf {\bibinfo {volume} {174}},\ \bibinfo {pages}
        {624} (\bibinfo {year} {1968})}\BibitemShut {NoStop}%
    \bibitem [{\citenamefont {Kwok}\ and\ \citenamefont
        {Schultz}(1969)}]{Kwok_1969}%
    \BibitemOpen
    \bibfield  {author} {\bibinfo {author} {\bibfnamefont {P.~C.}\ \bibnamefont
            {Kwok}}\ and\ \bibinfo {author} {\bibfnamefont {T.~D.}\ \bibnamefont
            {Schultz}},\ }\bibfield  {title} {\bibinfo {title} {Correlation functions and
            green functions: zero-frequency anomalies},\ }\href
    {https://doi.org/10.1088/0022-3719/2/7/312} {\bibfield  {journal} {\bibinfo
            {journal} {Journal of Physics C: Solid State Physics}\ }\textbf {\bibinfo
            {volume} {2}},\ \bibinfo {pages} {1196} (\bibinfo {year} {1969})}\BibitemShut
    {NoStop}%
    \bibitem [{\citenamefont {Suzuki}(1971)}]{Suzuki_1971}%
    \BibitemOpen
    \bibfield  {author} {\bibinfo {author} {\bibfnamefont {M.}~\bibnamefont
            {Suzuki}},\ }\bibfield  {title} {\bibinfo {title} {Ergodicity, constants of
            motion, and bounds for susceptibilities},\ }\href
    {https://doi.org/https://doi.org/10.1016/0031-8914(71)90226-6} {\bibfield
        {journal} {\bibinfo  {journal} {Physica}\ }\textbf {\bibinfo {volume} {51}},\
        \bibinfo {pages} {277 } (\bibinfo {year} {1971})}\BibitemShut {NoStop}%
\end{thebibliography}
\end{document}